\documentclass[a4paper,11pt,twocolumns]{article}

\usepackage{authblk}
\usepackage{graphicx}
\usepackage{mathtools}
\usepackage{amsmath}
\usepackage{amssymb}
\usepackage{bm}
\usepackage{xcolor}
\usepackage{epstopdf}
\usepackage{float}
\usepackage{comment}
\usepackage{soul}
\usepackage{dcolumn}
\usepackage{hyperref}
\usepackage[normalem]{ulem}

\title{Conceptual and practical approaches for investigating irreversible processes}

\author[1]{Dario Lucente}
\author[2]{Marco Baldovin}
\author[2,3]{Fabio Cecconi}
\author[2]{Massimo Cencini}
\author[4]{Niccolò Cocciaglia}
\author[2,5]{Andrea Puglisi}
\author[2,3]{Massimiliano Viale}
\author[4]{Angelo Vulpiani}

\affil[1]{Department of Mathematics \& Physics\\University of Campania “Luigi Vanvitelli”, Viale Lincoln 5, 81100 Caserta, Italy}
\affil[2]{Institute for Complex Systems CNR, P.le Aldo Moro 2, 00185, Rome, Italy}
\affil[3]{INFN, Sezione Roma1, P.le Aldo Moro, 2 00185 Rome, Italy}
\affil[4]{Department of Physics\\University of Rome “La Sapienza”, P.le Aldo Moro 5, 00185 Rome, Italy}
\affil[5]{INFN, Sezione Roma2, Via della Ricerca Scientifica 1, 00133, Rome, Italy}

\date{}

\begin{document}

\newcommand{\eref}[1]{Eq.~\eqref{#1}}
\newcommand{\fref}[1]{Fig.~\ref{#1}}

\newcommand{\lx} {\left}
\newcommand{\rx} {\right}
\newcommand{\eps} {\epsilon}
\newcommand{\veps} {\varepsilon}
\newcommand{\ave}[1] {\lx\langle #1 \rx\rangle}
\newcommand{\be} {\begin{equation*}}
\newcommand{\ee} {\end{equation*}}
\newcommand{\bea} {\begin{eqnarray*}}
\newcommand{\eea} {\end{eqnarray*}}
\newcommand{\bX} {\mathbf{X}}
\newcommand{\bY} {\mathbf{Y}}
\newcommand{\bU} {\mathbf{U}}
\newcommand{\bu} {\mathbf{u}}
\newcommand{\bx} {\mathbf{x}}
\newcommand{\by} {\mathbf{y}}
\newcommand{\bz} {\mathbf{z}}
\newcommand{\bh} {\mathbf{h}}
\newcommand{\bv} {\mathbf{v}}
\newcommand{\bA} {\mathbf{A}}
\newcommand{\bB} {\mathbf{B}}
\newcommand{\bsg} {\bm{\sigma}}
\newcommand{\bxi} {\bm{\xi}}
\newcommand{\bfeta} {\bm{\eta}}
\newcommand{\bsigma} {\bm{\sigma}}
\newcommand{\bzeta} {\bm{\zeta}}
\newcommand{\cL} {\mathcal{L}}
\newcommand{\cB} {\mathcal{B}}
\newcommand{\cZ} {\mathcal{Z}}
\newcommand{\cD} {\mathcal{D}}
\newcommand{\cE} {\mathcal{E}}
\newcommand{\cF} {\mathcal{F}}
\newcommand{\cJ} {\mathcal{J}}
\newcommand{\cH} {\mathcal{H}}
\newcommand{\cI} {\mathcal{I}}
\newcommand{\cK} {\mathcal{K}}
\newcommand{\cM} {\mathcal{M}}
\newcommand{\cW} {\mathcal{W}}
\newcommand{\cP} {\mathcal{P}}
\newcommand{\cN} {\mathcal{N}}
\newcommand{\cO} {\mathcal{O}}
\newcommand{\cT} {\mathcal{T}}
\newcommand{\cQ} {\mathcal{Q}}
\newcommand{\cU} {\mathcal{U}}
\newcommand{\cV} {\mathcal{V}}
\newcommand{\cS} {\mathcal{S}}
\newcommand{\Tdir} {\mathbf{X}_\rightarrow^{(\cT)}}
\newcommand{\Tinv} {\mathbf{X}_\leftarrow^{(\cT)}}

\newcommand{\xia}[1] {\xi^{(#1)}}
\newcommand{\xiaf}[1] {\widetilde{\xi}^{(#1)}}
\newcommand{\xiafo}[1] {\xiaf{#1}(\om)}
\newcommand{\xif} {\widetilde{\xi}}
\newcommand{\xifo} {\xif(\om)}
\newcommand{\om} {\omega}
\newcommand{\xf} {\widetilde{x}}
\newcommand{\xfo} {\xf(\om)}
\newcommand{\cLf} {\widetilde{\cL}}
\newcommand{\cLfo} {\cLf(\om)}
\newcommand{\cR} {\mathcal{R}}
\newcommand{\cRf} {\widetilde{\cR}}
\newcommand{\cRfo} {\cRf(\om)}
\newcommand{\cG} {\mathcal{G}}
\newcommand{\cGf} {\widetilde{\cG}}
\newcommand{\cGfo} {\cGf(\om)}
\newcommand{\cA} {\mathcal{A}}
\newcommand{\cAf} {\widetilde{\cA}}
\newcommand{\cAfo} {\cAf(\om)}
\newcommand{\cBf} {\widetilde{\cB}}
\newcommand{\cBfo} {\cBf(\om)}
\newcommand{\cBa}[1] {\mathcal{B}^{(#1)}}
\newcommand{\cBaf}[1] {\widetilde{\cB}^{(#1)}}
\newcommand{\cBafo}[1] {\cBaf{#1}(\om)}
\newcommand{\cGa}[1] {\mathcal{G}^{(#1)}}
\newcommand{\cGaf}[1] {\widetilde{\cG}^{(#1)}}
\newcommand{\cGafo}[1] {\cGaf{#1}(\om)}
\newcommand{\cSgf} {\widetilde{\Sigma}}
\newcommand{\cSgfo} {\cSgf(\om)}
\newcommand{\cC} {\mathcal{C}}
\newcommand{\cCf} {\widetilde{\cC}}
\newcommand{\cCfo} {\cCf(\om)}
\newcommand{\bQ} {\mathbf{Q}}
\newcommand{\bF} {\mathbf{F}}
\newcommand{\bG} {\mathbf{G}}
\newcommand{\bn} {\mathbf{n}}
\newcommand{\br} {\mathbf{r}}
\newcommand {\Pab} {\mathbf{G}}
\newcommand {\pa} {\bm{\Pi}}
\newcommand {\one} {\mathbf{1}}
\newcommand {\sa} {a}
\newcommand {\bsa} {\bm{a}}
\newcommand {\obsa} {\overleftarrow{\bm{a}}}
\newcommand {\osa} {\overline{a}}
\newcommand {\osga} {\overline{a'}}
\newcommand {\usa} {\underline{a}}
\newcommand{\av}[1]{\left\langle#1\right\rangle}
\newcommand{\cbr}[1]{\left(#1\right)}
\newcommand{\sbr}[1]{\left[#1\right]}
\newcommand{\bbr}[1]{\left\{#1\right\}}
\newcommand{\new}[1]{\textcolor{blue}{#1}}
\newcommand{\old}[1]{\textcolor{black}{\sout{#1}}}

\newcommand{\matx}[1] {\mathrm{#1}}

\DeclareRobustCommand{\vect}[1]{
  \ifcat#1\relax
    \boldsymbol{#1}
  \else
    \mathbf{#1}
  \fi}

\maketitle
\begin{abstract}
Current research in statistical mechanics mostly concerns the investigation of out-of-equilibrium, irreversible processes, which are ubiquitous in nature and still far from being theoretically understood. Even the precise characterization of irreversibility is the object of an open debate: while in the context of Hamiltonian systems the one-century-old proposal by M. Smoluchowski looks still valid ({\it a process appears irreversible when the initial state has a recurrence time that is long compared to the time of observation} \cite{Smoluchowski1916}), in dissipative systems, particularly in the case of stochastic processes, the problem is more involved, and quantifying the ``degree of irreversibility'' is a pragmatic need. The most employed strategies rely on the estimation of entropy production: this quantity, although mathematically well-defined, is often difficult to compute, especially when analyzing experimental data. Moreover, being a global observable, entropy production fails to capture specific aspects of irreversibility in extended systems, such as the role of different currents and their spatial development.
This review aims to address various conceptual and technical challenges encountered in the analysis of irreversibility, including the role of the coarse-graining procedure and the treatment of data in the absence of complete information. The discussion will be mostly based on simple models, analytically treatable, and supplemented by examples of complex, more realistic non-equilibrium systems.
\end{abstract}
\begin{flushright}
{\it At any time there is only a thin layer separating \\ what is trivial from what is impossibly difficult. \\ It is in that layer that discoveries are made...}\\
(Andrei N. Kolmogorov)\\
\end{flushright}

\section{Introduction} \label{sec:introduction}
Understanding non-equilibrium phenomena stands as a key frontier in
modern statistical physics \cite{degroot84,KTH91,livi2017nonequilibrium,maes2003time,esposito2010entropy,bialek2012statistical}. This
field has emerged from two distinct objectives. Firstly, there is the
extensive effort, rooted in Boltzmann's seminal contributions, to
address the long-standing and challenging problem of linking the
irreversible behavior of the macroscopic world with the reversible
microscopic dynamics governed by Hamilton’s equations
\cite{cercignani1998ludwig}. In addition, there is a practical need to
characterize the numerous non-equilibrium (irreversible) phenomena
pervasive in science, particularly in physics and chemistry, as well
as in various applications. Examples of such phenomena include
transport, diffusion, and thermomechanical effects
\cite{kreuzer1981nonequilibrium}.

One of the most important characteristics of out-of-equilibrium
systems is the presence of currents induced by external constraints or
fields, which lead to the breaking of time-reversal symmetry and
inhomogeneites between the system degrees of freedom (e.g. spatial
inhomogeneities in extended systems). Mathematically, we can say that
a system is considered out of equilibrium if detailed balance does not
hold, or equivalently, if the entropy production rate $\Sigma$ is positive
\cite{ruelle1996positivity,Sekimoto2010,seifertrev,pelipigo}. However,
entropy production is difficult to measure in real systems, and
can be explicitly calculated only for Markov processes, such as those
described by Langevin and Master equations~\cite{pelipigo}.  On
the other hand, even when the entropy production is known, being a
global quantity, it does not directly inform us about the physical
currents between degrees of freedom ~\cite{sarra2021response}.  For
instance, even a system of linear Langevin equations can have a
nonzero entropy production and be consistently classified as out of
equilibrium; in such a case, due to the absence of a spatial
structure, it is not trivial to individuate the currents
\cite{crisanti2012nonequilibrium, lucente2023revealing}.  Furthermore, in general, quantifying
entropy production requires detailed knowledge of the underlying
system, including all its degrees of freedom and their interactions~\cite{lucente2023revealing}. Conversely, in many practical situations, we only
have access to a limited set of observables, representing a projection
or coarse-graining of the system. Thus, it is natural to wonder in
which cases and how we can infer non-equilibrium properties and
characterize the breaking of time-reversal symmetry using such partial
information.

In this review, we discuss various aspects of a broad class of
non-equilibrium systems, from Markov chains to high-dimensional
chaotic systems, analyzing both temporal and spatial aspects of
non-equilibrium states. Our aim is not to provide an exhaustive
compendium of the many facets of non-equilibrium statistical
mechanics. Instead, we present a series of observations and ideas on
specific aspects that need to be addressed in everyday research.
We focus on the practical challenges and difficulties in
characterizing the breaking of temporal symmetry, its connection with
spatial structures, and the physical characteristics of the underlying
system. We will discuss the use of several tools to this end, such as
correlation functions, particularly those suited to detect
asymmetries \cite{pomeau1982symetrie,Joss}, response theory and fluctuation-dissipation relations \cite{ruelle1998,kubo1966fluctuation,FDreport},
recently introduced thermodynamic uncertainty relations \cite{barato2015thermodynamic,seifert2018stochastic}, and causation
analysis to characterize the irreversibility associated with
non-reciprocal interactions between degrees of freedom \cite{aurell2016causal,baldovin2020understanding}.

In the first part, we primarily focus on analytically treatable models
(mostly in the context of Markov processes), which allow us to introduce the
main tools of analysis and highlight key problems and subtleties
arising from coarse-graining and the lack of complete information in
data analysis.  In particular, we will pay attention to the consequences of
non-Gaussian perturbations, which can be relevant in small systems,
and the necessity of a finite scale resolution.  We will then
progressively consider more complex systems, drawing examples from
high-dimensional, spatially extended chaotic systems to realistic
simulations of turbulent flows \cite{Xu2014flightcrash,cencini2017time} or models for turbulence, where
non-equilibrium properties and irreversibility manifest over multiple
spatio-temporal scales \cite{cocciaglia2024nonequilibrium}.

\textcolor{black}{The paper is organized as follows.  Section \ref{sec:general}, starting from the distinction between transient and persistent out-of-equilibrium states,
summarises some fundamental aspects of non-equilibrium statistical mechanics. In particular, we discuss the conceptual ingredients that are needed to a suitable thermodynamic description of the system, such as the presence of many degrees of freedom, typicality and coarse-graining. Some difficulties encountered in the understanding of experimental data
and/or numerical calculations are also outlined. Finally, we introduce the  indicators that are most commonly used to quantify time-reversal symmetry breaking, namely time-asymmetric correlation functions and entropy production rate. 
Section \ref{sec:markov} is devoted
to Markov processes, with a focus on linear stochastic systems and
jump processes. Through a systematic use of the theory of stochastic processes, a general and well-defined mathematical formulation of systems in and out of equilibrium is provided. Fluctuation-Dissipation Theorems and their relations with equilibrium properties are extensively covered. Then, the problem of inferring the thermodynamic properties of a system from partial information is discussed in analytically treatable models. 
On the one hand, a no-go theorem stating the impossibility of inferring the equilibrium properties of a system by measuring a single stationary degree of freedom, valid for Gaussian processes, is discussed.
On the other hand, it is shown that a generalized response allowing the understanding of the thermodynamic nature of the underlying model can be computed by comparing experiments performed under different conditions.
Section
\ref{sec:empirical} focuses on general strategies for the estimation of
entropy production from data. More specifically, we discuss the thermodynamic uncertainty relations, which are lower bounds on entropy production obtained by analyzing the signal to noise ratio of a generic current, as
well as some numerical brute-force techniques. Conceptual subtleties such as the dependence on coarse-graining levels, or on the observed currents of these empirical proxies, are carefully scrutinized in simple models: this study shows that useful information can be usually obtained only in cases where a good understanding of the system is available {\it ab origine}. Finally, other approaches based on the exit-time statistics or hidden Markov modeling are briefly discussed. 
Section \ref{sec:causation} goes beyond entropy production rate examining the potential implications of causation indicators in the analysis of non-equilibrium systems.  The two causal indicators taken into account are transfer entropy, which is an information-theoretic measure of information fluxes between variables, and response function. As discussed there, the advantage of considering these indicators is that they not only discriminate between equilibrium and out-of-equilibrium systems, but also provide information on how the time-reversal symmetry is broken. Section \ref{sec:turbulence} discusses non-equilibrium in
turbulence: it is a remarkable case study where all the previous theoretical considerations naturally apply. In particular, it is shown that both correlations and responses of suitably defined observables not only are able to highlight the non-equilibrium nature of particles advected by turbulent velocity field, but also reveal other interesting aspects such as energy and/or enstrophy cascades. Finally, some conclusions are drawn in Section \ref{sec:conclusion}.}

\section{Irreversibility in transient and persistent  non-equilibrium phenomena 
\label{sec:general}}

When discussing non-equilibrium phenomena (NEP), a first distinction can be identified between {\em transient} and {\em persistent} non-equilibrium. The former displays a non-equilibrium behaviour for a limited amount of time only, before reaching their final equilibrium state, where the dynamics is reversible. Conversely, in persistent NEP
time reversal symmetry is continuously broken\footnote{We prefer to use the term ``persistent'' instead of ``stationary'', which is not completely general. For instance, an electric noisy circuit with a periodic forcing is in a persistent, but clearly not stationary, non-equilibrium state.}. They are kept out of equilibrium indefinitely (at least, with respect to the experimentally accessible times) by external drivings, or by intrinsically time-irreversible internal mechanisms. 

A prototypical example of transient NEP is gas diffusion. 
Consider a large number $N \gg 1$ of
particles, initially localized in a small region of the
available volume $V \sim L^d$ of a $d$-dimensional box of side length $L$. The particles of the gas
will uniformly distribute over $V$, in a characteristic time of 
$O(L^2)$. 
The reverse
process will never be observed within astronomical time scales, as a consequence of Kac's lemma~\cite{Cencini2009} (see discussion below): diffusion is therefore an irreversible process. Once the gas has occupied the whole available volume, the system undergoes an equilibrium dynamics, meaning that the NEP is transient. For an example of persistent NEP, one can think instead of the electric current $j$ in a conductor, 
produced by an externally imposed electric field $E$. 
The well-known Ohm's law, $j =
\sigma E$, defines the linear dependence of the two physical quantities, where the constant $\sigma$ is the electric
conductivity~\cite{ohm1827galvanische}. The presence of a preferential direction determined by
the electric field, and of a persistent current of charge carriers $j$, clearly indicates the  irreversible nature of the process.
Remarkably, Ohm's law straightforwardly
arises from linear response theory, specifically the Green-Kubo
relation: $\sigma$, a non-equilibrium quantity, can
be computed in terms of equilibrium properties, i.e., time
correlations \cite{KTH91,FDreport}.

This review will mostly focus on persistent NEP. However, in the following we also briefly revisit some general aspects of transient
NEP, which played a crucial role
in understanding the second law of thermodynamics. These concepts clarify how irreversibility arises in macroscopic systems
from their microscopically reversible Hamiltonian dynamics, and will also be useful for the discussion of persistent NEP. 

\subsection{Transient non-equilibrium phenomena}

\subsubsection{The role of the number of degrees of freedom}

Poincar\'e recurrence theorem states that a Hamiltonian system with $N$ degrees of freedom in a confined domain will reach again a state arbitrarily close to its
initial condition, after a very long time. Therefore, strictly speaking, transient NEP can be classified as non-equilibrium
only when observed over a ``short'' time, i.e., much shorter than Poincar\'e recurrence one. It is known from Kac's lemma that this time is of order $e^{a N}$: 
$a$ here is a strictly positive constant, whose precise value
depends on how close to the initial condition the system needs to be found at recurrence, and
is not really important for the following considerations.  The  exponential dependence on $N$ implies that, for a system made
of a small number of particles, also the recurrence time is relatively small, and irreversibility is not typical.

\begin{figure}[t!]
\centering
\includegraphics[width=1.\textwidth]{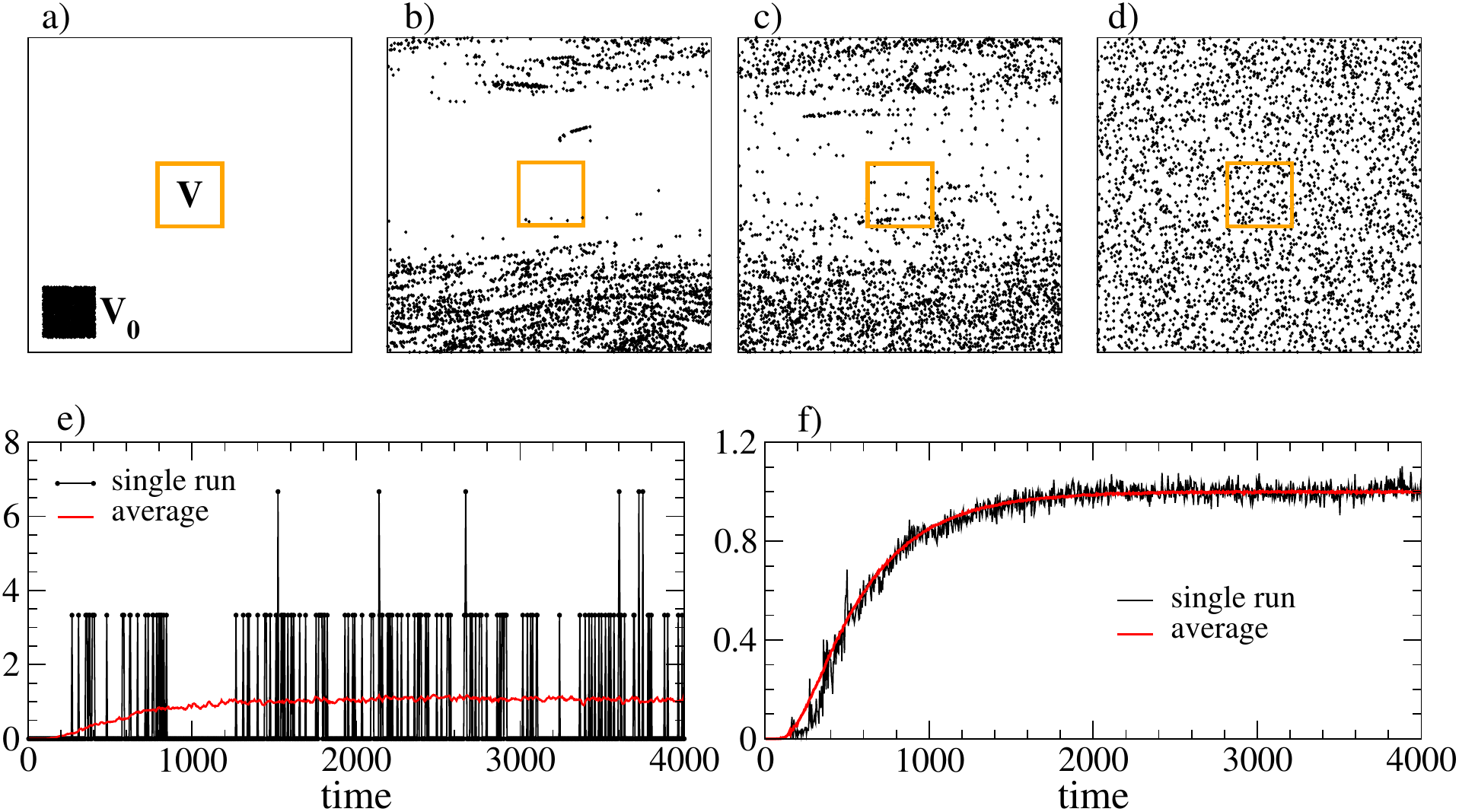}
\caption{ Irreversible spreading of an ink drop.  Panels (a-d): time
  evolution of the ink particles.  Ink particles start (a) uniformly
  distributed in $V_0$ and the instantaneous occupation $n(t)$ is
  monitored in the (orange) box $V$, panels (b-d) show the evolution
  till the ink particles are uniformly distributed in the whole
  contained. Panels (e) and (f) show the instantaneous occupation
  $n(t)/n_{eq}$ (black curve) and its average $\ave{n(t)}/n_{eq}$ (red
  curve) in the case of few $N_t=8$ and many $N_t=2.5 \,10^4 \gg 1$
  ink particles, respectively. In (e) the average is over $500$
  independent initial conditions starting from $V_0$. When the number
  of particles is large (f), the irreversible behavior is well evident
  even looking at a single realization of $n(t)$, this is not the case
  when their number is small (e). See Ref.~\cite{cerino2016therole}
  for further details.
\label{General:Fig_1}
}
\end{figure}

To explain
the relevance of the number of
degrees of freedom for the occurrence of irreversibility, we discuss the spreading of an
ink drop~\cite{cerino2016therole}, represented as a system of
$N_t$ tagged particles in a fluid, initially uniformly distributed in
a small region $V_0$.  We can study this phenomenon by considering a 
system of $N_t$ particles interacting among themselves, as well as with
the $N_s$ solvent particles ($N_s\gg N_t$).  In mathematical terms, the
ink drop and the solvent correspond to phase-space points evolving through a
symplectic dynamics that mimics the Hamilton equation (see
Ref.~\cite{cerino2016therole} for details).  A simple way to monitor
the mixing process of the ink amounts to counting the number of
ink particles $n(t)$ in a region $V$ at time $t$ (see
Fig.~\ref{General:Fig_1}(a-d)). In Fig.~\ref{General:Fig_1}(e,f), we
show the evolution of $n(t)$ in a single realization and its average
over many realizations in two different cases: (e) for a small number of
tagged particles, where in a single realization $n(t)$ does not
display any irreversible tendency to a final state, and only the
average $\langle n(t) \rangle$ shows clear irreversibility; (f) for a
large number of tagged particles ($N_t \gg 1$), where the irreversible
behavior is well evident even looking at a single realization of
$n(t)$.
In everyday experience, we  have usually access to a single
realization of a certain phenomenon. In order to decide whether it is reversible or not, we need the trajectory to be
\textit{typical}~\cite{goldstein2012typicality}: if this is the case, the (ir)reversible nature of the phenomenon is clear from a single realization. This is true only for macroscopic bodies, made of
many interacting degrees of freedom.

The example above  also exemplifies the conceptual
difference between the physical irreversibility in a single
(macroscopic) system $\mathbf{X}$, and the relaxation of a phase-space probability
distribution $\rho({\bf X},t)$ towards an invariant distribution.  The
latter is a property of an ensemble of initial conditions, which is verified whenever
for large $t$ one has $\rho({\bf X}, t) \to \rho_{inv}({\bf X})$,
independently of the initial density distribution $\rho({\bf X},0)$.
Although this property is quite important from a mathematical point of view, it is not the mark
of a genuinely irreversible behavior, as demonstrated by the above
example.

\textcolor{black}{The difference between relaxation of the probability distribution to an invariant one and the irreversibility in a unique system can be understood considering a
low-dimensional symplectic chaotic system, e.g. the Arnold cat map \cite{Arnold68}:
\begin{subequations}
\begin{eqnarray}
x_{t+1}&=&x_t+ y_t \,\, \text{mod} \, 1 \\
y_{t+1}&=&x_{t+1}+ y_t \,\, \text{mod} \, 1  \,.    
\end{eqnarray}    
\end{subequations}
This map is chaotic and mixing, i.e. $ \rho({\bm X}, t)\to \rho_{inv}({\bm X})$ (with $\bm X=(x,y)$), but at variance with the
behavior of a macroscopic system, it is impossible to observe any
qualitative difference between a single direct trajectory $ {\bm X}_0,
{\bf X}_1,\ldots,{\bf X}_{t-1}, {\bf X}_t$ and its time reversed ${\bf
  X}_t, {\bf X}_{t-1},\ldots, {\bf X}_{1}, {\bf X}_0$.} \textcolor{black}{Even considering an ensemble of initial conditions, each of them
evolving independently of the others, their behavior cannot represent the dynamics of a macroscopic body.}
\textcolor{black}{Having a large number of (interacting)
degrees of freedom is therefore a necessary condition for observing transient
irreversibility of macroscopic systems.}

\subsubsection{Irreversibility and typicality}

Consider a system with
$N \gg 1$ degrees of freedom, interacting in some way\footnote{\label{foot:inter} Although it is possible to have interesting results even in non interacting system~\cite{baldovin2023ergodic}, some (even weak) interaction among the particles is usually required to observe a
genuine thermodynamic behavior, and thus irreversibility. For
instance, consider $N\gg 1$ particles in a box, whose
velocities at the initial time are extracted from the Maxwell–Boltzmann distribution at
temperature $T_1$ for half of them and $T_2\neq T_1$ for the other
half. In the absence of interaction, the momentum of each particle is
conserved and, consequently, the time evolution of some macroscopic
observables (e.g. the fourth moment of the particle momenta) will not
attain the microcanonical equilibrium value.}, and an
observable ${\cal O}$ depending on all (or at least many) of them. It is generally expected that, if the initial condition is far enough from equilibrium, i.e.
\begin{equation}
\mathcal{O}(0)= \ave{\mathcal{O}}_{eq} +\delta \mathcal{O}(0)\,, \quad \text{with} \,\, |\delta \mathcal{O}(0)| \gg \sigma_\mathcal{O} \,, \label{eq:cond}
\end{equation}
where $\sigma_\mathcal{O}$ denotes the magnitude of the equilibrium
fluctuations of the observable $\mathcal{O}$, then almost all trajectories ${\cal O}(t)$ 
will be close to the average $\langle{\cal O}(t) \rangle$, excluding very unlikely
cases. In other words, the behavior of ${\cal O}(t)$ is expected to be typical.

\begin{figure}[t!]
\centering
\includegraphics[width=1\textwidth]{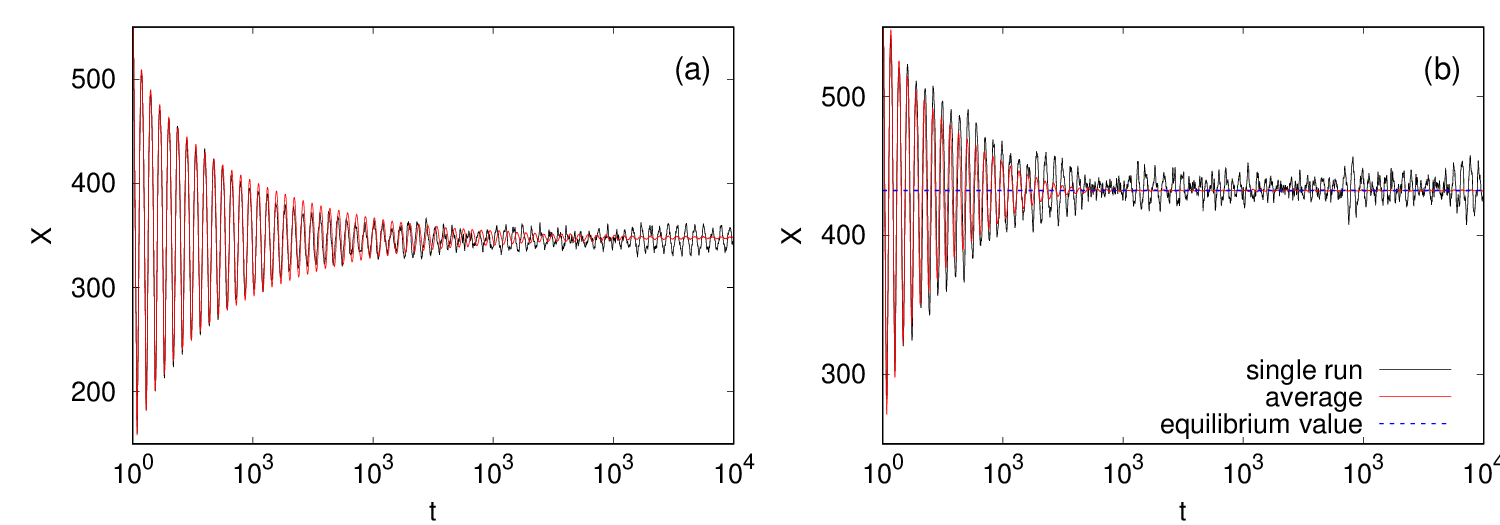}
\caption{Evolution of the piston position $X(t)$: (a) for the interacting particle (chaotic) model, (b) for the non-interacting (non-chaotic) particle model. In both cases the number of ``gas'' particles is $N=1024$ with $m=1$ while the piston has mass $M=10$, the initial temperature is $T_0=10$, and the initial displacement of the piston position is  $X(0)= X_{eq}+10 \sigma_{eq}$. 
Black curves denote  $X(t)$ in a single realization; red curves refer to the ensemble average $\langle X(t) \rangle$, and in (b) the blue dashed line is the analytical equilibrium value. The results are taken from Ref.~\cite{cerino2016therole}, where more details can be found.
}
\label{General:Fig_2}
\end{figure}

A general mathematical proof of the above statement is still
missing. However, the result can be proved
rigorously in certain stochastic processes (e.g. the
celebrated Ehrenfest model~\cite{baldovin2019irreversibility}) and  for dilute gases \cite{lanford1975time}, in
the so-called Grad-Boltzmann limit\footnote{Consider a system of hard
spheres of radius $\sigma$, with $N$ particles per unitary volume. The
Grad-Boltzmann limit corresponds to increasing the number of particles
while decreasing their size, in such a way that the collision rate
approaches a constant value, i.e. $N \to \infty$, $\sigma \to 0$ and
$N \sigma^2 \to \text{constant}$. By doing so, the volume occupied by
the particles scales as $N \sigma^3 \to 0$, consistently with the physical interpretation of dilute-gas
limit.}.  To exemplify this
result, without entering into mathematical detail, we discuss here some
numerical simulations  supporting the scenario
that for a generic macroscopic observable ${\cal O}$ satisfying
(\ref{eq:cond}), one has
\begin{equation}
\text{Prob}
\Big\{ 
{\cal O}(t) \simeq \langle{\cal O}(t)\rangle 
\Big\} \simeq 1\;.
\label{countP}
\end{equation}
The system we consider can be viewed as a simplified model of a piston~\cite{cerino2016therole}, i.e., a  channel
containing $N$ particles of mass $m$, closed by a fixed vertical wall
on one end, and by a frictionless moving wall of mass $M$ (the
piston itself) on the other.  We denote with $x_n(t)$ the coordinate of the $n-$th particle parallel to the channel, in the framework of the fixed wall, and with $X(t)$ the position of the
piston, so that $0\le x_n \le X \;\forall n$. If we assume a constant force $F$ to act on the
piston, and we take into account the interactions between the particles
inside the channel, the Hamiltonian of the total system reads:
$$
H= \frac{P^2}{2M} +\sum_ i \frac{|\bm p_i|^2}{2m} +\sum_{i<j}U(|{\bm q}_i -{\bm q}_j|) +
U_w({\bm q}_1,..., {\bm q}_N, X) +F X \, ,
$$ where $U$ is the interacting potential between the particles, and
$U_w$ denotes the interaction of the particles with the piston.  In
the case of non-interacting particles,\footnote{Notice that even if
the particles do not interact among each-other, their interaction is mediated by the
collisions with the moving wall (the piston). As a consequence energy is redistributed.} one has $U=0$, and $U_w$ is the hard-wall potential, yielding elastic collisions. The dynamics is
not chaotic, and it is easy to find the ``equilibrium'' position of
the piston, $\langle X \rangle_{eq}$, and its variance $\sigma_X^2$.
In the presence of interactions such as, e.g., $ U(r)=U_0/r^{12}$ and $U_w=U_0\sum_n |x_n-X|^{-12}$, it is not
possible to determine analytically the equilibrium statistical
properties of the system, however the
problem can be easily studied numerically (see Ref.~\cite{cerino2016therole} for details).
The system starts at $t=0$, with the piston at rest ($\dot{X}(0) = 0$) in
$X(0) = X_0$, and the initial microscopic state is set as an equilibrium
configuration of the gas in the volume imposed by the piston position
at a given temperature $T_0$.  When the initial state is far enough
from equilibrium, i.e. $|X_0-X_{eq} |\gg \sigma_{X}$, the evolution of
$X(t)$ is irreversible, as shown in Fig. \ref{General:Fig_2}: damped
oscillations around the equilibrium position are clearly
detectable. From a conceptual point of view the important result is
that the single trajectories are typical, i.e. close to the average,
both for the chaotic and the non-chaotic case (i.e., either with or without inter-particle interactions). We also stress that, as
one can directly inspect from the figure, the qualitative features of
the chaotic and non-chaotic system are essentially indistinguishable
with respect to irreversibility: chaos plays little role in
irreversible behaviors (as well as for other statistical properties~\cite{cencini2008role}).

\subsubsection{Coarse-grained description}

\textcolor{black}{An aspect that will become important in the remainder of the paper is the level of description adopted when studying a given phenomenon or system.
For the sake of simplicity, let us reconsider again the problem of diffusion. Similarly to the example of the ``ink'' drop of Fig.~\ref{General:Fig_1}, one can start from the microscopic (deterministic, Hamiltonian) dynamics of the particles of interest and of the solvent (in real world, the tagged or colloidal particles and the molecules of the gas or liquid in which they are embedded). Or, one can change the level of description and mimic the microscopic reversible dynamics in term of a suitable stochastic process.  For instance, the
diffusion phenomena in a box can be described with a stochastic model,
e.g. the overdamped Langevin equation \cite{G90}
 \begin{equation}
   \gamma {\frac{d\bx}{dt}} =  -  \nabla  U({\bf x}) + \gamma\sqrt{2 D} \bm \eta \, ,
\label{I.1}
\end{equation}
where $\eta$ is a white noise (i.e. $\langle \eta_i(t)\eta_j(s)\rangle = \delta_{ij}\delta(t-s)$), and the potential $U({\bf x})$  is zero inside the box and divergent on the boundary, so to confine the particles. Relaxing the overdamped approximation, one can use the (full) Langevin equation \cite{G90}
 \begin{equation}
\label{I.2}
 \begin{aligned}
\frac{d{\bf x}}{dt} &= {\bf  v} \,,\\
\frac{d{\bf v}}{dt} &= - \gamma {\bf v}  - \nabla  U({\bf x})   +  \gamma\sqrt{ 2 D} \eta \, .
 \end{aligned}
\end{equation}
Therefore, practically, there is the freedom to adopt different
mathematical descriptions that are, under some aspects, equivalent up to a
certain coarse-graining level. For the specific case of diffusion,
the underlying idea is that the effect of fast collisions with the
solvent molecules is modeled by the white noise term.  As for the
original, microscopic description, also in this case the
irreversibility of the diffusion process can appear only if we look at
a large number of particles, initially close, evolving with
(\ref{I.1}) or (\ref{I.2}).  Only in a few special cases, it is possible
to perform experiments with good control following a system with many
degrees of freedom and repeating many times the measurements with
different initial conditions and then computing averages: therefore,
usually the transient NEP can be described just at a qualitative
level.}

Let us notice that the overdamped description \eref{I.1} can be seen as a coarse-graining in time of the
underdamped one \eref{I.2}: the former dynamics is described by only $d$ variables, $\bx\in\mathbb{R}^d$, while the latter lives in $2d$ dimensions. Therefore, in general, the result of the computation of some quantities (e.g., entropy) depends on the used model, even if the physical phenomenon is the same. 
Such an unpleasant fact is an unavoidable aspect of the coarse-graining procedure.For a discussion of this problem under the point of view of entropy production see~\cite{bilotto2021excess}.

\subsection{Persistent non-equilibrium phenomena: entropy production rate}
\label{sec:perneq}

Consider
again the case of the  conductor subject to an externally imposed electric field. If we can follow the
position (or the velocity) of a charge carrier for a long time in a
single experiment, we can realize that a current is present.
Since there is a preferential direction of motion, and, hence, the time-reversal symmetry of the process is persistently broken, we can conclude that the
system is in a (persistent) non-equilibrium state. The larger the current, the farther the system is from equilibrium.

The characterization of the non-equilibrium nature of a system in the general case is, of course, not as simple as this: in fact, in many cases the detection of the currents, which are responsible for the irreversible nature of the process, is a very challenging task. The problem needs, therefore, to be formalized in precise mathematical terms, e.g. by means of a systematic use of stochastic processes. The aim is to provide a general criterion valid independently of the details of the physical system.

A natural attempt in this direction could be to
look at two observable functions of the state, $\vect{X}(t)$, of the system, $A(t) \equiv A(\vect{X}(t))$ and $B(t)\equiv B(\vect{X}(t))$, and to check whether their
correlations show time-reversal symmetry, i.e. if
$$
\ave{A(t)B(0)}=\ave{A(0)B(t)}\,.
$$
To identify a
non-equilibrium system, it would then be sufficient to find two functions
$A(t)$ and $B(t)$ for which the above relation does not hold. One might be
tempted to define a ``degree of irreversibility'' related to the quantity
$$
\Delta
C_{AB}(t)=\ave{A(t)B(0)}-\ave{A(0)B(t)}\,,
$$
such as, e.g., $\int_0^{\infty}dt |C_{AB}(t)|$,
 to quantify the
time-reversal symmetry breaking. In Ref.~\cite{pomeau1982symetrie,Joss} it was proposed for instance to use a correlation function of the form
\begin{equation}
\Delta C_{xx^2}(t)= \langle x(t) x^2(0)\rangle-\langle x(0)x^2(t)\rangle
\label{eq:poomeaugeneral}
\end{equation}  
where $x$ represents a generic observable of the system, but, of course, other choices are possible. Correlations functions of
this kind can detect asymmetries (\textit{viz.} irreversible
behaviors) and thus inform us about the non-equilibrium character of
the system from a single variable. However, the above approach has two main drawbacks.

First, $\Delta C_{AB}$ depends both on the choice
of observable and the reference frame, meaning it is not an intrinsic
quantity. Secondly, for processes characterized by Gaussian statistics, functions such as \eqref{eq:poomeaugeneral} may be trivially zero even if the system is out of equilibrium (see e.g. Sec~\ref{sec:markov}).

Another way, which has a rather solid mathematical base, to decide whether a system is in a
state of non-equilibrium (and to introduce an adequate characterization of
the distance from equilibrium) is based on the comparison between the probability of forward (or direct) and inverse (or backward) trajectories.
The idea is to recognize whether the direct trajectory is more typical than the inverse one.

The first step is to identify the direct and
inverse trajectory.  In a mechanical system, one has for the former
$$\Tdir =\lx\{\mathbf{Q}(t), \mathbf{P}(t)\rx\}_{0<t<\cT}
$$
where $\mathbf{Q}$ and $\mathbf{P}$ are the generalized coordinates and momenta. The inverse trajectory is then defined as 
$$
\Tinv =\lx\{\mathbf{Q}(\cT-t), -\mathbf{P}(\cT-t)\rx\}_{0<t<\cT}\,,
$$
taking into account the fact that, when the motion is reversed, the momenta $\mathbf{P}$ change sign because they are proportional to $d\mathbf{Q}/dt$.
In the general case, the state of the system is described by a vector
$\bx(t)= (x_1(t),x_2(t),...,x_D(t))$, and  the  direct trajectory reads
$$
\Tdir =\lx\{x_1(t),x_2(t),...,x_D(t) \rx\}_{0<t<\cT}
$$
while the inverse one is
$$
\Tinv =\lx\{\epsilon_1 x_1(\cT-t),\epsilon_2 x_2(\cT-t),...,\epsilon_D x_D(\cT-t) \rx\}_{0<t<\cT}\,,
$$ where $\epsilon_i= \pm 1$ denotes the parity of the variable $x_i$:
$+1$ ($-1$) for variables that are even (odd) with respect to time
reversal. Notice that sometimes it can be not trivial to determine the
parity \cite{kubo1957statistical}. \textcolor{black}{The task is particularly challenging when the variables represent internal degrees of freedom and their parity is generally determined based on physical intuition. This clearly has an impact on the definition of entropy production, as demonstrated by the debate about the equilibrium nature of active Orstein-Uhlenbeck particles~\cite{caprini2018comment,dabelow2019irreversibility}.} 

Once $\Tdir$ and $\Tinv$ are identified, to evaluate the
equilibrium properties of the system we need to compare the
$\text{Prob}(\Tdir)$ with $\text{Prob}(\Tinv)$.  To this aim it is convenient to define the entropy production rate~\cite{lebowitz1999gallavotti}
\begin{equation} 
\label{eq:entproddef}
\Sigma = \lim_{\cT\to\infty} \ave{ \frac{1}{\cT} \ln{\frac{\text{Prob}(\Tdir)}{\text{Prob}(\Tinv)}} } = \lim_{\cT\to\infty} \Sigma^{(\cT)}
\end{equation}
where the average $\ave{\cdot}$ is made with respect to $\text{Prob}(\Tdir)$.
The entropy production rate does not suffer from the same
limitation of $\Delta C_{AB}$: it is an intrinsic quantity, i.e it does not depend on the used variable $\bx$ 
since it has the form of a Kullback–Leibler divergence. In addition, it is 
bounded from below by zero ($\Sigma\ge0$) and it is equal to zero
($\Sigma=0$) if and only if the forward and the backward path have the
same probability, i.e. if $\Delta C_{AB}=0$ for every choice of the observables
$A, B$.  Remarkably, $\Sigma$ cannot increase
under coarse-graining procedures: for a discussion on this topic see~\cite{crisanti2012nonequilibrium}
 and~\cite{lucente2022inference}.
 
 However, although $\Sigma$ is a well-defined
quantity, it is often not sufficient to capture all the
non-equilibrium characteristics, in particular temporal or spatial asymmetries.
Furthermore, several practical difficulties have to be faced when
calculating $\Sigma$: among the others, incomplete knowledge of the state and/or
limited resolution of the measurement procedure. In such cases, one has to resort to alternative approaches, e.g. to study suitable (or generalized) response functions, or correlation functions able to detect the asymmetries. The identification of these indicators often relies on the knowledge of the physics of the system under investigation.

\textcolor{black}{As stated in the Introduction our aim is to review such practical difficulties for the characterization of the irreversibility and to exemplify possible way out on the basis both of simplified systems and more complex examples taken from practical research.
In particular, in the following Section we provide a detailed
discussion of the aforementioned problems in more specific situations,
within the framework of linear stochastic processes, beginning with
Markovian processes and then going beyond this class, e.g. considering
Gaussian processes with colored noise or stochastic differential
equations with Poisson noise.}

\section{About Non-Equilibrium in \\ Linear Stochastic Processes} \label{sec:markov}
In the case of Markov processes, it is possible to provide an expression for the entropy production rate $\Sigma$ in terms of the transition probability $\cW_t(\bx|\by)$, i.e. the conditional probability of having $\bx$ at time $t$ given $\by$ at time $0$, and of the stationary probability density $\pi(\bx) = \lim_{t\to\infty}\cW_t(\bx|\by)$ which satisfy the chain rule $\pi(\bx)=\int d\bx\,\pi(\by) \cW_t(\bx|\by)$.
The result is (see Appendix \ref{subsec:entgau})
\begin{align}\label{eq:entprodmarkov}
    \Sigma &= \lim_{t \to 0} \frac{1}{t}\int d\bx\;\pi(\bx) 
    \int d\by\;\cW_t(\by|\bx)\ln{\frac{\cW_t(\by|\bx)}{\cW_t(\bx|\by)}} = \\
    &= \lim_{t \to 0} \frac{1}{t}\int d\bx\;d\by\; P_t(\bx,\by) \ln\frac{P_t(\bx,\by)}{P_t(\by,\bx)}
\end{align}
where $P_t(\bx,\by)=\pi(\bx)\;\cW_t(\by|\bx)$ is the joint probability of having $\bx$ at time $0$ and $\by$ at time $t$.
The last row of \eref{eq:entprodmarkov} makes it explicit that, in the case of Markov processes, the equilibrium/non-equilibrium nature of the process can be decided by looking at the violation of the detailed balance condition~\cite{G90}, i.e
\begin{equation} \label{eq:detail_balance}
    P_t(\bx,\by) = \pi(\bx)\cW_t(\by|\bx) = \pi(\by){\cW}_t(\bx|\by) = P_t(\by,\bx)\,,
\end{equation}
which is indeed the condition that establishes the invariance under time reversal of the transition $\bx \leftrightarrow \by \; \forall t>0$.
Note that Eq.\eqref{eq:entprodmarkov} is formal and require explicit knowledge of both the stationary distribution $\pi(\bx)$ and the propagator $\cW_t(\by|\bx)$. Thus, despite its good mathematical properties, entropy production is often hardly accessible, because the analytical computation of Eq.\eqref{eq:entprodmarkov} is not always feasible, and moreover its measure in experiments is usually hard as it is based on estimation of probability distributions in possible high-dimensional spaces.

\subsection{Equilibrium Condition in Gaussian  Processes}
Gaussian stochastic processes, particularly time-continuous ones, are commonly employed as useful effective models for describing, at least under some conditions, the dynamics of various physical and biological phenomena~\cite{ciliberto2013heat,scalliet2015cages,maggi2015multidimensional,caprini2019entropy,gilson2023entropy,sekizawa2024decomposing}.
Given their amenability to analytical calculations, these processes are excellent cases for testing physical theories. 
In order to clarify some non-trivial peculiarities inherent to this class of processes, we summarize here some well-known aspects of their equilibrium properties and we discuss the problem of unveiling temporal irreversibility from experimental signals.
Let us consider the  stochastic differential equation for the Ornstein-Uhlenbeck process in $D$ dimensions:
\begin{eqnarray}\label{eq:OU}
\dot{\bx} + \matx{A} \bx = \bxi + \bh(t) \qquad \ave{\xi_i(t)\xi_j(t')} = \matx{D}_{ij}\delta(t-t')\,,
\end{eqnarray}
where the real part of eigenvalues of $\matx{A}$ (a $D \times D$ real matrix) is positive (so that the system relaxes to a stationary probability density), $\matx{D}$ is the covariance matrix of the noise $\bxi$ and $\bh$ is an external field introduced just to make the response function explicit.
For simplicity, we are considering only even variables under time-reversal, although most considerations are true for odd variables as well.
By direct integration of \eref{eq:OU} we have
\begin{align} \label{eq:linevol}
\bx(t) &= e^{-(t-t')\matx{A}}\bx(t') + \int_{t'}^t ds\, e^{-(t-s)\matx{A}} \lx[\bxi(s) + \bh(s)\rx]
\end{align}
from which a simple computation \cite{lucente2022inference} leads to
\begin{align} \label{eq:ourescorr}
\cR_{ij}(t-t') &= \lx.\frac{\partial \ave{x_i(t)}}{\partial h_j(t')}\rx|_{\bh=0} &\cR(\tau) &= \begin{cases} e^{-\tau \matx{A}} & \tau \geq 0 \\ 0 & \tau < 0 \end{cases} \\
\cC_{ij}(t-t') &= \lx.\ave{x_i(t)x_j(t')}\rx|_{\bh=0} &\cC(\tau) &= \begin{cases} e^{-\tau \matx{A}} \matx{C} & \tau \geq 0 \\ \matx{C} e^{-|\tau|\matx{A}^T} & \tau < 0 
\end{cases} \\
\cC(0) = \matx{C} &= \int_0^\infty ds \, e^{-s\matx{A}} \matx{D} e^{-s\matx{A}^T} \iff &\matx{D} &= \matx{C}\matx{A}^T+\matx{A}\matx{C}\,.
\end{align}
Note that the equations above, since they involve just average values, hold not only for Gaussian processes, but for any type of $\delta$-correlated noise.
However, in the case of Gaussian noise, we can add an explicit and compact expression for entropy production rate $\Sigma$ too, which reads (see Appendix \ref{subsec:entgau})
\begin{align}\label{eq:entprodlin}
\Sigma &= \text{Tr}\lx\{ \lx(\matx{C}\matx{A}^T-\matx{A}\matx{C}\rx) \matx{D}
^{-1}\matx{A}\rx) = \text{Tr}\lx\{(\matx{A}^T\matx{D}^{-1}-\matx{D}^{-1}\matx{A})\matx{A}\matx{C}\rx\}\,.
\end{align}
For equilibrium systems, i.e. $\Sigma=0$, one recovers the celebrated Onsager reciprocal relations
\begin{equation}
\matx{A}\matx{C}=\matx{C}\matx{A}^T \iff \matx{A}^T\matx{D}^{-1} = \matx{D}^{-1} \matx{A} \iff \matx{D} \matx{A}^T = \matx{A} \matx{D}\,.
\end{equation}\label{eq:onsager}
Note that, since the paths distribution is Gaussian, i.e.
\begin{equation}
\text{Prob}(\Tdir) \sim \exp{-\frac{1}{2}\sum_{ij}\int dt \int dt' x_i(t) \cD_{ij}(t-t') x_j(t')} 
\end{equation}
using the property
\begin{equation}
\lx(\int ds\, \cD(t-s)\cC(s-t')\rx)_{ij} = \delta_{ij} \delta(t-t')\,,
\end{equation}
it is possible to show that the condition 
\begin{equation}
    \text{Prob}(\Tdir)=\text{Prob}(\Tinv)
\end{equation}
is equivalent to $\cC(t)=\cC(-t)$.
Let us stress the fact that in order to understand if the system is in equilibrium it is necessary the knowledge of all dynamical variables in the system, a practical difficulty in real-world experiments.
At equilibrium, a very important relation holds between correlation and response:
since $\matx{C}\matx{A}^T=\matx{A}\matx{C}$ and $\matx{C}\matx{A}^T+\matx{A}\matx{C}=\matx{D}$, one has $2\matx{A}\matx{C} = \matx{D}$; therefore, the derivative of $\cC(t)$ must satisfy, $\forall t > 0$,
\begin{align} \label{eq:lineqt}
\frac{d\cC(t)}{dt} &= -\frac{1}{2}\cR(t) \matx{D}\,. 
\end{align}
Such equation expresses the equilibrium condition for a linear Gaussian process in the familiar form of a fluctuation-dissipation theorem.

It is interesting to study the above relation in the reference frame that has the eigenstates of the symmetric matrix $\matx{D}$ as a basis, or, equivalently, when the covariance matrix of the noise is diagonal, i.e. $\matx{D}_{ij}=\ave{\xi_i(t)\xi_j(t')}= 2T_i \delta_{ij}\delta(t-t')$. In this case the contribution of the noise can be interpreted as the effect of $D$ thermal baths, with temperatures $\{T_i\}_{i=1,...,D}$. If we look at the $i$-th single dynamical variable only, we find 
\begin{align}\label{eq:fludisi}
    \frac{d\cC_{ii}(t)}{dt} = -T_i \cR_{ii}(t)\,.
\end{align}
Note that we obtained an equilibrium condition of the form \eqref{eq:lineqt} despite having $D$ different temperatures. This is possible with a non-symmetric drift matrix $\matx{A}$ which satisfies the condition $\matx{A}_{ij} T_j = \matx{A}_{ji} T_i$, condition that implies a very special property for $\matx{A}$, i.e. $\matx{A}_{ij}\matx{A}_{jk}\matx{A}_{ki}=\matx{A}_{ij}\matx{A}_{kj}\matx{A}_{ji}\; ,\, \forall i<j<k$ \cite{lucente2022inference}.
In the same way then, the violation of  equality \eref{eq:fludisi} can be exploited as a measure of non-equilibrium of a system. In these cases typically the expression above is written in terms of their Fourier transforms $\widetilde{\cC}(f)$ and $\widetilde{\cR}(f)$ (see \cite{lucente2022inference} for details about the derivation), i.e.
\begin{align} \label{eq:lineqf}
\widetilde{\cC}(f) &= 2\text{Re}\lx\{\widetilde{\cR}(f)\rx\} \matx{C} = -\frac{1}{2\pi f} \text{Im}\lx\{\widetilde{\cR}(f)\rx\} \matx{D}\,.
\end{align}
These equilibrium conditions turn out to be fundamental in some cases, for example when a Markovian process is projected onto a small space.
In this case, the projected dynamics is, in general, not Markovian anymore, and it contains memory terms. A precise definition of non-equilibrium is therefore more tricky. 
Indeed, as we will discuss in the next Sections, it is possible to design non-Markovian out-of-equilibrium processes with time-reversal symmetry and vanishing entropy production by simply projecting a multidimensional Markov process onto a space of smaller dimension.
In these situations, one possible mathematical formulation of non-Markovian equilibrium system relies on the generalized fluctuation-dissipation theorem which relates the response to external forcing $\cR$ to the time-derivative of correlation functions ($\cR\propto\dot{\cC}$).
Before entering this topic, however, we will briefly review some known results on the fluctuation-dissipation relations, and about the possibility to exploit them to infer the equilibrium properties of a system.

\subsection{Fluctuation-Dissipation Relations}\label{sec:fdr}
The first general fluctuation-dissipation relation has been derived by Kubo for Hamiltonian 
systems~\cite{kubo1957statistical}. 
In a nutshell, the idea is to consider weak perturbations of an equilibrium system whose dynamics is encoded 
in the unperturbed Hamiltonian $\mathcal{H}_0(\mathbf{x})$. 
The perturbed Hamiltonian 
$$
\mathcal{H}(\bx,t)=\mathcal{H}_0(\bx)-\mathcal{F}(t)A(\bx)
$$
can be used to derive an expression for the average variation of a generic observable $B(\bx)$ due to the perturbation $A(\bx)$, modulated by the time-dependent function $\mathcal{F}(t)$. 
Without entering into the details of the derivation (which are discussed in~\cite{kubo1957statistical,kubo1966fluctuation,KTH91,FDreport,sarracino2019fluctuation}), we just recall the Kubo formula
\begin{equation}
    \ave{\Delta B(t)}=\ave{B(t)}_\mathcal{H}-\ave{B}_{\mathcal{H}_0}=\int_{-\infty}^{t} {\rm d}\,t'\cR_{AB}(t-t')\mathcal{F}(t')\,,
\end{equation}
where $\cR_{AB}(t)$ is defined as
\begin{equation}
    \cR_{AB}(t)=\beta \ave{B(t)\frac{dA(0)}{dt}}_{\mathcal{H}_0}=-\beta\ave{\frac{dB(t)}{dt}A(0)}_{\mathcal{H}_0}=-\beta\frac{d\cC_{AB}(t)}{dt} \,,
    \label{eq:fdr_classical_kubo}
\end{equation}
and $\beta=1/T$ is the inverse of temperature. 
Equation \eqref{eq:fdr_classical_kubo} describes the response of the system to an infinitesimal impulsive perturbation and can be reformulated as  
\begin{equation}
\chi_{AB}(t)=\int_{0}^t {\rm d} \,t'\,\cR_{AB}=-\beta\lx\{\cC_{AB}(t)-\cC_{AB}(0)\rx\}\,,
\end{equation}
where $\chi_{AB}(t)$ is the susceptibility or admittance.
This derivation is valid for systems close to equilibrium, since it relies on the assumption that the unperturbed stationary distribution is the canonical one ($\pi(\mathbf{x})\propto e^{-\beta \mathcal{H}_0(\mathbf{x})}$). 
Several attempts to generalize the above formula to non-equilibrium systems have been made. For instance, a generalized fluctuation relation for a large class of systems admitting a non-singular invariant measure has been derived in~\cite{FDreport,ruelle1998}. This result relates the response to an initial perturbation $\delta x_j(0)$ with a properly defined correlation function, i.e. 
\begin{equation}
    \cR_{ij}(t)=\lim_{\delta x_j(0) \to 0}\frac{\ave{\delta x_i(t)}}{\delta x_j(0)}=-\ave{x_i(t)\frac{\partial \ln\pi(\mathbf{x})}{\partial x_j}}\,,
    \label{eq:generalized_fdr}
\end{equation}
where the average is performed on the unperturbed dynamics.
The variation of a generic observable can be written as
\begin{equation}
    \ave{\Delta B(t)}=-\sum_j\ave{B(t)\frac{\partial \ln\pi(\mathbf{x})}{\partial x_j}\bigg\rvert_{t=0}\delta x_j(0)}\,.
\end{equation}
 Despite its generality, sometimes \eref{eq:generalized_fdr} is not very practical, because it requires the explicit knowledge of the stationary measure $\pi(\mathbf{x})$. For this reason, other forms of generalized fluctuation-dissipation relations involving derivatives of the propagator only have been derived for stochastic systems driven by white noise~\cite{caprini2021generalized,caprini2021fluctuation,baldovin2021handy}.
It should be noted that \eref{eq:generalized_fdr} is a functional relation between correlations and responses that does not depend on the equilibrium nature of the system. 
This is particularly evident in linear systems where it takes the form $\cC(t)=\cR(t)C(0) \;\forall t>0$ meaning that the temporal evolution of the correlation is ruled by the deterministic part of the dynamics.

Some authors, starting from the failure of \eref{eq:fdr_classical_kubo} in non-equilibrium systems, introduced the concept of effective temperatures~\cite{cugliandolo2011effective}. For a critical discussion of this topic see~\cite{puglisi2017temperature,villamaina2009fluctuation}.
While the violation of a generalization of \eref{eq:fdr_classical_kubo} for stochastic processes effectively discriminates between equilibrium and non-equilibrium systems, the interpretation in terms of effective temperature is not always able to provide relevant indications regarding the system under investigation (see~\cite{crisanti2012nonequilibrium,villamaina2009fluctuation} for details).


To understand the usefulness of fluctuation-dissipation relations in discriminating equilibrium and non-equilibrium, it is convenient to discuss the relation between response and correlation in the case of a Brownian particle evolving through a generalized Langevin equation for the velocity of the particle. Such relation is found by means of the Mori-Zwanzig formalism to derive effective stochastic equations, and its general form reads~\cite{mori1965transport,mori1965continued}
\begin{equation}
\dot{v}+\int_0^t dt'\, \gamma (t-t') v(t') = \xi(t) \quad t>0 \,,
\label{eq:gle}
\end{equation}
where the memory kernel $\gamma(t-t')$ is a delayed friction force and the correlated noise verifies
$$
\ave{\xi(t)\xi(t')}=\nu(t-t')\,.
$$
We highlight that if 
$$
\gamma(t)=\gamma_0\delta(t)+\Theta(t)\sum_{i=1}^D \gamma_i e^{-\lambda_i t}\,,
$$
where $\Theta(t)$ is the Heaviside step function, the equation can be regarded as a $D-$dimensional linear system projected onto a one-dimensional space (an example will be discuss in \ref{subsec:gyrator}, see   \eref{eq:effdyn}). When $\gamma$ is not a simple combination of exponential functions, \eref{eq:gle} can still be interpreted as a projection of an infinite dimensional linear system. 

Independently of the physical interpretation, \eref{eq:gle} can be conveniently studied in Fourier space, i.e.
\begin{align}
    \widetilde{v}(f)=\frac{\widetilde{\xi}(f)}{i2\pi f+\widetilde{\gamma}(f)}=\widetilde{\mu}(f)\widetilde{\xi}(f)\,,
\end{align}
having defined the complex mobility as $\widetilde{\mu}(f)=\lx(i2\pi f+\widetilde{\gamma}(f)\rx)^{-1}$. 
The relation between velocity correlation $\cC(t)$, noise correlation $\nu$ and mobility $\mu$ takes now the form
\begin{equation} \label{eq:gle_ft} 
\widetilde{\cC}(f)=|\widetilde{\mu}(f)|^2\widetilde{\nu}(f) \,.
\end{equation}
At the same time, if we multiply \eref{eq:gle} by $v(0)$, performing an average over the noise we get \cite{mori1965transport,mori1965continued}
\begin{align} \label{eq:gle_flt}
\widetilde{\cC}(f) = 2 \ave{v^2}\lx|\widetilde{\mu}(f)\rx|^2 \text{Re}\lx\{\widetilde{\gamma}(f)\rx\}\,,
\end{align}
so that, by comparing \eref{eq:gle_ft} with \eref{eq:gle_flt}, we  obtain the relation to be satisfied by friction $\gamma$ and noise $\nu$ in an equilibrium process~\cite{kubo1957statistical,kubo1966fluctuation,KTH91}, namely
\begin{equation}
\frac{\widetilde{\nu}(f)}{2\text{Re}\lx(\widetilde{\gamma}(f)\rx)}=\ave{v^2}=T\,,
\end{equation}
or equivalently
\begin{align}\label{eq:gle_fdr}
\cC(t) &= \cC(-t) = 2 T \mu(t) \qquad t \geq 0\\
\ave{\xi(t)\xi(t')} &= \nu(t-t') = T \gamma(|t-t'|)\,, \nonumber
\end{align}
where the equality $\ave{v^2}=T$ follows from equipartition.
Note that the first equation, which links correlation to response, is completely analogous to \eref{eq:lineqt} and, by simple time-differentiation, to  \eref{eq:fdr_classical_kubo}.
The previous relation can also be inverted to obtain the so-called first- and second-kind Fluctuation-Dissipation Relations \cite{kubo1966fluctuation}
\begin{align}
     \text{Re}\lx\{\widetilde{\mu}(f)\rx\}&=\frac{1}{2T}\int dt\,e^{-i 2\pi f t}\cC(t)\,,\\
     \text{Re}\lx\{\widetilde{\gamma}(f)\rx\}&=\frac{1}{2T} \int dt\,e^{-i2\pi f t}\nu(t)\,, \nonumber
\end{align}
where the first can be regarded as an extension of the Einstein relation between mobility and diffusion coefficient, while the second one 
corresponds to a generalization of Nyquist results connecting dissipation and noise correlation~\cite{KTH91}.
We stress that while \eref{eq:generalized_fdr} has been derived under general hypotheses and holds also out of equilibrium, the relations (\ref{eq:fdr_classical_kubo}) and (\ref{eq:gle_fdr}) are instead only valid at equilibrium. 
For this reason, violations of \eref{eq:gle_fdr} have been used by Harada and Sasa to quantify the average rate of energy dissipation in a class of Langevin equations~\cite{harada2005equality,harada2006energy}.

\subsection{A No-Go Theorem for Gaussian  Processes}
To highlight how information on all the variables which describe the system under investigation, as well as on the response, is essential to infer the equilibrium properties of a system, we now discuss a no-go theorem that holds for every Gaussian stochastic process. We set the discussion in the framework of time-continuous processes, but it can be easily generalized to any kind of process with Gaussian statistics for all relevant probabilities.
First of all, the linearity of \eref{eq:OU} allows us to integrate over some components to get an integro-differential stochastic equation for the remaining components of the process (see example in \ref{subsec:gyrator} \eref{eq:effdyn}), i.e.
\begin{equation} \label{eq:intediff}
\cL \bx =  \bxi
\end{equation}
where $\cL = \lx\{\cL_{ij}\rx\}_{i,j=1,D}$ is a set of linear operators (e.g. differentiation or integration) acting on the subset of dynamical variables and the noise $\bxi(t)$ will be in general colored with zero mean $\ave{\xi_i(t)}=0$ and covariance matrix $\ave{\xi_i(t)\xi_j(t')}=\nu_{ij}(t-t')$.
The above formalism is a compact notation for discussing both under- and over-damped Markov linear systems as well as their projections onto low-dimensional sub-spaces which in general result in non-Markovian dynamics~\cite{lucente2022inference}: note however that \eref{eq:intediff} represents the most general class of stochastic linear processes. 
We recall the fundamental fact that, 
since the statistics of processes described by \eref{eq:intediff} is Gaussian (multivariate in both time and variables) only averages and correlations are needed to fully specify the process. 

The explicit stationary properties of \eref{eq:intediff} can be obtained in Fourier space, where the operator $\widetilde{\cL}(f)$ can be easily computed, and the correlation function $\widetilde{\cC}(f)$ can be computed as $\ave{\widetilde{\bx}(f)\widetilde{\bx}^\dagger(f)}$.
In this way response and correlations read \cite{lucente2022inference}
\begin{align}
\label{eq:nogo}
    \widetilde{\cC}(f) &=  \widetilde{\cR}(f) \widetilde{\nu}(f)\widetilde{\cR}(f)^\dagger \,, \nonumber \\
    \widetilde{\cR}(f) &=\widetilde{\cL}(f)^{-1}
\end{align}
where $\matx{M}^\dagger$ denotes the conjugate transpose of matrix $\matx{M}$.
From the above formula, one immediately realizes that - apart from special cases - it is not possible to infer both the response $\widetilde{\cR}(f)$ and the noise correlation $\widetilde{\nu}(f)$ from the knowledge of $\widetilde{\cC}(f)$ only.
As a consequence we have that, in the case of Gaussian noise, since the process is completely determined by its first two moments, one cannot discriminate between the infinitely many models sharing the same correlation $\widetilde{\cC}(f)$.
For instance, it is immediately evident that it is not possible to find a unique model for the following correlation function:
\begin{align}
 \widetilde{\cC}(f) &=\frac{\nu}{((2\pi f)^2+\lambda^2)((2\pi f)^2+\mu^2)} \qquad \mu>\lambda \\
 \cC(t) &= \frac{\nu}{2(\mu^2-\lambda^2)} \lx( \frac{e^{-\lambda |t|}}{\lambda} - \frac{e^{-\mu |t|}}{\mu}\rx) \,,
 \label{eq:no-go-3-examples}
\end{align}
Simple computations show that the three following processes are all compatible with $\widetilde{\cC}(f)$ in \eqref{eq:no-go-3-examples}
\begin{align*}
\text{I} \lx\{
\begin{array}{l}
\dot{x} + \lambda x = \xi \\
\ave{\xi(t)\xi(t')} = \nu e^{-\mu|t-t'|}/2\mu \\
\cR(t) = \Theta(t)e^{-\lambda t}
\end{array}
\rx.
\text{II} \lx\{
\begin{array}{l}
\dot{x} + \mu x = \xi \\
\ave{\xi(t)\xi(t')} = \nu e^{-\lambda|t-t'|}/2\lambda \\
\cR(t) = \Theta(t)e^{-\mu t}
\end{array}
\rx.
\end{align*}
\begin{align*}
\text{III}
\lx\{ 
\begin{array}{l}
\ddot{x} + (\lambda+\mu) \dot{x} + (\lambda\mu) x = \xi \\
\ave{\xi(t)\xi(t')} = \nu \delta(t-t') \\
\cR(t) = \Theta(t)\lx(e^{-\lambda t} - e^{-
\mu t}\rx)/\lx(\mu-\lambda\rx)\,.
\end{array}
\rx.
\end{align*}
Note that any two-dimensional matrix with trace equal to $\lambda+\mu$ and determinant equal to $\lambda \mu$ could be considered a drift compatible with the correlation function above by providing a suitable covariance for the noise.
This would not be an issue if all compatible models had the same nature (i.e. if they were all equilibrium/non-equilibrium processes), but unfortunately this is not the case: by time-differentiating $\cC(t)$
\begin{align*}
 \frac{d\cC(t)}{dt} &= -\frac{\nu}{2(\lambda+\mu)}\lx(\frac{e^{-\lambda t}-e^{-\mu t}}{\mu-\lambda}\rx) \,,
 \qquad t \geq 0 
\end{align*}
we can note that only the stochastic process III satisfies the fluctuation-dissipation relation \eref{eq:fludisi}, once the temperature $T$ is fixed in such a way to satisfy the usual relation between friction and noise variance $2T(\lambda+\mu) = \nu$.
\subsubsection{An Example: The Brownian Gyrator} \label{subsec:gyrator}
In the following we discuss in some details a system described by \eref{eq:OU} focusing on the situation where just one variable is experimentally accessible.
It is natural to wonder whether such information is sufficient to decide about the equilibrium/non-equilibrium nature of the underlying model. As shown in the following this is impossible.
Here we consider the case of the so-called Brownian gyrator~\cite{filliger2007brownian,ciliberto2013heat,ciliberto2013statistical,argun2017experimental}, 
consisting of the two-dimensional linear system
\begin{equation}\label{eq:BG}
\left\{
\begin{array}{l}
\dot{x} + a x = b y + \sqrt{2T_x}\xi_x \\
\dot{y} + d y = c x + \sqrt{2T_y} \xi_y \\
\end{array}
\right.\,.
\end{equation}
Explicit computation shows that, whenever Onsager reciprocal relations are not satisfied, i.e.
$$
\matx{C}\matx{A}^T-\matx{A}\matx{C} = \Delta \propto b T_y-c T_x \neq 0\,,
$$
the system experiences a systematic torque: defining $\theta = \arctan{(y/x)}$, a rotational current 
$$
j_\theta = \ave{\dot{\theta}} \simeq \ave{\dot{x}y-\dot{y}x}/\ave{x^2+y^2} \propto \Delta
$$  arises and the entropy production rate is proportional to the square of this current, $\Sigma \propto j_\theta^2$.
The ``effective'' dynamics of the accessible variable $x$ is non-Markovian
\begin{align}\label{eq:effdyn}
\lx\{
\begin{array}{l}
\dot{x}+ax-bc\int_{-\infty}^t e^{-d(t-t')}x(t') = \eta(t) \\
\ave{\eta(t)\eta(t')} = 2\lx(T_x\delta(t-t')+T_ye^{-d(t-t')}/d\rx)
\end{array}
\rx.
\end{align}
and its correlation function $\cC_{x}(t)=\ave{x(0)x(t)}$ in Fourier space reads
\begin{align*}
    \widetilde{\cC}_{x}(f) = \frac{c_0+c_1 (2\pi f)^2}{\lx(\cD- (2\pi f)^2\rx)^2+\cT^2 (2\pi f)^2}
    \quad\text{where}\quad    
    \lx\{
    \begin{array}{l}
    \cT = a + d \\
    \cD = ad-bc \\
    c_0 \propto T_x d^2 + T_y b^2 \\
    c_1 \propto T_x \\
    \Delta \propto b T_y - c T_x 
    \end{array}
    \rx. \,.
\end{align*} 
Thus, once we choose $\cC_{x}(t)$ and $\Delta$, we have only $5$ equations to determine the parameters: two for the trace $\cT$ and the determinant $\cD$ of the drift matrix, two for the coefficients $c_0$ and $c_1$ of the correlation function, and one for the expression of $\Delta$. \textcolor{black}{Therefore, being the problem underdetermined, it is possible to build models with different entropy production values.} 
\textcolor{black}{In other words, the time-series of $x$ will appear always invariant under time-reversal (note that $\cC_{x}(t)=\cC_{x}(-t)$ for any 1-D Gaussian process). Consequently, since the violation of the generalized fluctuation theorem is the only indicator of non-equilibrium conditions, perturbation-response experiment is needed to separate the contributions of the drift from that of the noise and infer the equilibrium properties of the system.}

The above consideration does not exclude the possibility to obtain some guesses by running experiments where the measure of response is meant in a broader and more general way.
\textcolor{black}{In some cases, it is also possible to measure the entropy production rate of a system performing experiments under slightly different conditions.}
As an explicit example, we consider again the case of the Brownian Gyrator  \eref{eq:BG} and we assume to be able to manipulate the temperature $T_x$ of one thermal bath. In this case, useful information are obtained by comparing the $xx$ correlation function at temperatures $T_x^{(1)}$ to $T_x^{(2)}$ ($\cC_{x}^{(1)}(t)$ and  $\cC_{x}^{(2)}(t)$ respectively).
Indeed, from the correlation functions we can fit the relaxation times $1/\lambda$ and $1/\mu$ and the four coefficients $c_{\lambda,\mu}^{(1,2)}$ which are functions of the system parameters $a,b,c,d,T_x^{(1,2)}$ and $T_y$, i.e.
\begin{align*}
\lx\{
\begin{array}{l}
\cC_{x}^{(1,2)}(t) = c_\lambda^{(1,2)}e^{-\lambda t} +c_\mu^{(1,2)}e^{-\mu t} \\
r^{(1,2)} = \lambda \mu \lx(\lambda c_\mu^{(1,2)} + \mu c_\lambda^{(1,2)} \rx) \\
\end{array}\rx.
\rightarrow
\lx\{
\begin{array}{l} 
T_x^{(1,2)} = \lambda c_\lambda^{(1,2)} + \mu c_\mu^{(1,2)} \\
d = \sqrt{\lx(r^{(2)}-r^{(1)}\rx)/(T_x^{(2)}-T_x^{(1)})} \\
b^2 T_y = r^{(1,2)} - T_x^{(1,2)} d^2 \\
a + d = \lambda+\mu \\
bc = (\lambda + \mu - d )d-\lambda\mu\\
\end{array}\rx.\,.
\end{align*}
Although we cannot uniquely determine the model, we get a peculiar combination of parameters that enables us to compute the exact value for $\Sigma$ in both cases:
\begin{align*}
\Sigma^{(1,2)} = \frac{\lx(c T_x^{(1,2)}-b T_y\rx)^2}{2(a+d) T_x^{(1,2)} T_y} = \frac{\lx( (bc) T_x^{(1,2)} -b^2 T_y\rx)^2}{2(a+d) T_x^{(1,2)} (b^2T_y)}
\end{align*}
\begin{figure}[ht!]
\includegraphics[width=0.98\textwidth]{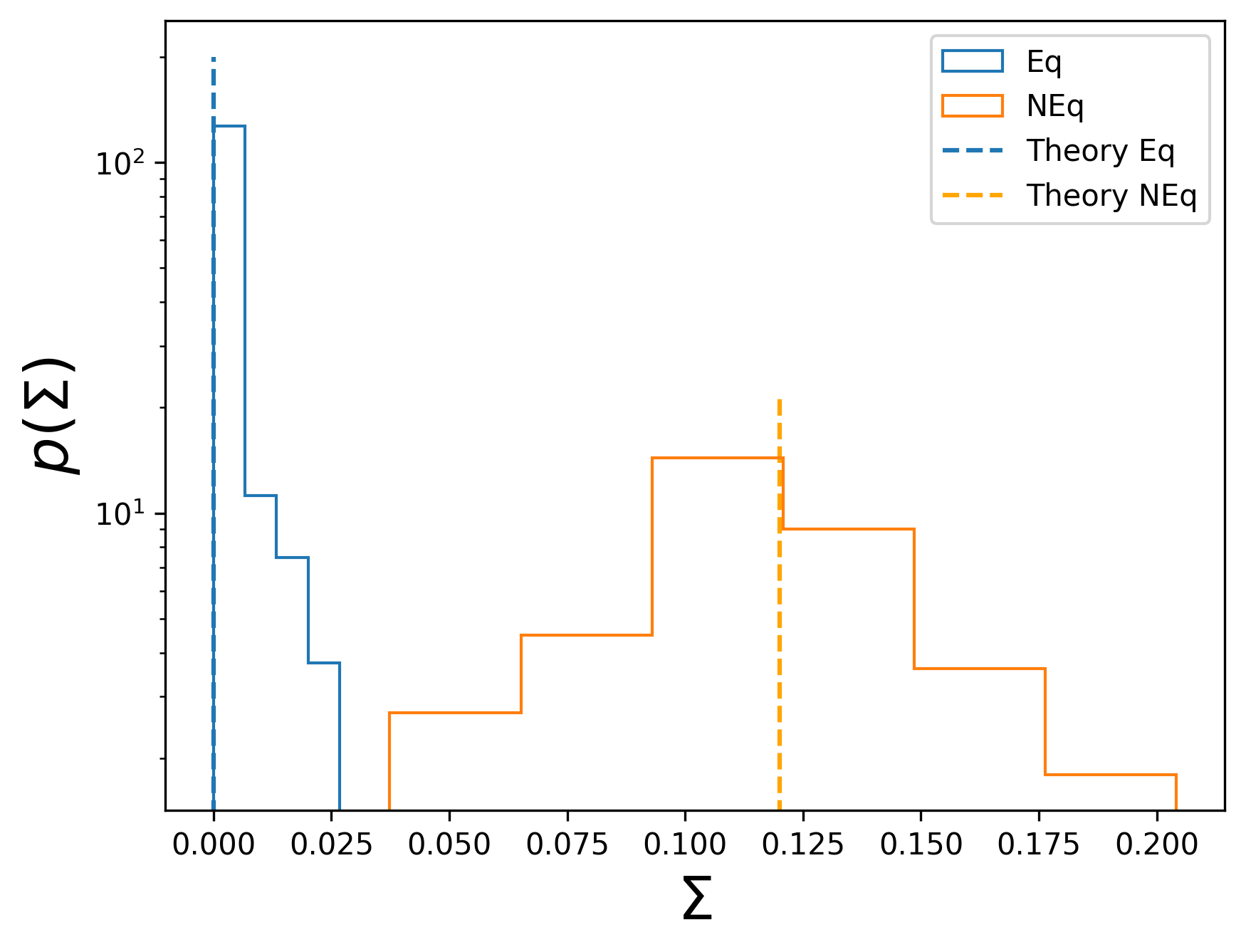}
\caption{
Histogram of entropy production rate estimated with our approach from $40$ independent realizations of process \eref{eq:BG} simulated at $2$ different system parameters.
The distributions of the results are correctly centered around the theoretical values (dots line) both for equilibrium and out-of-equilibrium systems \cite{lucente2023poisson}. The parameters used for numerical simulations are $a=\frac{8}{3}$, $b=c=-d=-\frac{2}{3}$, $T_y=\frac{1}{2}$ and $T_x^{(1)}=\frac{1}{5}$ (orange) or $T_x^{(2)}=\frac{1}{2}$ (blue).}
\label{Markov:Fig_1}
\end{figure}
\fref{Markov:Fig_1} shows an application of the formula above on model \eqref{eq:BG} simulated for $2$ different values of $T_x$.
\subsection{Non-Equilibrium induced by Poisson Noise}
\begin{figure}[ht!]
    \includegraphics[width=0.48\textwidth]{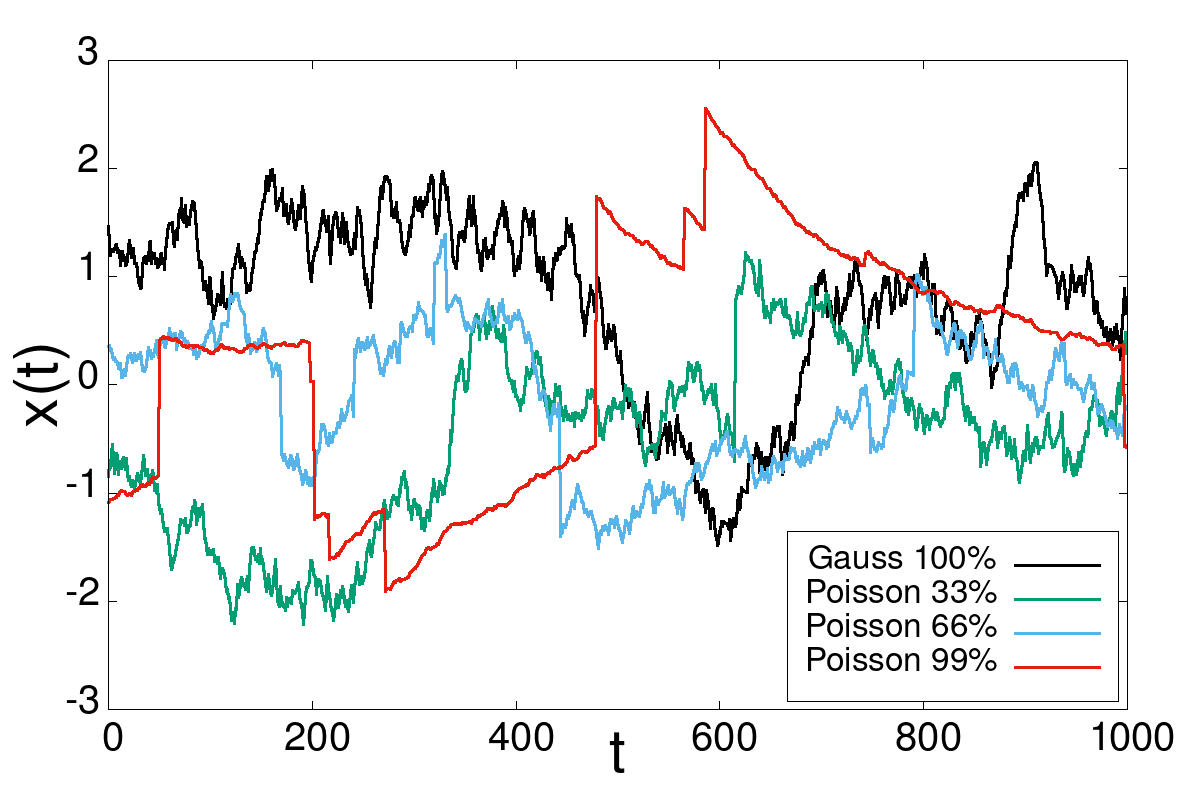}
    \includegraphics[width=0.48\textwidth]{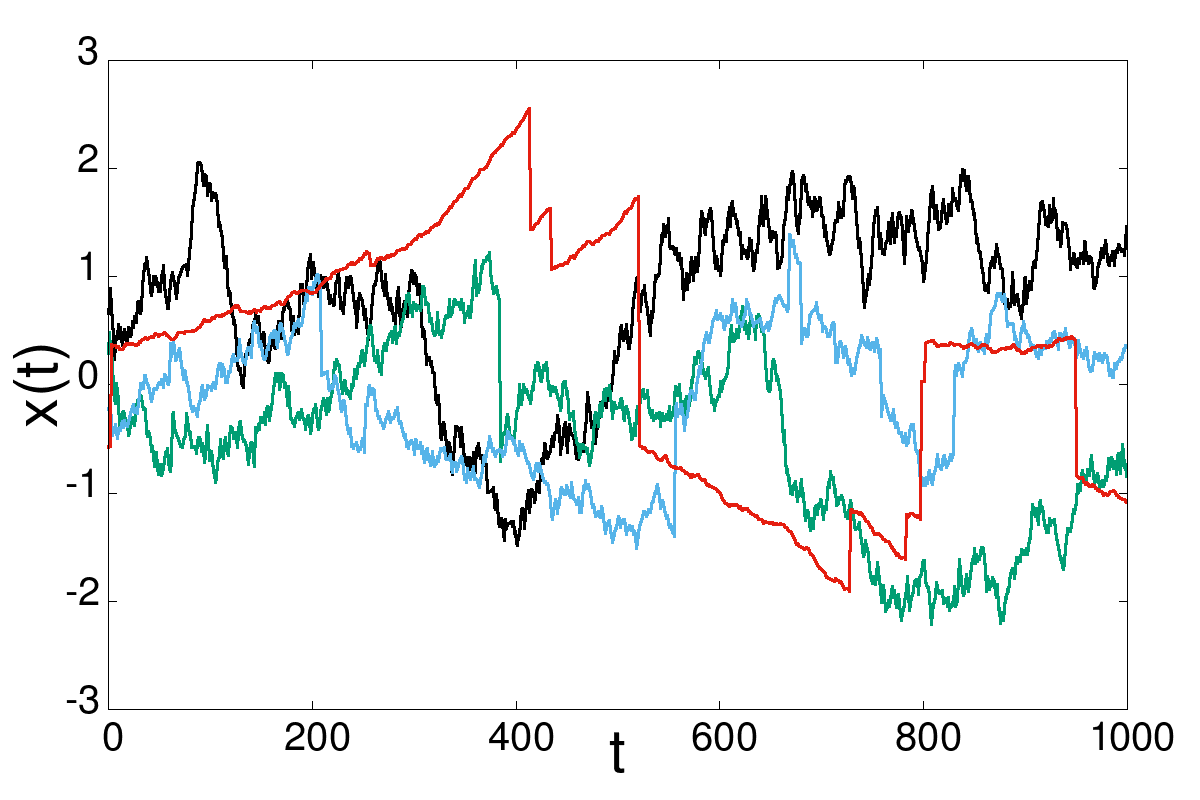}
  \caption{Examples of direct (left) and time-reversed (right) trajectories for processes driven by a Gaussian and Poisson noise, the amplitude of the jumps are sampled from Gaussian distribution with covariance matrix $\Gamma$. We keep the variance of the total noise $\nu'=\nu+\lambda\Gamma$ constant and we vary the fraction of Poissonian noise of the process. It is evident that, as the Poissonian contribution increases, the time-reversed trajectory becomes more and more incompatible with the direct one, i.e. it is difficult to find any piece of the latter in the former.} \label{Markov:Fig_2}
\end{figure}

Although the use of Gaussian noise has rather obvious motivations, in some contexts, e.g. in small systems, this assumption appears inadequate, and other kinds of noise must be considered. 
We now examine another wide class of stochastic process with independent and stationary increments, which naturally arises in certain physics experiments: the compound Poisson process \cite{ken1999levy,kyprianou2014fluctuations,schilling2016introduction,kanazawa2012stochastic,kanazawa2015minimal,kanazawa2015asymptotic,kanazawa2017statistical}.
In this dynamics jumps of random amplitude occur at random times, distributed according to a Poissonian statistics. 
We are interested in understanding how the equilibrium/non-equilibrium nature of the system changes, with respect to models characterized by Wiener noise only, when this additional noise is taken into account.
Consider for instance the following linear equation
\begin{align} \label{eq:stopoi}
\lx\{
\begin{array}{l}
\dot{\bx} + \matx{A}\bx = \bxi(t) + \bzeta(t) \\
\bzeta(t) = \sum_{k} \bu^{(k)}\delta(t-t_k)
\end{array}
\rx.
\lx\{
\begin{array}{l}
\bxi \sim \cG_\nu(\bxi) \\
\bu^{(k)} \sim \cP\lx(\bu^{(k)}\rx) \\
t_k-t_{k-1} \sim \cQ_\lambda(t_k-t_{k-1}) = \lambda e^{-\lambda(t_k-t_{k-1})}
\end{array}
\rx.
\end{align}
where $y \sim \mathcal{P}(y)$ means that $\mathcal{P}(y)$ is the PDF of the stochastic process $y$, and $\bxi(t)$ is the usual Wiener process with $\ave{\xi_i(t)\xi_j(t')}=\matx{D}_{ij}\delta(t-t')$. The amplitude of the jumps $\underline{\bu}=\lx\{\bu^{(k)}\rx\}_k$ are i.i.d., drawn from a generic distribution $\cP(\bu)$ with covariance matrix $\ave{u_i^{(k)} u_j^{(k')}}=\Gamma_{ij}\delta_{kk'}$, while the intervals $\Delta t = t_{k}-t_{k-1}$ between to consecutive jumps are i.i.d. and extracted from an exponential distribution $\lambda e^{-\lambda \Delta t}$. 
It is easy to understand that a system driven only by $\bzeta(t)$ cannot be at equilibrium. Consider the time interval between two jumps: in the direct path, the system relaxes toward its mean value, while in the time-reversal one, the system moves away from its stationary state. 
The impossibility of observing the reverse paths is quite obvious and it is well illustrated in Fig.~\ref{Markov:Fig_2} showing direct (left) and inverse (right) trajectories of a one-dimensional Ornstein-Uhlenbeck process driven by Gaussian or Poissonian noise.

\begin{figure}[ht!]
    \includegraphics[width=0.95\textwidth]{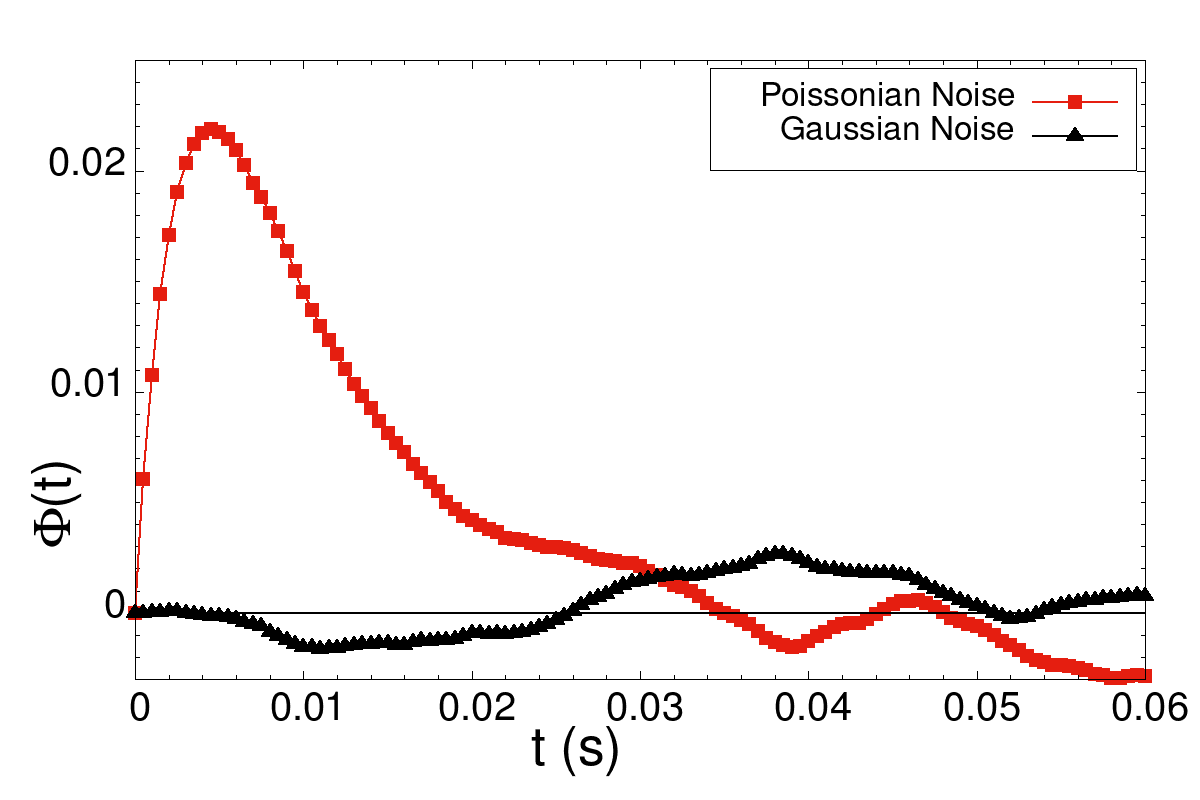}
  \caption{Degree of irreversibility $\Phi(t)$ for Gaussian (black) and Poissonian (red) process for a $2-$dimensional linear process. Using the same nomenclature of \eqref{eq:BG} for elements of the drift matrix $A$, the parameters used for numerical simulations are $a=122.9$, $b=1$, $c=0$, $d=26.1$. Regarding the noise, in the Poissonian case we consider $\lambda=1261$, $\Gamma=\lx(\begin{array}{cc}
      2.002225 & 0 \\
      0 & 0
  \end{array}\rx)$ and $D=\lx(\begin{array}{cc}
      0 & 0 \\
      0 & 1647\\
  \end{array}\rx)$ while in the Gaussian regime $D=\lx(\begin{array}{cc}
      2524.8 & 0 \\
      0 & 1647\\
  \end{array}\rx)$. 
  Note that parameters are chosen to make the two-point correlation function $\cC(t)$ identical for both noises and their values have been obtained by fitting a real experimental signal of a granular system (see~\cite{lucente2023revealing} for further details).}
  \label{Markov:Fig_3}
\end{figure}

More formally, one can prove that such processes lack  detailed balance. Indeed, this property must be satisfied separately by the jump process and by the continuous part.
Regarding the discontinuous part, it takes the form 
\begin{equation}
    \pi(\bx)\cP(\by-\bx)=\pi(\by)\cP(\bx-\by)\,.
\end{equation}
Thus, assuming a symmetric distribution of jump amplitudes $\cP(\bu)=\cP(-\bu)$ and a generic spatially non-uniform stationary measure ($\pi(\bx)\neq\pi(\by)$ for $\bx\neq\by$), the detailed balance condition cannot be satisfied.
Interestingly, despite its non-equilibrium nature, the relationship between correlation $\cC(t)$ and responses $\cR(t)$ has the same structure of a Gaussian process (i.e. $\cC(t)=\cR(t)\matx{C}=e^{-t\matx{A}} \matx{C}$) but with different noise matrix $\matx{D}' = \matx{D}+\lambda\Gamma=\matx{C}\matx{A}^T+\matx{A}\matx{C}$.
This implies that the usual equilibrium relations valid for Gaussian systems (Onsager $\matx{A}\matx{C}=\matx{C}\matx{A}^T$ and generalized fluctuation-dissipation relations $\dot{\cC}(t)=-\cR(t)\matx{D}'$) are not sufficient anymore to conclude that the system is in equilibrium. 
The above results are just a consequence of the linear structure of the system and the absence of moments of degree higher than the second in the expressions of $\cC(t)$ and $\cR(t)$.
When instead we take into consideration quantities that depend on such higher moments, we can better appreciate the differences with respect to a purely Gaussian process.
For instance, the entropy production rate $\Sigma$, as proven in the appendix of \cite{lucente2023poisson}, reads 
\begin{align} \label{eq:epr_poisson_linear}
    \Sigma=\text{Tr}\lx[\lx(\Delta + \lambda \Gamma\rx) \matx{D}^{-1}\matx{A}\rx] \qquad \Delta = \matx{C}\matx{A}^T-\matx{A}\matx{C}\,.
\end{align}
The expression is formally analogous to \eref{eq:entprodlin}, but it should be noted that only the Gaussian ``temperatures'', described by the matrix $\matx{D}$, enter the expression of $\Sigma$, while the Poisson noise contributes through the stationary covariance $\lambda\Gamma$, see~\cite{lucente2023poisson}. 
This observation clarifies why such processes are often called ``athermal''\cite{kanazawa2012stochastic,kanazawa2015asymptotic,kanazawa2017statistical,kusmierz2018thermodynamics}.
A direct inspection shows that $\Sigma>0$ and the minimum is attained for $\matx{A}\matx{C}=\matx{C}\matx{A}^T$, i.e. in correspondence of ``classical'' equilibrium $\Delta=0$ as defined by Onsager \cite{lucente2023poisson}.
Finally, since the system is not Gaussian, its non-equilibrium nature can be inferred from higher-order correlation functions even from a single 
time-series.
As mentioned in Section \ref{sec:perneq}, Pomeau suggests to compare $\ave{x(0)x^3(t)}$ and $\ave{x^3(0)x(t)}$ \cite{pomeau1982symetrie}: this choice allows us to estimate the degree of irreversibility of a process.
For instance,~\fref{Markov:Fig_3} shows the differences between Gaussian and Poissonian noise for the quantity 
$$
\Phi(t)=\ave{x(0)x^3(t) - x(0)^3x(t)}/\ave{x^4}
$$
in the case of a $2-$dimensional linear process.
\section{Estimates of Entropy Production} \label{sec:empirical}
\subsection{Scale-dependent entropy production \texorpdfstring{$\Sigma(\eps,\Delta t)$}{}} \label{subsec:scaent}
Measuring entropy production represents a formidable challenge, both because it requires knowledge of all the variables of the system and because of the considerable amount of data to obtain reliable estimations. Here we discuss the role played by resolution, i.e. the relevance of time sampling and coarse graining, introducing the concepts of scale-dependent entropy production $\Sigma(\epsilon,\Delta t)$ (in analogy with the $\epsilon$-entropy~\cite{abel2000exit,gaspard1993noise,boffetta2002predictability} used in dynamical system). In a nuthshell, the idea is:
\begin{enumerate}
\item Introduce a partition 
$\lx\{B_i^{(\eps)}\rx\}_{1\le i \le K}$ of size $\epsilon$ of the phase space.
\item Define an empirical Markov chain 
\begin{align*}
\pi_i &= \overline{\mathbf{1}_{B_i^{(\eps)}}(x(t))} \\
\pi_i P_{ij} &= \overline{\mathbf{1}_{B_j^{(\eps)}}(x(t+\Delta t)) \cdot \mathbf{1}_{B_i^{(\eps)}}(x(t))}
\end{align*}
where we have dropped from $\pi_i$ and $P_{ij}$ the $\eps$ and $\Delta t$ dependence and $\mathbf{1}_A(x)$ is the characteristic function of set $A$. i.e.
$$
\mathbf{1}_A(x) = \begin{cases} 1 & \text{ if } x \in A\\ 0 & \text{otherwise}\end{cases}
$$
and $\overline{f(x)}$ is the time average of $f(x)$, i.e 
$$
\overline{f(x)} = \frac{1}{n} \sum_{k=1}^n f(x(t_k))
$$
where $t_k = t_0 + k \Delta t$ and $n \gg 1$.
\item Add a regularization for the missing reverse transitions\\ (e.g. if we have $P_{ij}>0$ but $P_{ji}=0$ then we impose a very little offset $P_{ji}=\delta <1/t_\text{max}$).
\item Compute the Entropy Production Rate of such Markov chain as
\begin{align*}
\Sigma(\epsilon,\Delta t)=\frac{1}{\Delta t}\sum_{ij}\pi_i P_{ij}\log\lx(\frac{P_{ij}}{P_{ji}}\rx).
\end{align*}
\item Take the limit $\epsilon \to 0$ and $\Delta t \to 0$ of $\Sigma(\epsilon,\Delta t)$.
\end{enumerate}
In the limit of infinite data and for $\epsilon \to 0$ and $\Delta t\to 0$ one has $\Sigma(\epsilon,\Delta t)\to\Sigma$~\cite{lucente2023poisson}. However, we stress that $\Sigma(\eps,\Delta t)$ in general is just a proxy, neither a lower nor an upper bound for the entropy production rate $\Sigma$. The reason is that the coarse-grained process is not Markovian, while the scale-dependent entropy estimation relies on a Markovian approximation and therefore can exceed the entropy production of the microscopic system~\cite{van2023time}.
Let us highlight the relevance of scale resolution for the behavior of $\Sigma(\epsilon,\Delta t)$ through an example. 
Consider the one-dimensional system
\begin{align}
    &\dot{x}=-\nabla_x V(x)+f + \xi(t)+\zeta(t)\, ,\\
    & V(x)=\frac{L V_0}{2\pi}\lx(1-\cos\frac{2\pi x}{L} \rx)\,.
\end{align}
which describes a particle moving on a tilted periodic potential subject to Gaussian ($\ave{\xi^2}=2T$) and Poisson ($\ave{\zeta^2}=2\lambda\Gamma$) noises. Such a system is widely employed as a minimal model for transport phenomena \cite{luczka1997symmetric,kusmierz2018thermodynamics,bialas2020colossal,bialas2022periodic,bialas2023mechanism,spiechowicz2013absolute,spiechowicz2014brownian}, and several properties have been established. The pulling force $f$ induces a stationary current $j_s$ (even in the Gaussian case) and $\Sigma$ is positive. In Gaussian systems, the relation between entropy production rate and current is \cite{Sekimoto2010,seifertrev,cocconi2020entropy}
\begin{equation}
\Sigma\propto \frac{j_s^2}{T}\,,
\end{equation}
while for the system driven by Poisson noise it takes the form
\begin{equation}
\Sigma=\frac{j_s}{T}f + \Delta \Sigma_p
\label{eq:relation_epr_current}
\end{equation}
where
\begin{equation}
\Delta \Sigma_p = \frac{\lambda V_0 L}{2\pi T}\ave{\cos{\frac{2\pi x}{L}}} \lx(1-e^{-2\lx(\pi \Gamma/L\rx)^2}\rx)
\end{equation}
and $\ave{\cdot}$ is the average over the stationary measure.
In order to understand the effect of scale-resolution on the entropy production measurements, it is important to identify the characteristic scales of the system. 
The deterministic part of the dynamics has two relevant time-scales: the relaxation time $\tau_r$ inside each well and the average exit time $\tau_e$. The other characteristic times come from Poisson noise, which has an intrinsic temporal scale $\tau_p=1/\lambda$ (the average inter-events time) while the average amplitude size $\sigma_p\sim\sqrt{u^2}$ fixes the characteristic length scale. For coarse-graining resolutions $\eps$ greater than $\sigma_p$, the scale-dependent entropy production misses the contribution of Poisson noise since transitions to different cells due to jumps are unlikely. Therefore, if the temporal resolution $\Delta t$ is much bigger than $\tau_p$ the contribution of Poisson noise to the scale-dependent entropy production $\Sigma(\eps,\Delta t)$ cannot be appreciated. However, regarding the temporal behavior of $\Sigma(\eps,\Delta t)$ one should discuss separately the two cases $\tau_r<\tau_p$ and $\tau_r>\tau_p$. In the former, the noisy part of the dynamics is dominated by Poisson noise and hence the dynamics never resembles its Gaussian counterparts. In the latter, instead, since in a time-interval of order $\tau_r$ a large number of jumps occur, central limit theorem applies and the statistic of $\zeta$ on this time-scale is well described by a Gaussian process. Thus, for $\tau_r\gg\Delta t \gg \tau_p$ the scale-dependent entropy production $\Sigma(\eps,\Delta t)$ takes values close to the ones computed for Gaussian system, i.e. $\Sigma(\epsilon,\Delta t)\sim j_s^2/T_{eff}$ with $T_{eff}=T+\lambda\Gamma$, while for $\Delta t\ll\tau_p$ the contribution of Poisson noise is correctly taken into account. 

\begin{figure}[t]
\centering
    \includegraphics[width=0.48\textwidth]{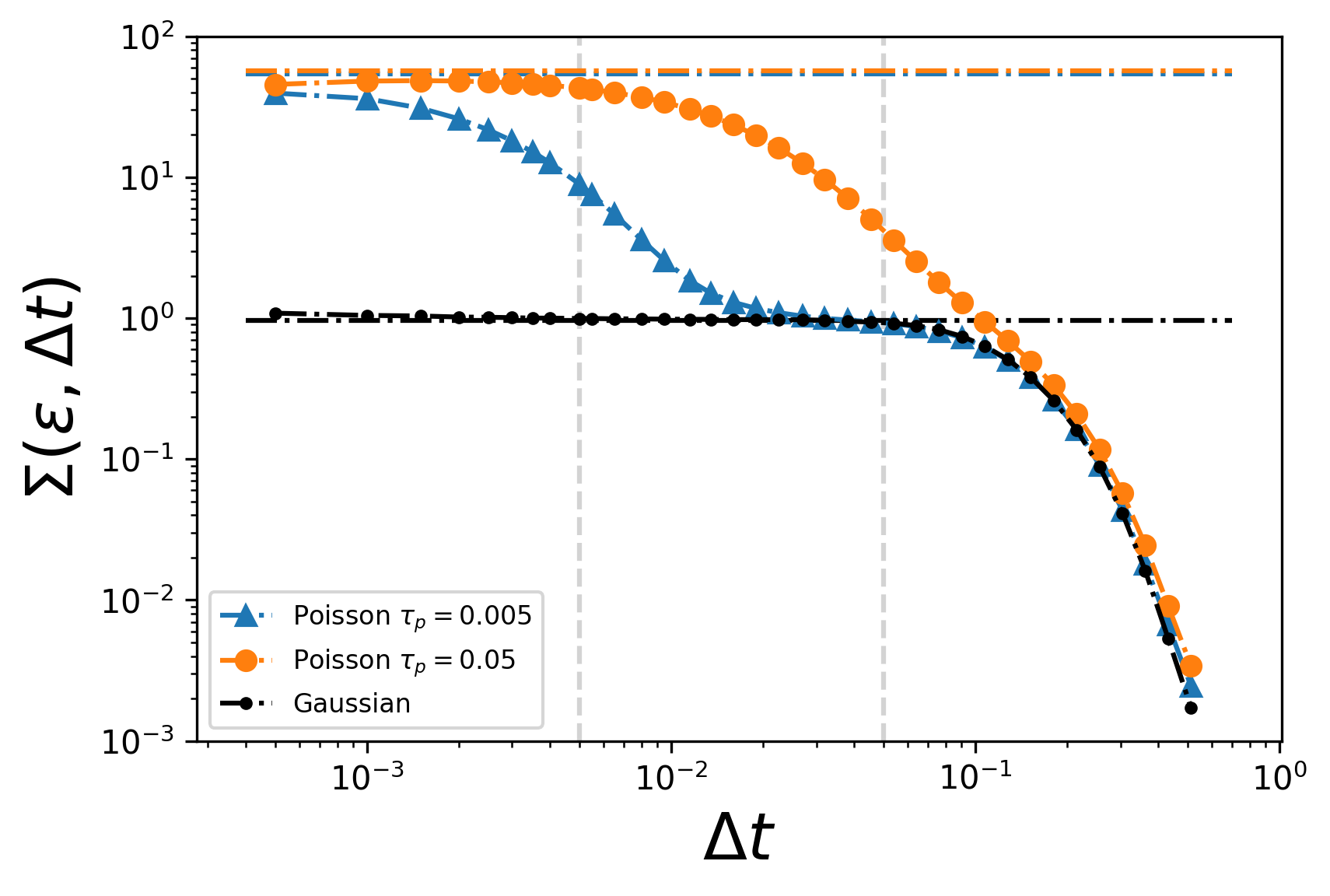}
    \includegraphics[width=0.48\textwidth]{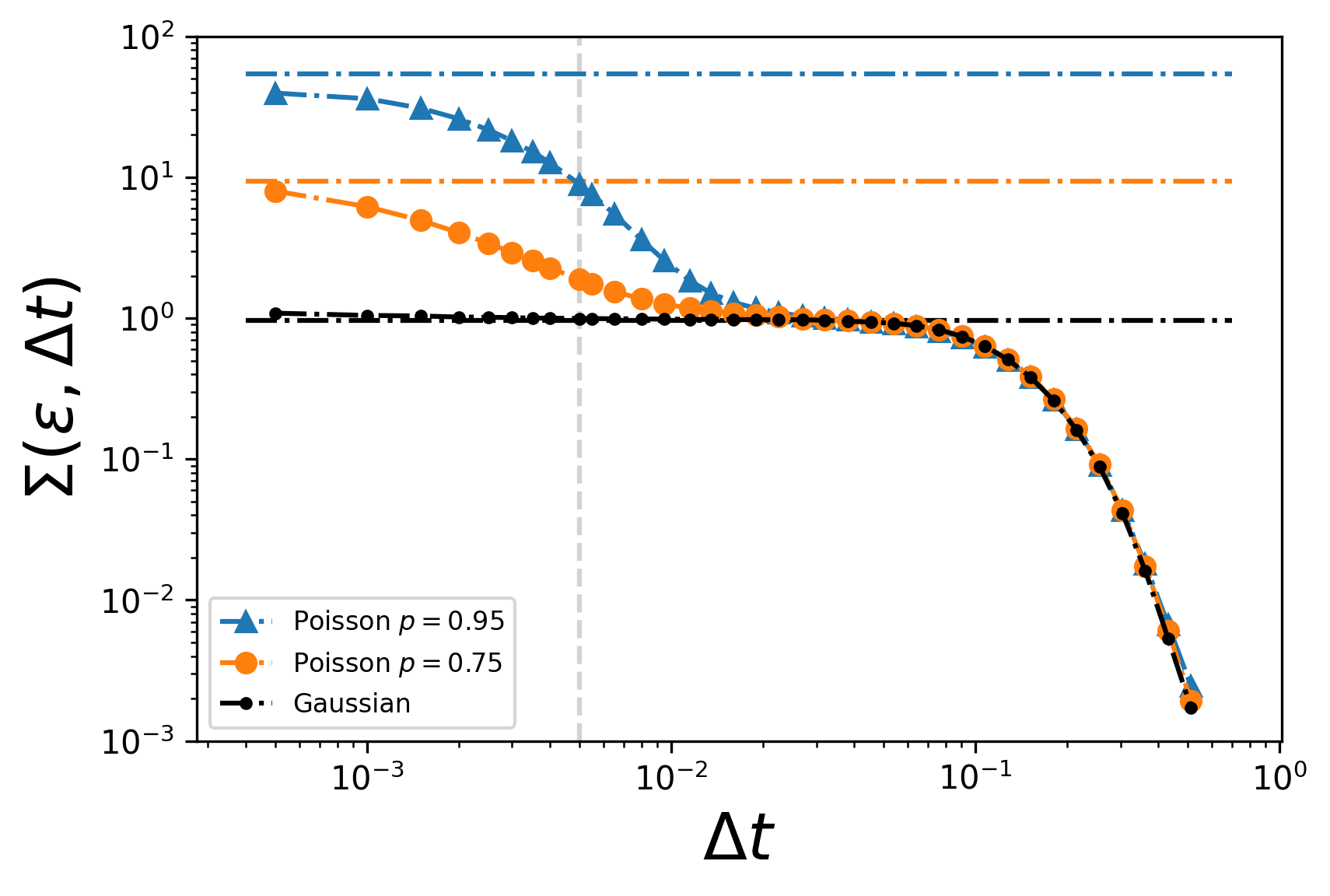}
    \caption{Empirical entropy production rate $\Sigma(\eps,\Delta t)$ as a function of $\Delta t$ for a particle in a symmetric periodic potential $V(X)$ and pulled by a constant force $f$, $\eps \simeq 4\cdot10^{-3}$. Left panel show $\Sigma(\eps,\Delta t)$ for two different Poisson jump rates $\tau_p$ ($\tau_p=0.05$ red, $\tau_p=0.005$ blue). Right panel shows the convergence of $\Sigma(\eps,\Delta t)$ towards the theoretical values (horizontal lines) for different levels of Poisson noise ($75\%\,,95\%$) for $\tau_p=0.005$ (left).} \label{Empirical:Fig_1}
\end{figure}
The above considerations are supported by numerical simulations, as shown in Fig.~\ref{Empirical:Fig_1}. The left panel shows that the Gaussian plateau arises only for intermediate temporal resolution ($\tau_r\gg\Delta t\gg\tau_p$). From a close inspection of the right panel, it becomes evident that differences between Gaussian and non-Gaussian cases (as well as differences between different Poisson noises) only arise for $\Delta t < \tau_p $. 

\subsection{Thermodynamic Uncertainty Relations \label{sec:tur}}
We have already seen that entropy production $\Sigma$ in its most general definition is a powerful concept but it has several disadvantages, particularly when $\Sigma$ has to be estimated from experimental/numerical data. 
In the cases where the full dynamical equations  are known, $\Sigma$ can take the form of a time-integrated functional of some complicated function of - in principle - {\em all the degrees of freedom} involved in such a dynamics. Therefore, a partial empirical observation cannot faithfully estimate the entropy production: in fact, not even its average rate. 
The estimation of $\Sigma$ is even more challenging for those  physical systems that do not benefit from an
accurate theoretical modeling: as we have already seen in Sec.\ref{subsec:scaent}, general recipes starting from experimental data of a few observables - possibly at a coarse resolution in time and space - are hardly useful for approximating the entropy production rate. 
In recent years, several works have been devoted to understanding the relation between the average entropy production rate (or other quantities related to it) and the currents crossing non-equilibrium systems,
especially in its steady state~\cite{seifert2019stochastic}. It is clear that one cannot hope to get - in general -  information about the total entropy production rate starting from any average current.
Indeed complex systems, typically involving many time-scales and several relevant variables, can be traversed by many physical currents and the total entropy production rate is somehow the result of the combination of all of them~\cite{puglisi2010entropy}. The clearest example of such a principle is found in irreversible thermodynamics \cite{degroot84}, where 
\begin{equation} \label{dec0}
\Sigma=  \sum_{\alpha=1}^n A_\alpha J_\alpha
\end{equation}
where $A_\alpha$ is the $\alpha$-th affinity or thermodynamic force, while $J_\alpha$ is the associated average current. A detailed treatment of this decomposition principle can be found in the Schnakenberg network theory~\cite{schnakenberg1976network}, that decomposes a non-equilibrium Markov process (living in a discrete space of states) into fundamental cycles (such as those in an electric circuit) each crossed by its own current:
\begin{equation}\label{dec1}
\Sigma=  \sum_{\alpha=1}^n A(\overset{\rightarrow}{C}_\alpha)J(\overset{\rightarrow}{C}_\alpha)
\end{equation}
where $\overset{\rightarrow}{C}_\alpha$ are the $n$ fundamental cycles of the graph associated to the process, $A(\overset{\rightarrow}{C}_\alpha)$ is the affinity or thermodynamic force that acts directly in cycle $\alpha$ (a kind of total asymmetry in the transition rates of the edges of that cycle), while $J(\overset{\rightarrow}{C}_\alpha)$ is the average net current in that cycle. Close to equilibrium the currents are linear combinations of the affinities with coefficients that compose the symmetric Onsager matrix. We do not intend to delve into the details of such a theory, but it is clear that Eqs.~\eqref{dec0} and~\eqref{dec1} require lot of information to retrieve a valid estimate of $\Sigma$. An interesting alternative to this direct measurement is represented by the so-called Thermodynamic Uncertainty Relations (TUR)~\cite{horowitz2019thermodynamic}, which 
establish a link
between the entropy production and the first two cumulants of the fluctuations of any kind of current measured in the system (instead of knowing the average of all of them and also the corresponding affinities). 

The most common TUR discussed in the recent literature provides a lower bound for the integrated (in a time $t$) entropy production $\Sigma_t$ in the form of a ``precision rate'' for the fluctuations of {\em any} non-equilibrium current integrated for the same time $t$, $J_t$, in the system (in the following we take $k_B=1$ for the Boltzmann constant):
\begin{equation} \label{tur0}
\Sigma_t \ge 2\; \frac{\langle J_t \rangle^2}{\text{Var}(J_t)}.
\end{equation}
The relation~\eqref{tur0} was first derived for some models in \cite{barato2015thermodynamic} and then 
generalized to Markov processes in steady states in \cite{gingrich2016dissipation}. 
We note that in a steady state for large $t$, one has $\Sigma_t=t \Sigma$, $\langle J_t\rangle=t J$ and $\text{Var}(J_t)\sim 2 D_J t$, 
where $D_J$ is the diffusivity associated to the current rate whose average we denote by $J$, leading to a rate version of Eq.~\eqref{tur0}:
\begin{equation} \label{tur1}
\Sigma \ge  \frac{ J^2}{D_J}.
\end{equation}

The power of this relation comes from its generality: it stands true for a very wide set of situations and physical systems, and it involves any observable current. 
Its downside, obviously, is in the fact that it only provides us with an inequality which cannot, in general, be proved to be tight. 

In the few years after the TUR was first proved, several follow-up results have better enlightened the origins of the inequality and its applicability to non-equilibrium thermodynamics, see \cite{horowitz2019thermodynamic} for a first review with perspectives. 
To better understand the meaning of Eq.\eqref{tur0}, it is interesting to discuss what happens in the close-to-equilibrium limit: in that case, as mentioned above, the linearity between currents and affinities makes the entropy production rate take a bilinear form  $\Sigma=\sum_{\beta,\gamma} L_{\beta\gamma}A_{\gamma}A_{\beta}$~\footnote{This form is much more general than the case discussed above of Markov processes on a network, it encompasses all cases where a discrete set of currents can be identified in a system close to equilibrium.}.  
The Einstein relation implies that the diffusion coefficient for the fluctuations of the time-integral of the current $J_\alpha$ is $D_\alpha=L_{\alpha\alpha}$, therefore the TUR relation for the $\alpha$ current reads~\cite{barato2015thermodynamic} 
\begin{equation} \label{tur2}
\frac{\Sigma}{ J_\alpha^2/D_\alpha} =\frac{L_{\alpha\alpha}\sum_{\beta,\gamma}L_{\beta\gamma}A_\beta A_\gamma}{\sum_{\beta,\gamma}L_{\alpha\beta}L_{\alpha\gamma}A_\beta A_\gamma} \ge \left(1 +\frac{\sum_{\beta,\gamma \neq \alpha} G_{\beta\gamma} A_\beta A_\gamma}   {J_\alpha^2}\right) \ge 1,
\end{equation}
where in the last passage we have used the fact that $G_{\beta\gamma}=(L_{\alpha\alpha}L_{\beta\gamma}-L_{\alpha\beta}L_{\alpha\gamma})$ can be proven to be a positive semi-definite matrix (this is a consequence of the fact that the Onsager matrix $L$ is also positive semi-definite). 
From these few lines of calculations, one learns that, in the equilibrium limit, the equality is obtained if $A_\beta=0$ for each $\beta \neq \alpha$. 

Generalizations of the TUR and its applications to various inference problems, particularly for the maximum efficiency of molecular motors and for the minimal number of intermediate states in enzymatic networks, are discussed in \cite{seifert2019stochastic}. 
Several techniques have been employed to derive TURs, including large deviation theory \cite{gingrich2016dissipation,gingrich2017inferring},
bounds to the scaled cumulant generating function (see for instance \cite{dechant2018current,koyuk2020thermodynamic} and \cite{koyuk2020thermodynamic,dechant2020fluctuation}), and other approaches. 
 An interesting way to derive it, is the application of the generalized Cram\'er-Rao inequality \cite{liu2020thermodynamic}, which includes quantifying
the Fisher information of the Onsager-Machlup measure of the path and a virtual ``tilt'' of the original dynamics. 
However, such a strategy cannot be directly applied to a system with underdamped dynamics, raising the necessity of alternative treatments, see \cite{hasegawa2019uncertainty,lee2021universal}. 
The Cram\'er-Rao approach has been used to derive generalized TURs that are optimized to give a lower bound for diffusivity of a tracer particle under the action of non-linear friction and non-equilibrium baths with multiple 
time-scales and multiple temperatures \cite{plati2023thermodynamicbounds,plati2024thermodynamicuncertainty}. 
A recent more direct derivation of TUR, with a discussion about how it can be saturated, has been proposed in \cite{dieball2023direct}. 

As discussed in \cite{gingrich2017inferring}, the TUR bound is most useful if it is tight, but there are two main reasons why it might be loose: 
(1) the distribution of the current fluctuations is non-Gaussian and 
(2) the choice of the current can be sub-optimal, i.e. it does not contain enough information about the total entropy production.
First applications to the problem of inferring $\Sigma$ from the study of currents fluctuations were obtained in works \cite{gingrich2017inferring,li2019quantifying}. 

In order to give an idea of the limits of the TUR strategy in the entropy production inference, in the following we discuss some interesting attempts to evaluate it.
In~\cite{gingrich2017inferring}, a diffusion process in two dimensions is studied, with an external field driving the system across the four wells of a landscape. 
The transitions among the four quadrants, each containing one of the wells, are measured to play the role of $J$ as mentioned earlier. 
This task requires measuring {\em all the degrees of freedom}, i.e. the process in its full dimensionality. 
The authors concluded that the TUR can underestimate the entropy production by a factor which is between $0.2$ and $0.8$ and, interestingly, the error on the estimate is not seriously affected by the strength of the external driving, but rather, it is affected by the depth of the wells. 
The estimated power of the TUR increases when the wells are deeper, likely because the coarse-graining into four quadrants is closer to the physics of the process. 
In \cite{li2019quantifying}, a two-dimensional ``bead-spring'' model, in practice a Brownian Gyrator with conservative coupling and different temperatures, was studied and the authors compared two approaches to estimate $\Sigma$: 1) an empirical direct measurement of the entropy production from a {\em vectorial} time-series of the numerical solution of the model (an approach that not only requires knowledge of all relevant degrees of freedom, but even when all d.o.f. can be measured it is  doomed to fail as the dimensionality of the problem increases); 2) a measure of the average and variance of a scalar current obtained from some projection of all the d.o.f.: these two cumulants are easy to obtain with a small amount of data and can be plugged into Eq.~\eqref{tur1} in order to get an estimate of $\Sigma$; the authors first proved that an optimal choice (leading to estimates close to the real value of $\Sigma$ in the same order of magnitude) of the measured current is $j_F=\int d{\mathbf x}\, \mathbf{F}({\mathbf x}) \cdot \mathbf{j}({\mathbf x})$ where $\mathbf{F}({\mathbf x})$ is the thermodynamic force acting on the system at point $\mathbf{x}$ and $\mathbf{j}({\mathbf x})$ is the local  current, where both quantities are empirically determined (by averaging over a long trajectory) but, again, they require the knowledge of all the degrees of freedom of the system.  Further investigation about the power of inference of the TUR was conducted in \cite{vanvu2020entropy}, where a deterministic method to estimate entropy production based on the TUR for classical  Markovian dynamics was suggested, by computing an "optimal" current that maximizes the lower bound. In this context, it was shown that the optimal current saturates the TUR for overdamped Langevin dynamics driven by Gaussian white noise, but, as it will become clear in the following, for general Markovian processes this is not always granted.
Inference, however, relies on a wise choice of a basis for currents (in principle any functional basis) and again the measurement of all the relevant degrees of freedom. Thus, real applications in~\cite{vanvu2020entropy} are limited to simple systems such as a 4 states Markov jump process, a driven Brownian particle that circulates on a ring with a periodic potential, finally a multidimensional bead spring model (a Gaussian continuous process), where the knowledge of all the variables is needed. Notably, other strategies relying upon the measurement of an optimal current have been proposed~\cite{manikandan2020inferring,otsubo2020estimating}. In \cite{manikandan2020inferring}, the authors build a large set of currents as random linear combinations of empirical microscopic currents and define the optimal choice as the one that maximizes the bound in the limit of infinitely short trajectories. It is shown that such a procedure provides estimates of the average entropy production rate arbitrarily close to the real value and, moreover, can also be used to obtain its probability distribution. In~\cite{otsubo2020estimating} instead, the authors consider the possibility of finding the best estimate (largest bound) by using machine learning procedures. 
All the methods discussed so far improve the bound through an optimization scheme for the observable to be measured. In \cite{dechant2021continuous}, the authors adopt a completely different point of view.
Introducing a continuous family of stochastic dynamics the authors derive a tighter version of the TUR connecting entropy production rate and currents between different members of the family. In addition, it is shown that this bound is saturated by an appropriate choice of the observable. A more practical approach consists in improving TURs by considering more information besides the fluctuations of a current: this obvious concept has been put in an interesting form in \cite{dechant2021improving}, where it has been shown that 
\begin{align}
\eta_J + \chi_{J_t,Z}^2 &\le 1\\
\eta_J&=\frac{2\langle J_t \rangle^2}{\text{Var}(J_t)\Sigma_t}\\
\chi_{J_t,Z}&=\text{Cov}(J_t,Z)/\sqrt{\text{Var}(J_t)\text{Var}(Z)}
\end{align}
where $\chi_{J_t,Z}$ is the Pearson correlation coefficient between $J_t$ and any variable $Z$, while $\text{Cov}(a, b)$ is the covariance between variables $a$ and $b$. Note that the usual TUR states that $\eta_J\le 1$ while the term $\chi_{J_t,Z}$ is bounded ($ -1 \le \chi_{J_t,Z} \le 1$) and actually improves the estimate. 
It is also important to mention that, recently, another relation, named Variance Sum Rule, connecting the entropy production to force fluctuations has been derived \cite{diterlizzi2024variance}. The power of this relation is that it is an {\em equality}, rather than an inequality 
 and it has been successfully applied to determine the entropy production in experiments with an optically trapped colloid and with in-vivo red blood cells. At first glance, the approach requires the explicit knowledge of the process in its full dimensionality and the forces acting on the system; however the novelty of this recipe makes it difficult to estimate its future applications.

From the previous discussion we understand that the usefulness of the TUR bound for systems with many degrees of freedom and several currents is questionable. There is an interesting case, however, where the TUR could be useful in a "re-normalised" form also in the case of several degrees of freedom. That is the case where a single macroscopic current is measured, of the form
\begin{equation}
J_{t}^{macro} =\frac{1}{n} \sum_{\alpha=1}^{n} J_t^\alpha,
\end{equation}
and the $J_\alpha$ are all related to {\em equivalent} degrees of freedom. A typical example of this condition is the case of flagella in microscopic living systems, for instance the tail of a sperm cell or the two flagella making a C. reinhardti algae swim (both examples belong to the same category of flagella that are constituted by a so-called "axoneme" with a very conserved structure, occurring also in other cells or living beings)~\cite{gilpin2020multiscale}. In these flagella, a travelling wave produces the noisy periodic beating - which is responsible for swimming under viscous conditions. The accumulation of periods (phase) of the travelling wave represents an integrated current $J_{macro}$ whose fluctuations could be used in the TUR to estimate the entropy production of the flagellum. However such a wave is produced by the concurrence of $n$  molecular motors ("dyneins") innervating the axoneme, with $n \sim 10^3 - 10^5$ (depending on the  length of the flagellum, which on its turn depends upon the organism and/or its age)~\cite{goldstein2011emergence}. Each molecular motor performs its own periodic motion whose accumulated phase represents a microscopic current $J_\alpha$: the dynein is known to be close to optimal in the TUR sense, as the bound is smaller than the real dissipation rate by a factor $\eta \approx 0.2-0.5$ (see~\cite{hwang2018energetic}). 
If the organism dissipates energy almost only in the ATP consumption for feeding the molecular motors, as it happens for a sperm cell, then the total dissipation rate of the structure is of order $n$ times the dissipation rate of a single molecular motor. While the average of $J_t^{macro}$ is the same as the average of the molecular motor currents, i.e. $\langle J_t^{macro} \rangle=t J_{macro}=t J_\alpha$ for any $\alpha$ (all motors are equivalent), its diffusivity $D_{macro}$ can take values between a minimum $D_\alpha/n$ in the case of totally asynchronous motors, and a maximum $D_\alpha$ in the extreme case of totally synchronized motors. Then, in the  case of totally asynchronous motors, the TUR is close to be saturated (with similar efficiency $\eta$) because $J_{macro}^2/D_{macro}=n J_\alpha^2/D_\alpha$ and the macroscopic dissipation rate is $n$ times larger than the motor one. On the contrary, in the extreme case of totally synchronized motors, the real total entropy production rate is $n$ times larger than the TUR bound  which take a similar value to the bound for the single molecular motor  $J_{macro}^2/D_{macro}=J_\alpha^2/D_\alpha$. Recent experiments with sperms flagella and a study of synchronization models with different kinds of noise suggest that the second situation is more likely to happen in real axonemes, corroborating a conjecture of strong coupling between adjacent molecular motors, see \cite{maggi2023thermodynamic,costantini2024thermodynamic}.
 \begin{figure}[ht!]
     \centering
      \includegraphics[width=10cm]{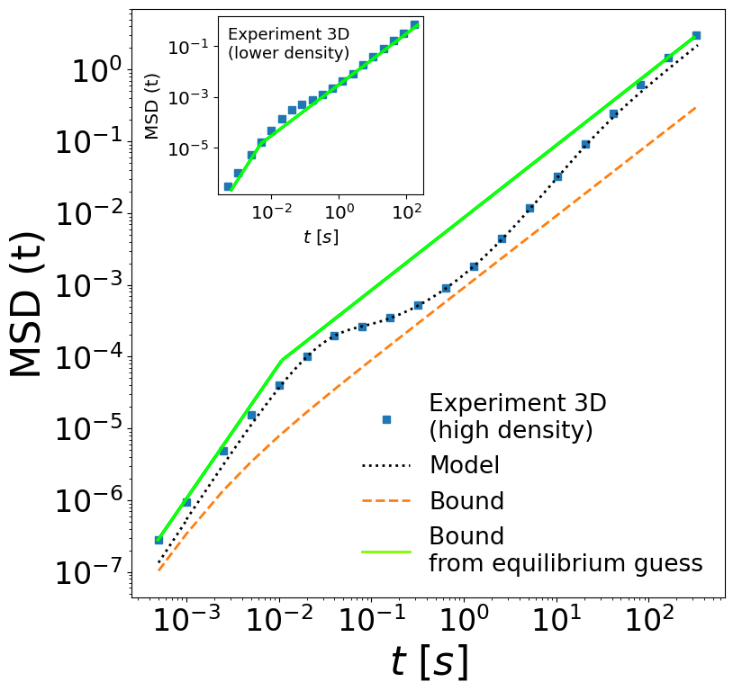}
     \caption{MSD of a large intruder immersed in a vibrated granular fluid at high density, see~\cite{plati2023thermodynamicbounds}. The equilibrium guess is constructed by connecting the two slopes of the ballistic and the diffusive regime following what we would expect at equilibrium from Eq.~\eqref{msd2}. The experimental data are below the equilibrium guess and therefore Eq.~\eqref{msd2} is violated, i.e. the data are not compatible with equilibrium. The inset shows an MSD (in the dilute regime) whose form is compatible with thermodynamic equilibrium. \label{Empirical:Fig_2}}
 \end{figure}
 
As discussed before, when the statistics of the process under investigation is Gaussian and a limited set of variables can be measured (smaller than the complete set of relevant degrees of freedom) the discrimination of equilibrium from non-equilibrium is basically impossible on a general ground, since the same empirical time series is compatible with both equilibrium and non-equilibrium models. However, in particular cases, one may invoke additional hypothesis and assumptions that restrict the field of compatible models, making the discrimination possible. For instance in \cite{plati2023thermodynamicbounds} it is seen that a TUR can be proven for the diffusion of the position $\theta(t)$ of a tracer particle under the influence of multiple baths, taking the form:
\begin{equation} \label{msd}
\langle \Delta \theta(t)^2 \rangle \ge \frac{2 \langle \Delta \theta(t) \rangle^2}{\Sigma_t^{ext}+\mathcal{I}}
\end{equation}
where $\Sigma_t^{ext}$ is the part of the entropy production (integrated along the time $t$), which only originates from the presence of an external driving force (i.e. excluding the entropy produced because of heat flowing between different thermal baths), while $\mathcal{I}$ is related to the Fisher information for a linear perturbation of the dynamics \cite{Cover2006}. When the multiple baths are at thermal equilibrium,  it appears that Eq.~\eqref{msd} reduces to a simpler expression
\begin{equation} \label{msd2}
\langle \Delta \theta(t)^2 \rangle \ge \frac{t^2}{a t +b}
 \end{equation}
 where $a=1/(2D)$ being $D$ the diffusivity of the tracer particle, while $b=m/T$ being $m$ and $T$ the mass and temperature of the tracer particle: in this way the full bound can be deduced by observing the mean squared displacement in the long time limit ($\sim Dt$ with $D$ the diffusivity) and in the short one ($\sim T/m t^2$) and therefore an immediate evaluation of the validity or violation of inequality~\eqref{msd2} can be done, see an example with experimental data in Fig.~\ref{Empirical:Fig_2}.
 If the inequality is violated, then the equilibrium hypothesis can be immediately ruled out. If the inequality is not violated, however, nothing can be said about the equilibrium or non-equilibrium character of the system. A similar situation has been discussed in the recent~\cite{dechant2023thermodynamicconstraints}, where constraints on the power spectrum of a continuous stochastic process can be used in the same way: it can exclude equilibrium in specific situations.

\subsection{An application for TUR and \texorpdfstring{$\Sigma(\eps,\Delta t)$:\\}{}the Poissonian-Brownian Gyrator}
\begin{figure}[h]
    \includegraphics[width=0.5\textwidth]{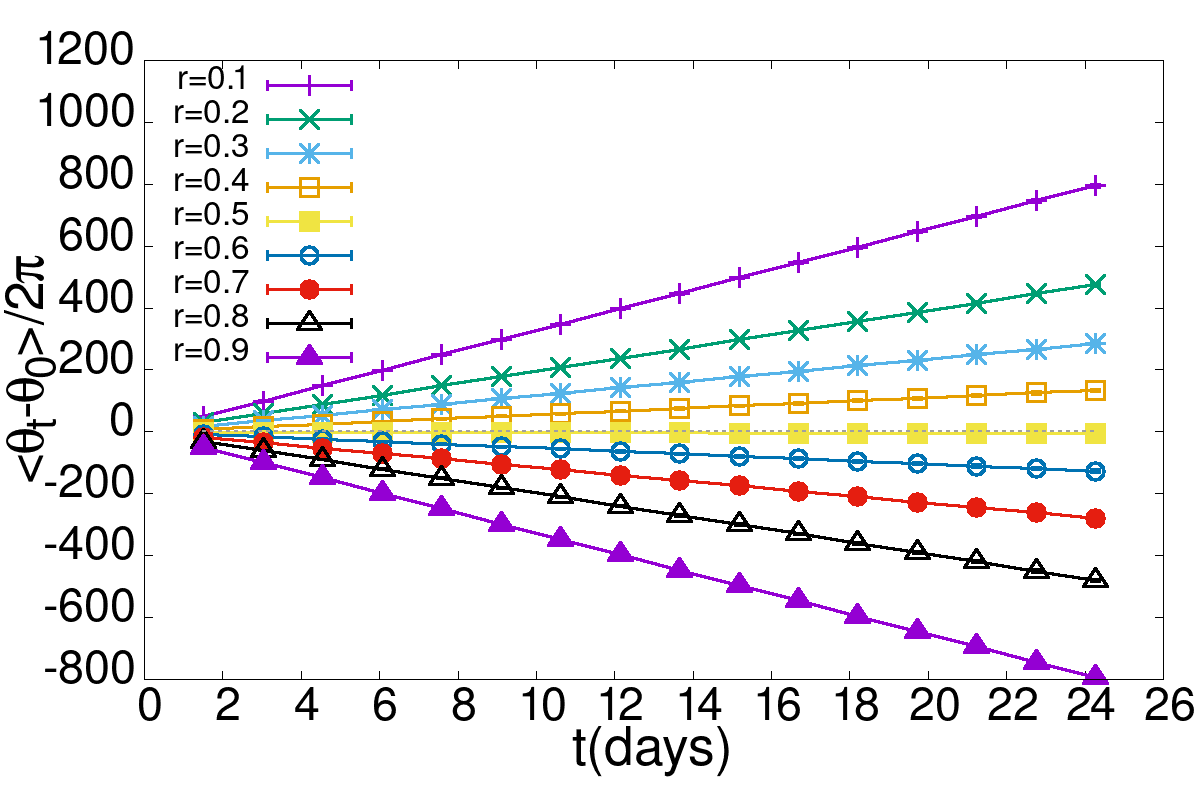}
    \includegraphics[width=0.5\textwidth]{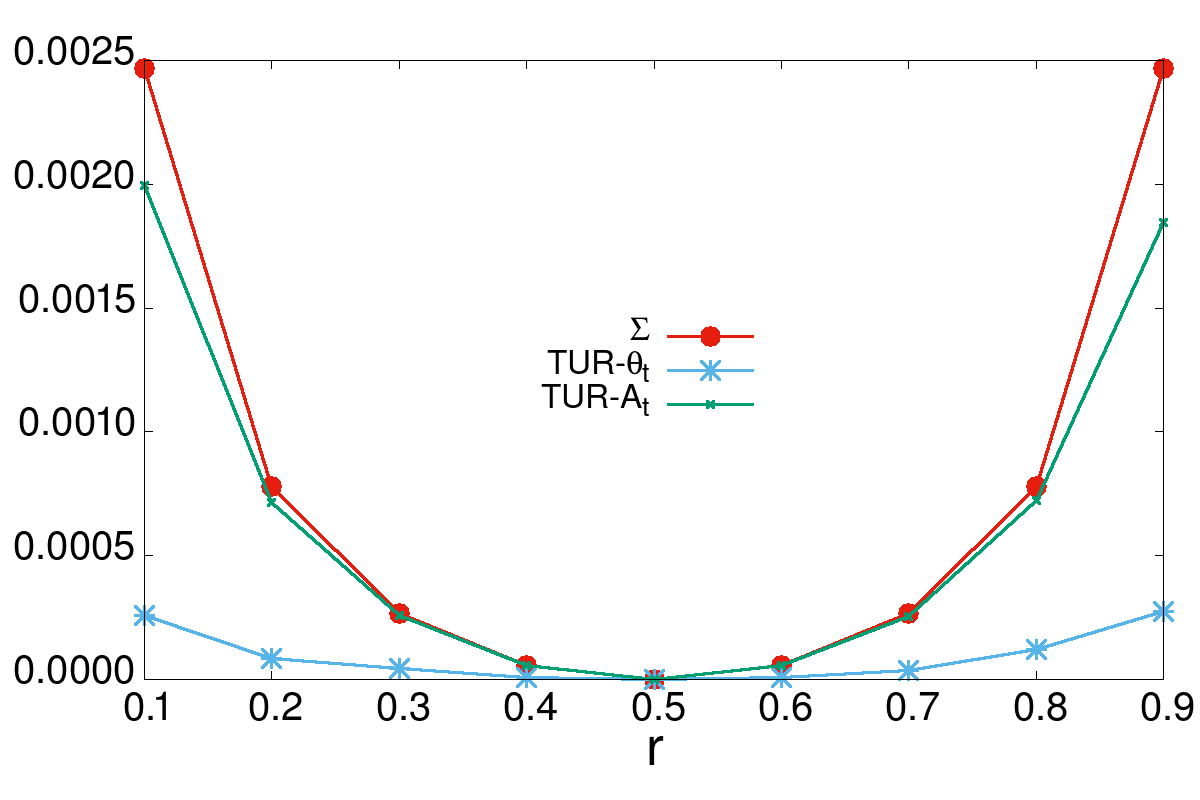}
    \caption{Currents in Brownian Gyrator. Left: the average rotation angle traveled by the system as a function of time, different plots represent different values of $r$. 
    The angular velocity of the system can be obtained by looking at the slope of the lines. Right: lower bounds obtained by looking at the TUR computed with the "integrate currents" $A(t)$ (green) and $\theta(t)$(blue) compared to the true value of entropy production (red).
    See footnote for details of the numerical simulations that produced the figures of this section.}
    \label{Empirical:Fig_3}
\end{figure}

\begin{figure}[ht!]
    \includegraphics[width=0.5\textwidth]{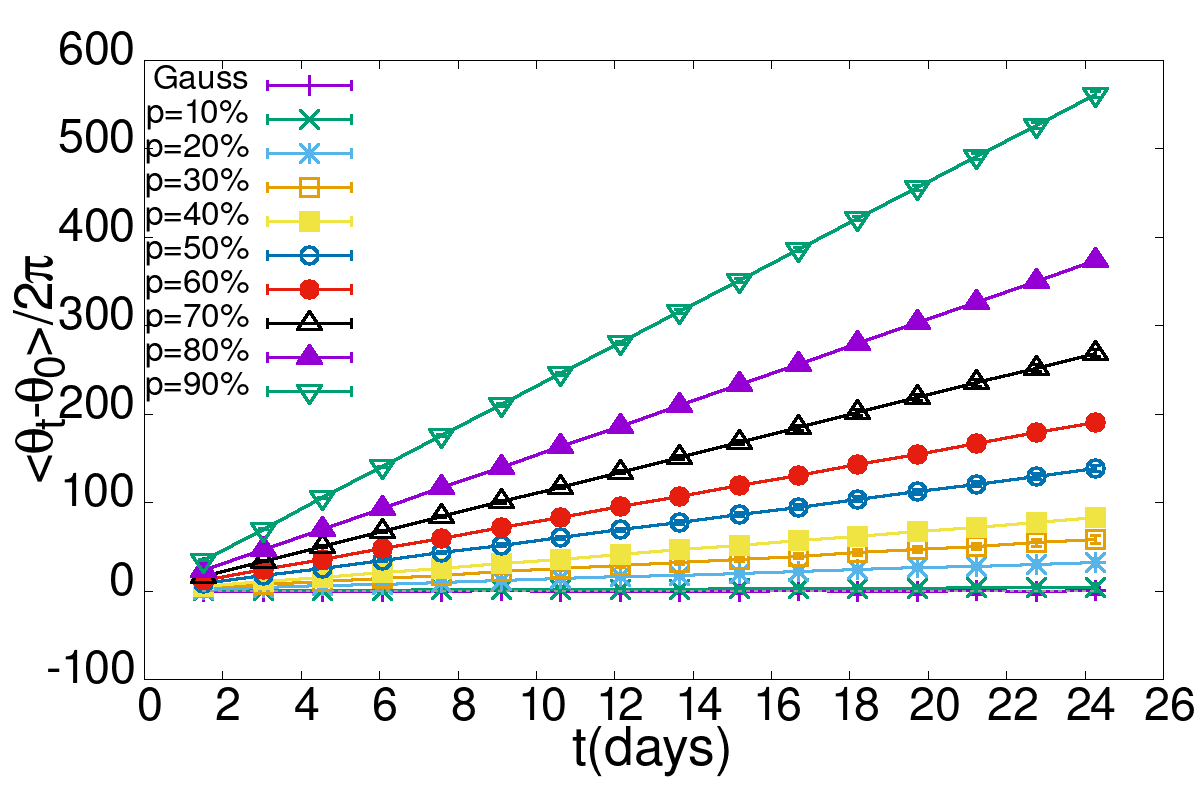}
    \includegraphics[width=0.5\textwidth]{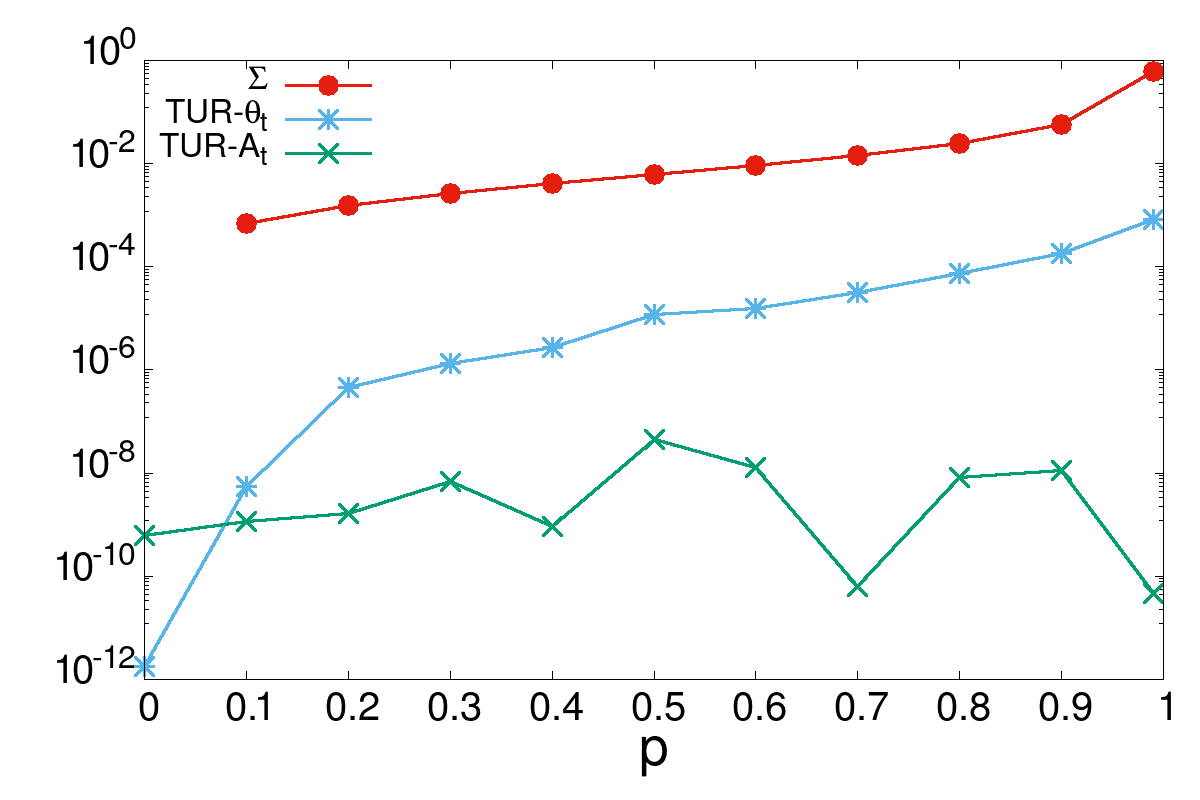}
    \caption{Currents in Poissonian Gyrator when Onsager equilibrium condition holds ($r=0.5$). Left: average rotation angle for increasing Poisson noise contribution $p \in [0.1,0.9]$. We can note that, although the average of angular momentum $\ave{l(t)}$ vanish, the system is clearly rotating around the origin faster and faster as $p$ increases.
    Right: Comparison between the TURs and the true value of entropy production. The TUR computed with $A(t)$ it's not significant, the one computed with $\theta(t)$ is better but, however, two orders of magnitude smaller than the true value.}
    \label{Empirical:Fig_4}
\end{figure}

Let us now illustrate with a practical example some critical issues related to the techniques described above.
Consider the case of Brownian gyrators defined in Sec. \ref{subsec:gyrator} by \eref{eq:BG} when the drift is isotropic ($d=a$ and $c=b$) and add a Poissonian jump process with rate $\lambda$ just along the $x$ component, i.e.
\begin{align}\label{eq:PBG}
&\lx\{
\begin{array}{ll}
\dot{x} + ax = by + \xi + \sum_k u_k \delta(t-t_k) & \ave{\xi^2} = 2T r (1-p) \\
\dot{y} + ay = bx + \xi' & \ave{\xi'^2} = 2T (1-r)\\
\lambda\sigma^2 = 2T r p & \ave{u_k^2}=\sigma^2 \; \forall k \\
\end{array} \rx.
\end{align}
We have seen that, for a purely Gaussian process, when the quantity $\Delta = b(T_y-T_x)$ is different from zero (in our model when $p=0$ and 
$r \neq 1/2$), such system exhibits a non-zero probability current that makes it rotate around the origin. We can better appreciate this by noting that the average of the angular momentum $\ave{l(t)}=\ave{x(t)\dot{y}(t)-y\dot{x}(t)} \propto \Delta$ is different from zero. Therefore, it seems natural to consider the integral of the angular momentum $A(t)=\int_0^t dt' l(t')$ to estimate the entropy production through the TUR shown in Sec. \ref{sec:tur} \eref{tur0}.
As \fref{Empirical:Fig_3} shows \footnote{All numerical simulations are based on the exact algorithm described in the Appendix \ref{subsec:sims} and  the main parameters are $a=6.25\cdot10^{-2}$, $b=2.18\cdot10^{-2}$, $T=10^{-2}$ and $\lambda=1/960$. In this way the correlations are $\mathcal{O}(1)$ and the two typical time scales due to the drift are approximately $\tau_<=120$ and $\tau_>=240$ steps (our time step is fixed to $1$), well below the average time between two jump $1/\lambda=960$.}, this estimate (green line) is quite good.
For comparison we show the TUR estimate obtained using a different current, i.e. the rotation angle $\theta(t)=\int_0^t dt' \omega(t')$ where $\omega(t')=\frac{d}{dt'}\arctan{[y(t')/x(t')]}$  (blue line), which appears considerably worse. Note that, at equilibrium for $r=0.5$, rotation is absent and $\Sigma = 0$.
But, what does it happen when we add jumps to the equation?
Once $r$ and $\lambda$ are fixed, we can study the contribution of Poissonian noise by varying  parameter $p$ in the interval $[0,1]$. In this way the two-time correlations do not change as $p$ varies and this means that, for example, the average of the angular momentum $\ave{l(t)}$ will not depend on the fraction $p$ of Poissonian noise present in the system.
We know from the previous Section that the addition of a Poissonian noise, even in the case of Onsager equilibrium (for which $r=0.5$ corresponds to $\mathbb{A}\mathbb{C}-\mathbb{C}\mathbb{A}^T=\Delta=0$), brings the system out of equilibrium and with an entropy production rate which is strictly positive $\Sigma \propto p/(1-p)$. Because of the lack of sensitivity of $A(t)$ to the Poissonian noise, the TUR computed with the integrated angular momentum $A(t)$ will  not be significant, although a rotation of the system is observed and increases as the Poissonian noise increases (\fref{Empirical:Fig_4}). This is because the statistic of the rotation $\theta$ depends on all the moments of the path distribution, not only on the second one, making it well affected by the amount of  Poissonian noise. As clear from \fref{Empirical:Fig_4}, the estimate obtained with the TUR is now better if $\theta(t)$ is used instead of $A(t)$, however in both cases it is not particularly good. The worst results with TURs in the case of Poissonian noise is coherent with what discussed above, i.e. that optimal estimates can be reached only in the case of Gaussian fluctuations, provided that proper currents are measured.
\begin{figure}[ht!]
    \includegraphics[width=0.5\textwidth]{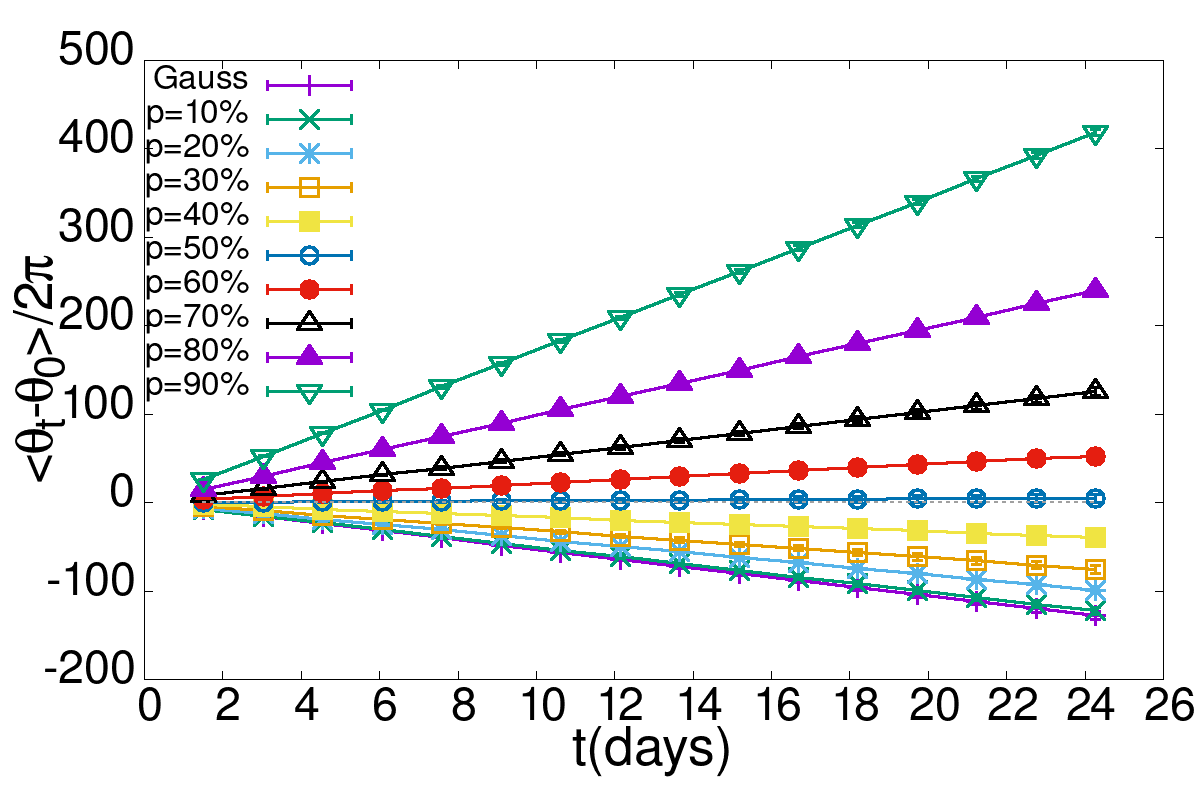}
    \includegraphics[width=0.5\textwidth]{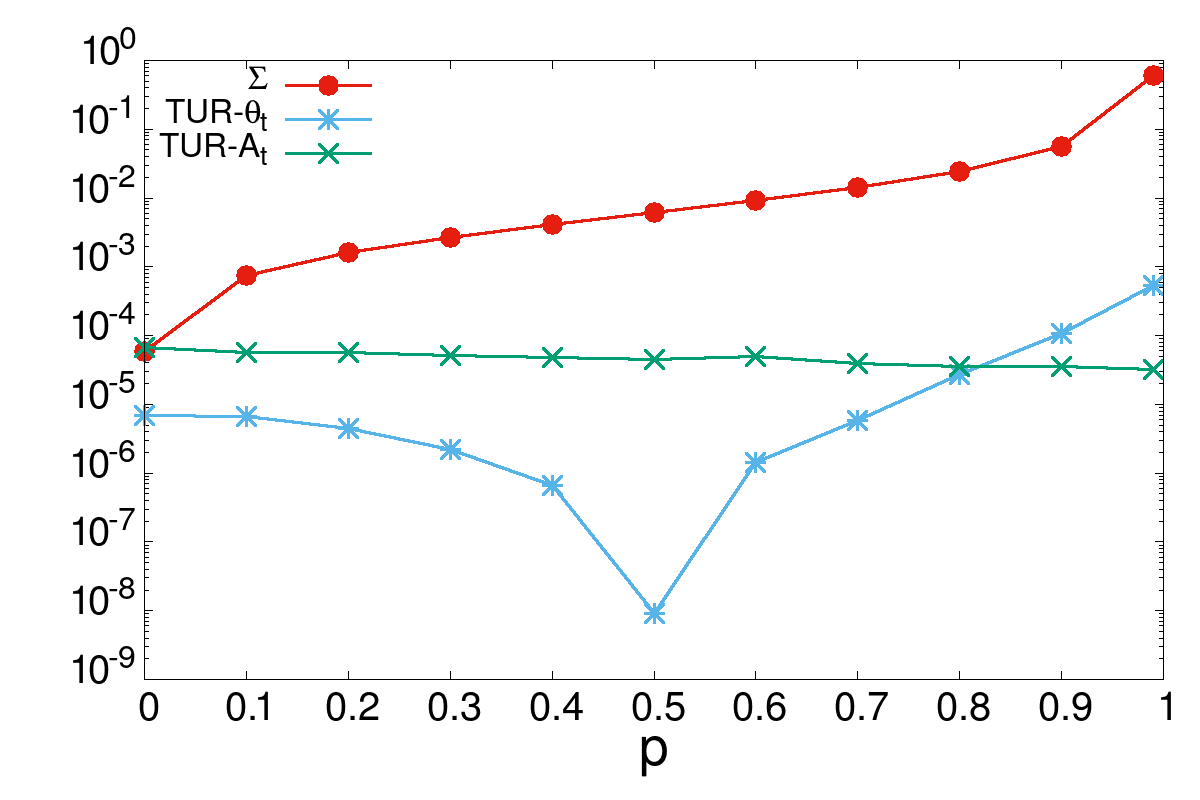}
    \caption{Currents in Poissonian Gyrator when Onsager equilibrium condition is broken ($r=0.6$). Left: as the Poissonian contribution increases the angular current changes sign. Right: TUR estimates are quite bad for both $A(t)$ and $\theta(t)$. Note that, for each value of $p$, the one obtained with $A(t)$ is more or less the value that the entropy production rate would have in the case of pure Gaussian noise ($p=0$).}
    \label{Empirical:Fig_5}
\end{figure} 
One can also verify the possibility of stall points: 
with the addition of Poissonian noise, just as the system rotates even when $\Delta=0$, in the same way the system may stop rotating even when $\Delta > 0$.
As we show in the left panel of \fref{Empirical:Fig_5}, although we weakly break the Onsager condition ($r=0.6$), we can find a value of $p$ (in our case $p=0.5$) for which there is no rotation at all. The rotation current changes sign when going through this stall point. 
Obviously, as we show in the right panel of \fref{Empirical:Fig_5}, these points have  destructive consequences on the TUR estimates.

\begin{figure}[ht!]
    \includegraphics[width=0.5\textwidth]{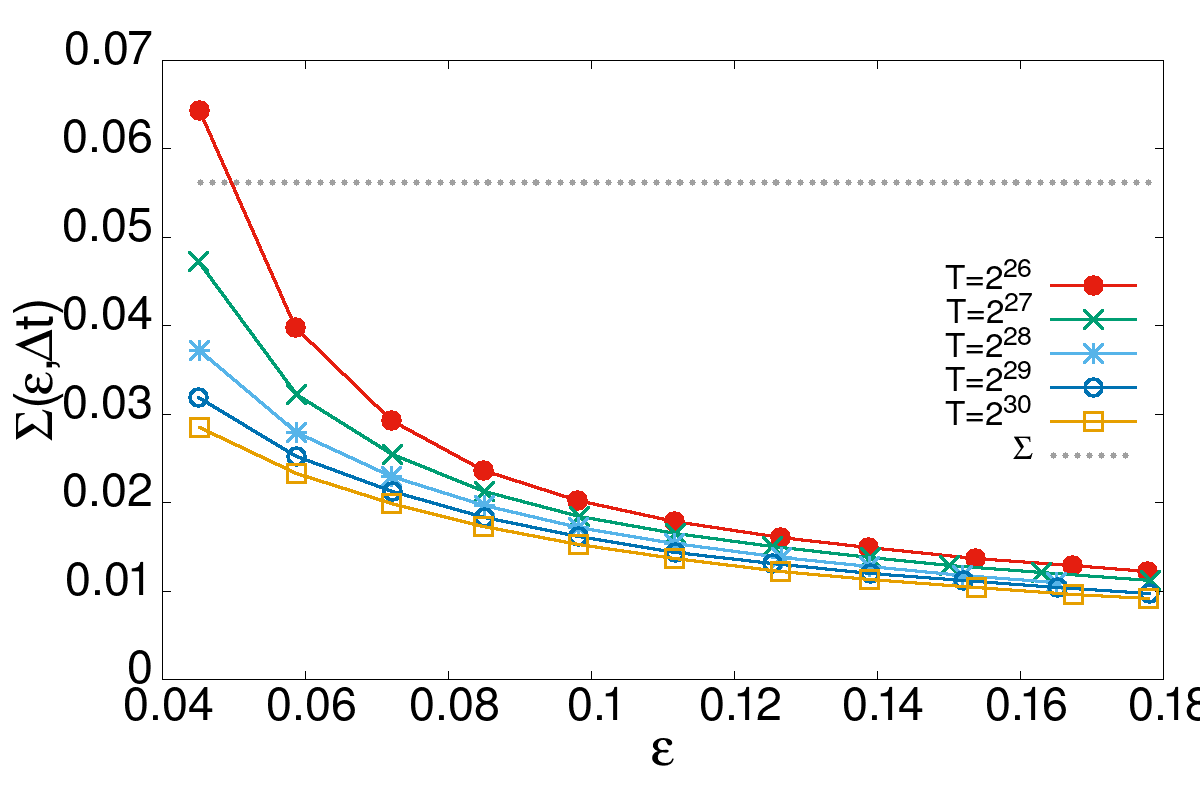}
    \includegraphics[width=0.5\textwidth]{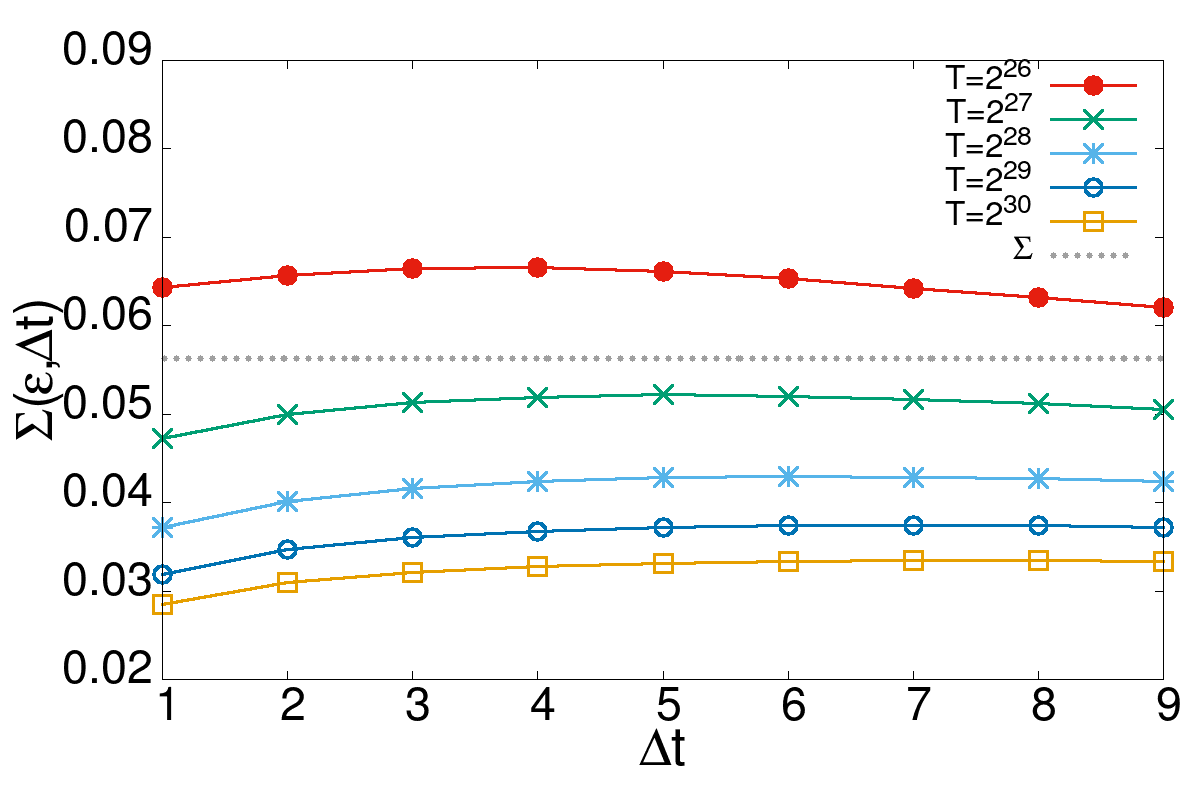}
    \caption{$\Sigma(\eps,\Delta t)$ for the Poissonian Gyrator~\eqref{eq:PBG}. Left: we fix $\Delta t=1$ and we look at the trends with $\eps \to 0$. Right: we fix $\eps=4.5 \cdot 10^{-2}$ and we look at the trends with $\Delta t \to 0$. The simulations have $r=0.5$ and $p=0.9$. The typical displacement for $\Delta t = 1$ is order $\sqrt{\ave{dx^2 + dy^2}} \simeq 0.1$}
    \label{Empirical:Fig_6}
\end{figure}

We conclude this Section with a comparison of TUR-estimates with the numerical algorithm described in the Section \ref{subsec:scaent}.
Compared to the examples of \fref{Empirical:Fig_1}, for which we had to partition a simple one-dimensional ring, in the case of gyrator we have to deal with an unlimited two dimensional phase space.
This requires additional approximations (e.g. to partition the system only up to a certain distance from the origin and neglect the dynamics that occur outside) that could spoil the effectiveness of our method which, when informed with a finite statistics, tends to be too doped and sensitive to several details, e.g. to the regularization necessary to manage the large number of the missing reverse transitions: it seems practically impossible to sample effectively the dynamics  and extrapolate $\Sigma$ from $\Sigma(\eps,\Delta t)$ for $\eps,\Delta t \to 0$.
In \fref{Empirical:Fig_6} we show the dependence on $\eps$ and $\Delta t$ of $\Sigma(\eps,\Delta t)$ for different lengths $T$ of the samples used to estimate the transition probabilities, from some millions to a billion of steps. We can note that, although the order of magnitude seems reasonable, a precise value of $\Sigma$ cannot be deduced from the trend we obtained even with very expensive numerical simulations.
\begin{figure}[ht!]
    \includegraphics[width=0.5\textwidth]{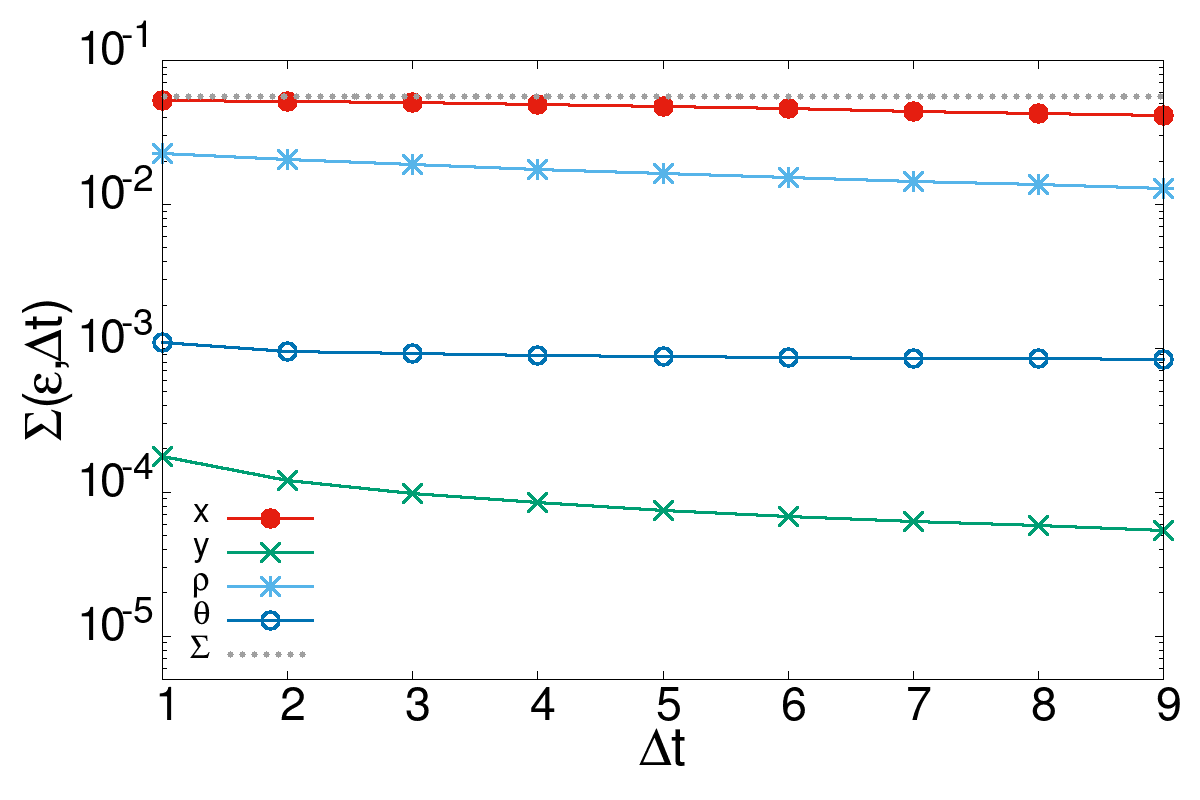}
    \includegraphics[width=0.5\textwidth]{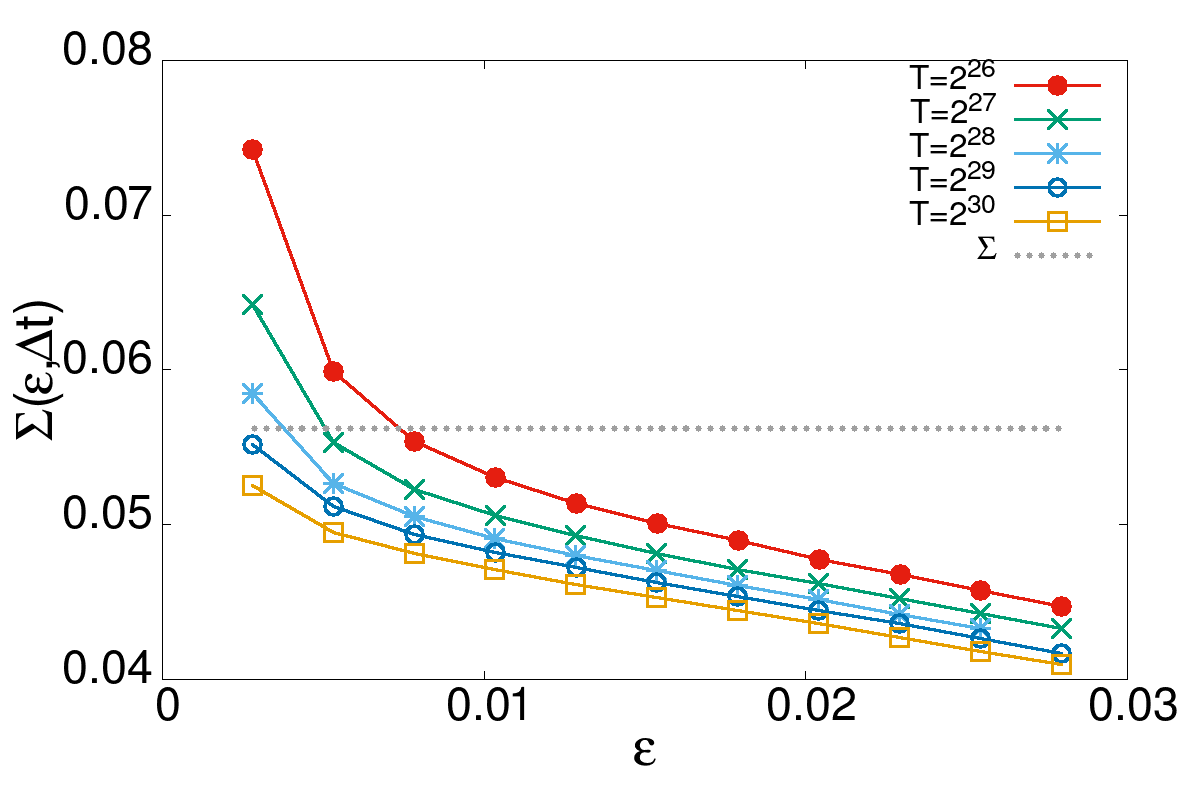}
    \caption{$\Sigma(\eps,\Delta t)$ computed with 1D signals. Left: dependence on $\Delta t$ at fixed little $\eps$ of $\Sigma(\eps,\Delta t)$ for several observable. Right: dependence on $\eps$ at fixed $\Delta t=1$ of $\Sigma(\eps,\Delta t)$ computed with the $x$-component only.}
    \label{Empirical:Fig_7}
\end{figure}

Finally, in order to mitigate at least the problem of under-sampled dynamics, we can ask ourselves what we would have obtained by looking at a single variable of the system, one of the two components $x$ or $y$, the distance from the origin $\rho=\sqrt{x^2+y^2}$ or the angle $\theta=\arctan{y/x}$. 
The results are shown in \fref{Empirical:Fig_7} and they indicate once again how important it is the choice of the observable. Even ignoring the problem of the  limit for $\eps \to 0$, see the right panel of \fref{Empirical:Fig_7}, we see that the best entropy production estimate is obtained by considering the signal $x(t)$. This is due to the fact that variable feels directly the effect of the jumps. The estimates provided by the signals $\rho$ and $\theta$ are in between those provided by $x$ and $y$ separately as $\rho$ and $\theta$ are non-linear combination of these two signals. 

The above analysis shows, in general, how difficult it is to have a correct estimate of the entropy production rate from experimental or numerical data, even in  simple cases.

\subsection{Exit times statistics and  hidden Markov modelling}
An alternative strategy that has been proposed in recent years consists in using other observables, rather than current fluctuations, in order to prove or disprove the hypothesis of an equilibrium model underlying the empirical data. Such a strategy is typically not aimed to retrieve an exact value for the entropy production rate, but again some kind of lower bound which should be sufficient, when not zero, to put in evidence the non-equilibrium nature of the system. One of the underlying ideas, here, is to exploit the information-theoretic formulation of entropy production to decompose it into different contributions depending on the experimentally accessible information.
These approaches are interesting because they have unveiled non-trivial aspects of the theory of stochastic processes, but at the same time they suffer from the same limitations encountered in the TURs, i.e. the general and obvious fact that partial information cannot account for the full information required to characterise the entropy production of a given dynamics. 	

To be more specific, let us consider a Markov process $s$ taking values in finite (countable) space $\Omega=\{1\cdots N\}$. According to Eq.\eqref{eq:entproddef}, the entropy production of such process reads
\begin{equation}
\Sigma^{\lx(\Omega\rx)} = \lim_{t\to\infty}\frac{1}{t}\sum_{\mathbf{s}^{(t)}}P\lx(\mathbf{s}^{(t)}\rx) \log\lx(\frac{P\lx(\mathbf{s}^{(t)}\rx)}{P\lx(\mathbf{s}_\leftarrow^{(t)}\rx)}\rx).   
\end{equation}
Generally, in an experiment the micro-space $\Omega$ is not accessible and only some macroscopic quantities $a(s)$ can be measured. Denoting $\Gamma$ the state-space corresponding to these macroscopic observables, a general result in information theory \cite{Cover2006,crisanti2012nonequilibrium} guarantees
\begin{equation} \label{entrbound}
\Sigma^{\lx(\Omega\rx)} \ge \Sigma^{\lx(\Gamma\rx)}= \lim_{t\to\infty}\frac{1}{t}\sum_{\mathbf{a}^{(t)}}P\lx(\mathbf{a}^{(t)}\rx) \log\lx(\frac{P\lx(\mathbf{a}^{(t)}\rx)}{P\lx(\mathbf{a}_\leftarrow^{(t)}\rx)}\rx).   
\end{equation}
Starting from the above formula, several bounds can be established depending on the assumptions on the dynamics in the state-spaces $\Omega$ and $\Gamma$.\\
For example, in \cite{roldan2012entropy} the authors provide a semi-analytical formula for an efficient estimation of the right-hand-side of Eq.\eqref{entrbound} based on the theory of products of random matrices. Their results rely on the assumption that the set of "macroscopic" variables $a$ consists in a sub-set of the micro-state $s$, for instance $s=\{s_1,s_2\}$ and $a=s_1$. Moreover, their approximation requires the knowledge of the dynamics behind the observations and therefore is not practical for modelling real-world experiments. 
In \cite{martinez2019inferring}, a new approach for estimating the entropy production from time-series measurements is proposed. The central result consists in an analytical expression for the entropy production of a semi-Markov process, i.e. a generalization of Markov processes where the waiting-time distributions can be non-Poissonian. It is shown that the entropy production rate has two different sources. One source accounts for the irreversibility generated by transitions between the states of the process regardless of the transition-times distribution, while the second contribution is due to the non-exponential distributions of the exit times. Treating molecular motors and partially hidden Markov networks, the authors provided evidence that their method is able to detect time irreversibility even in the absence of observable currents. In \cite{lynn2022decomposing,lynn2022emergence} the problem is approached from a different perspective. Motivated by models used in neuroscience, the authors consider the case of multipartite dynamics. In short, each micro-state $s=(s_1,\cdots,s_D)$ is a list of $D$ degrees of freedom corresponding to separate units in the system, and for sufficiently short sampling times two successive micro-states $s=(s_1,\cdots,s_D)$ and $s^{\prime}=(s_1^{\prime},\cdots,s_D^{\prime})$ differ by the value of one unit only ($s_i=s_i^{\prime}$ for all $i\neq j$). For this systems, the authors show that the entropy production rate can be decomposed into two contributions $\Sigma=\Sigma_{ind}+\Sigma_{int}$. The term $\Sigma_{ind}$ is related to single unit time series ($\mathbf{s}_1,\cdots \mathbf{s}_D$) considered independently one from each other taking the form
\begin{equation}
    \Sigma_{ind}=\lim_{t\to\infty}\frac{1}{t}\sum_{i=1}^{D}\sum_{\mathbf{s}_i}P(\mathbf{s}_i)\log\lx(\frac{P\lx(\mathbf{s}_i\rx)}{P\lx(\mathbf{s}_{i,\leftarrow}\rx)}\rx)
\end{equation}
while the term $\Sigma_{int}=\lx(\Sigma-\Sigma_{ind}\rx)\ge 0$ accounts for the interaction between units. Moreover, exploiting a hierarchy of bounds on entropy production, obtained by including more and more macroscopic variables $a$ in the coarse-grained description (see \cite{roldan2021quantifying} for further details), it is also shown that the interaction term can be further decomposed in contributions which accounts separately for $n-$th order interactions (interactions among $n \le D$ units). 
Another particularly fascinating approach in this framework consists in studying the transition statistics of the system, i.e. for instance the distribution of the waiting times for the return or the passage of some observables through a given value.
This approach has been studied for instance in \cite{skinner2021estimating,skinner2021improved}. Differently from \cite{martinez2019inferring}, the authors do not make additional assumption on the coarse-grained dynamics, and instead they 
 look for the Markov model with hidden variables, compatible with the observed waiting times distribution, that minimises the entropy production. If this minimum entropy production is non-zero, then the observed data must originate from a non-equilibrium system. In Skinner \cite{skinner2021improved} this approach has been applied to experimental data for gene regulatory networks, mammalian behavioral dynamics, and numerous other biological processes such as the heartbeat regulation in humans, dogs, and mice.  As for \cite{martinez2019inferring}, the interest of this approach is that it can retrieve some non-zero lower bound for the entropy production even with data that are time-reversal symmetric, e.g. when there are no visible currents. The bound coming from jump rates, however, can be quite loose, especially in these last cases. An alternative approach is presented in \cite{ehrich2021tighest}, that uses all observable data (not only transition rates) to find an underlying hidden Markov model responsible for generating the observed non-Markovian dynamics.  
Recently, some authors realized that considering non-conventional coarse-graining procedures in which macroscopic variables $a$ are identified with microscopic transitions (i.e. transitions between microstates $s$ and $s'$) leads to semi-Markovian dynamics for the coarse-grained processes~\cite{harunari2022learn}.
 Thus, in complete analogy with ~\cite{martinez2019inferring,ehrich2021tighest}, a lower bound for the entropy production rate is derived that equals the sum of two non-negative contributions, one due to the statistics of transitions (conditional probabilities of observed occurrence of the system in some state) and a second due to the statistics of intertransition times. The applications to experimentally validated biophysical models of kinesin and dynein molecular motors, and in a minimal model for template-directed polymerization reveals, again, that this strategy is suitable for detecting irreversibility even in the absence of net currents in the transition time series. A general theory encompassing all the above-mentioned cases was developed in \cite{vandermeer2022thermodynamic,van2023time}. This theory does not make any kind of assumptions about the dynamics of coarse-grained variables, but instead considers the observables in a very general way as a joint collection of events and waiting times. By exploiting these generalized waiting times distribution, the authors formulate entropy estimators resembling the formulas appeared in~\cite{martinez2019inferring,ehrich2021tighest,harunari2022learn} whose efficiency can be evaluated from the data themselves. 
 Furthermore, in certain cases, these estimators allow one to retrieve information regarding the topology of the underlying network. As with other estimators based on waiting times, with this approach it is possible to obtain non-zero entropy production even if all macroscopic events are invariant under time-reversal. To do so, however, the  joint distribution $\psi(t_1,\cdots,t_k)$ of the waiting times should not be invariant under time-reversal, i.e. $\psi(t_1,\cdots,t_k)\neq \psi(t_k,\cdots,t_1)$ for some $\{t_1,\cdots,t_k\}$, and, as clearly explained in~\cite{van2023time}, the difference $\psi(t_1,\cdots,t_k)- \psi(t_k,\cdots,t_1)$ play the role of a steady current.
To conclude the Section, we recall that in certain cases serious limitations appear for the above approaches. Imagine that the states-space $\Omega$ contain the states of a discrete-time Markov chain while the coarse-grained space $\Gamma$ contains just two elements $\{a_0,a_1\}$ with $a_0=\{1\}$ and $a_1=\{2,\cdots,N\}$. As shown in~\cite{lucente2022inference}, the resulting process is a $2$-state semi-Markov process and the entropy production vanishes, regardless of whether the Markov chain defined on $\Omega$ satisfies or not detailed balance. For this class of processes it therefore seems impossible to determine the thermodynamic nature of the system without making further assumptions on the models that generate the observations.
 

\section{Lack of equilibrium and causation indicators 
\label{sec:causation}}
As discussed in the previous sections, entropy production quantifies the degree of irreversibility in 
the dynamics of non-equilibrium systems. Due to its global nature, it lacks sensitivity to the structural details of a system, such as, inhomogeneity in temperature and chemical gradients, or non-reciprocal interactions~\cite{loos2020irreversibility, loos2023nonreciprocal}, such as asymmetries in the couplings between different sites or degrees of freedom.
Therefore, if one is interested in the specific description of the internal currents driving the system out of equilibrium, it is mandatory resorting to more microscopic or local observables. 
To this goal, promising candidates are the Transfer Entropy~\cite{schreiber2000measuring, barnett2009granger} and Response functions \cite{KTH91,chandler_book88}, which are usually employed to detect causal relationships between system variables \cite {sarra2021response,cecconi2023correlation}. 
In fact, it is intuitively expected that the presence of a physical current - usually a signature of time-irreversibility - determines an information flow and, therefore, causal relations among different parts of the system.
In this Section, we show by means of simple models that local causal indicators can provide insights into how spatial asymmetries and non-reciprocal interactions drive the system towards non-equilibrium states.

Before delving into the definitions of transfer entropy and response functions, it is essential to make a brief excursion through the notion of causation that stands as a cornerstone concept across various disciplines, including philosophy, natural and social sciences and engineering.
Also, it helps the comprehension of our everyday experience, allowing us to make informed decisions.
Cause-effect principles are implicitly or explicitly employed when quantitative theories are developed in terms of equations
involving parameters and quantities.
Nowadays, in the era of Big Data and Artificial Intelligence, the concept of causation is gaining even more importance.

The philosopher David Hume was among the pioneers in formalizing the concept of causation \cite{coventry2006hume},
by proposing the idea of simultaneous recurrence of 
events: a causal relationship between $A$ and $B$ (represented symbolically as $A\rightarrow B$) can be inferred when the consistent observation of $B$ is always preceded by $A$.
In more precise words, finding causation between a set $\{X_1,X_2,\ldots,X_M\}$ of events, variables, or observables etc., means to construct a directed graph where each $X_i$ represents a node and the causal connections are oriented links pointing from $X_i \mathrm{(cause)} \rightarrow X_j \mathrm{(effect)}$.
In this respect, causal indicators, algorithmic procedures and statistical tests become fundamental tools to establish quantitatively the connections among nodes.

Two different strategies can be followed in the definition of causal indicators.
The first one, deriving directly from Humes' view, can be referred to as {\em observational}, because
causality is detected only by observation of time series of events or data.
In a nutshell, the goal is to understand from data whether, and to what extent, the knowledge of a certain variable is useful to the actual determination of present and future values of another one.
In other words, if one observes that knowledge of the past states of variable $Y(t)$ improves the accuracy of forecasting future values of $X(t)$, it can be deduced that $Y$ has a certain influence on $X$: in symbols, $Y\rightarrow X$.
This is the spirit of Granger Causality \cite{granger1969investigating} as well as of Transfer Entropy \cite{schreiber2000measuring}.

The other approach, termed {\em interventional}, assumes that two variables are in a cause-effect relationship if an external action on one of them changes the observed value of the other. J.~Pearl formalized this idea through the notion of ``Do'' operation~\cite{pearl2009causality}, used to express interventions where a variable is set to a specific value to see the system's response.
In this context, causality coincides with the possibility to predict the result of the intervention. 
The interventional definition is a more physics-inspired interpretation of cause-effect relationships that can be quantified by a well-known observable, the response function~\cite{Cencini2009}. This approach to the study of causation has been recently analyzed and used in a series of works~\cite{aurell2016causal, baldovin2020understanding,baldovin2022extracting, falasca2024data, cecconi2023correlation, giorgini2024response}.

In the following, we briefly provide a mathematical definition of these two causal indicators, also leading 
to their operational employment.

\subsection{Transfer entropy}
Transfer entropy (TE) from process $X_t$ to process $Y_t$ is a concept borrowed from information theory 
introduced by Schreiber \cite{schreiber2000measuring} in the context of stochastic processes and dynamical systems 
and then reformulated by Palu\v{s} et al. \cite{palus2001} as conditional mutual information. 
Generally speaking, the entropy transfer from the evolution of the degree of freedom $x_j(t)$ to the evolution of the degree of freedom $x_i(t)$ is defined as the information (uncertainty) that we gain (lose) on the future states of $x_i$, if we not only consider the history of $x_i$, but we also include the past of $x_j$. 
It quantifies the causal influence of $x_j$ on $x_i$, in formulae,
\begin{equation}
\mathrm{TE}_{j\to i}(\tau) = 
\ave{\ln{\frac{P\lx[x_i(t)\,\big|\,{\bf X}^{(\tau)}_i,{\bf X}^{(\tau)}_j\rx]}{P\lx[x_i(t)\,\big|\,{\bf X}^{(\tau)}_i\rx]}}}.
\label{eq:defTRE}
\end{equation}
Here $t$ is a time index, ${\bf X}^{(\ell)}_i = \{x_i(t-1),\ldots,x_i(t-\tau)\}$
and 
${\bf X}^{(\tau)}_j = \{x_j(t-1),\ldots,x_j(t-\tau)\}$ are the past states
of variables $x_i(t)$ and $x_j(t)$ respectively.
The angular-brackets denote the average over the joint probability 
density $P\lx[x_i(t),{\bf X}^{(\tau)}_i,{\bf X}^{(\tau)}_j\rx]$, 
while 
$P\lx[x_i(t)\,\big|\,{\bf X}^{(\tau}_i,{\bf X}^{(\tau}_j\rx]$ and  $P\lx[x_i(t)\,\big|\,{\bf X}^{\tau}_i\rx]$ 
are the probability densities of $x_{i}(t)$ conditioned to the past histories.
Notice that TE identically vanishes for $i=j$ and is by definition asymmetric, $\mathrm{TE}_{j\to i} \neq \mathrm{TE}_{i\to j}$, 
thus naturally incorporating a direction of the entropy/information transfer 
from $x_j \to x_i$, that is generally different from $x_i \to x_j$. 
Note that the asymmetry is a natural consequence of the non-interchangeability between 
conditioning and conditioned events.

As clear from \eref{eq:defTRE}, TE measures how much information is gained 
on the future of $x_i$ when taking into account the past history, ${\bf X}^{(\tau)}_j$, 
\begin{equation}
\mathrm{TE}_{j\to i} = H\lx[x_i(t)|{\bf X}^{(\tau)}_i\rx] - 
H\lx[x_i(t)|{\bf X}^{(\tau)}_i,{\bf X}^{(\tau)}_j\rx],
\label{eq:defTE}
\end{equation}
where $H\lx[A|B\rx]$ indicates the conditional Entropy of the 
state $A$ given the state $B$. 
Note that Eq.\eqref{eq:defTE} implicitly assumes stationary processes.

The TE among degrees of freedom of a multivariate linear Gaussian Markov system (Ornstein-Uhlenbeck
process often employed in this review), 
$$
\dot{\bx} = -\matx{A}\bx + \bxi\,,
$$
can be easily expressed in terms of their time correlations 
$\cC_{ij}(t)=\ave{x_i(t) x_j(0)}$,
\begin{equation}
\mathrm{TE}_{j\to i}(t) =  -\frac{1}{2} \ln\lx(1 - \frac{\alpha_{ij}(t)}{\beta_{ij}(t)}\rx)\label{eq:TRE}
\end{equation}
where  
\begin{align}
\alpha_{ij}(t) =& \lx(\matx{C}_{ii} \cC_{ij}(t) - \matx{C}_{ij} \cC_{ii}(t)\rx)^2,\\
\beta_{ij}(t)  =& \lx(\matx{C}_{ii} \matx{C}_{jj} - \matx{C}^2_{ij}\rx)\lx(\matx{C}^2_{ii} - \cC^2_{ii}(t)\rx),
\end{align}
see refs.\cite{sarra2021response,cecconi2023correlation} for the derivation,
we indicate with $\matx{C}_{ij}=\cC_{ij}(0)$ the correlations matrix at zero lag (i.e. the covariance matrix).
The asymmetry $\alpha_{ij}(t)\ne \alpha_{ji}(t)$ and 
$ \beta_{ij}(t) \ne \beta_{ji}(t)$ is a straightforward consequence of the TE asymmetry emerging also in the Gaussian formulation. 

It is possible to show that $\mathrm{TE}_{j\to i}(\infty) = 0$, either by definition Eq.\eqref{eq:defTRE} invoking the independence of events far away in time, or using the correlation decay at large times in Eq.\eqref{eq:TRE} implying that $\alpha_{ij}(\infty)\to 0$.
Analogously, one expects $\mathrm{TE}_{j\to i}(0) = 0$. 
As a consequence, in many cases, TE is expected to have a skewed bell-shaped curve as a function of the time lag $t$.

\subsection{Response function}
The other causal indicator that could be useful to employ in out-of-equilibrium systems is the response function, which belongs to the interventional framework: 
indeed, the coordinate $x_j$ causally influences the coordinate $x_i$, if a perturbation of $x_j$ results in a variation of the measured value of $x_i$. 
In formulae, we say that $x_i$ influences $x_j$, if  
\begin{equation}
 \label{eq:resp_obs}
 \cR_{ji}(t) = \lim_{\delta x_j(\tau) \to 0}\,
\frac{\overline{\delta x_i(\tau+t)}}{\delta x_j(\tau)} 
\ne 0 \quad \quad \quad \mbox{for some }t>\tau\,,
\end{equation}
i.e. a small perturbation on $x_j(\tau)$ at time $\tau$ results in a non-zero future variation on the average of $x_i(t+\tau)$ over its unperturbed evolution. 
In Eq.\eqref{eq:resp_obs}, we again assume statistically stationary dynamics as in Eq.\eqref{eq:defTE}. 
If $\delta x_j$ is small enough, it is well known that 
the quantity \eqref{eq:resp_obs} can be related to the spontaneous 
correlations in the unperturbed dynamics by one of the pillars of non-equilibrium statistical mechanics, the fluctuation-response theorem (FRT)~\cite{FDreport}, also known as fluctuation-dissipation theorem (FDT).
When the process $\mathbf{x}(t)$ is stationary with invariant probability density function (PDF) $P_s(\mathbf{x})$, the response \eqref{eq:resp_obs} assumes the clear and general expression \eref{eq:generalized_fdr} seen in Sec.\ref{sec:markov}. 


It should be remarked that Eq.\eqref{eq:generalized_fdr} holds for systems with an invariant PDF and in general cases, it expresses the response in terms of complicated multivariate correlation functions.      
However, in systems governed by stochastic linear dynamics, even with no Gaussian noise, the response turns out to be related only to the two-time correlation function, \cite{baldovin2020understanding}
\begin{equation}
\cR(t) = \cC(t) \matx{C}^{-1}\;.
\label{eq:fdr}
\end{equation}
as we already proven in Sec.\ref{sec:markov} \eref{eq:ourescorr}.

\subsection{A toy model with non-reciprocal interactions} 
\begin{figure*}
    \centering
    \includegraphics[width=0.98\textwidth]{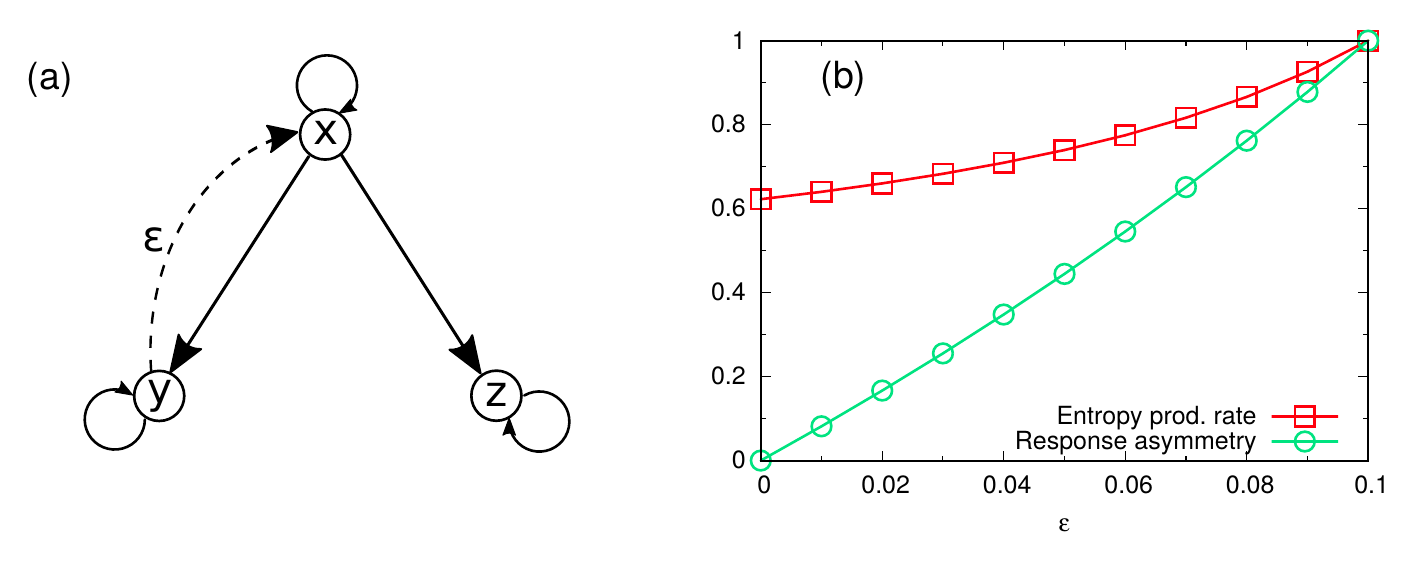}
    \caption{Entropy production rate and response asymmetry in a simple toy model with non-reciprocal interactions. The system is sketched in panel (a), where solid lines refer to a value a=0.5 in the interaction matrix (see Eq.~\eqref{eq:chocolate}) and the link corresponding to the dashed line is taken as a varying parameter $\varepsilon$. Panel (b) shows the entropy production rate (red) and the response asymmetry $\widetilde{R}_{zy}-\widetilde{R}_{yz}$ (green). Both quantities are rescaled by their values at $\varepsilon=0.1$.}
    \label{Causation:Fig_1}
\end{figure*}

The link between non-reciprocal interactions, causation and lack of equilibrium can be appreciated by considering linear systems, which are 
fully analytically solvable. 
As a first example, we consider the minimal non-equilibrium model discussed in Ref.~\cite{baldovin2020understanding}: a system of three variables $x_t$, $y_t$ and $z_t$, whose values are updated at discrete times according to the rule
\begin{subequations}
\begin{align}
  x_{t+1} =a&x_t + \varepsilon y_t  +  \eta^{(x)}_t \\
  y_{t+1} =a&x_t + a y_t +  \eta^{(y)}_t \\
  z_{t+1} =a&x_t + a z_t +  \eta^{(z)}_t\,,
\end{align}
\label{eq:chocolate}
\end{subequations}
where, $\eta_t$'s are independent Gaussian variables with zero mean and unitary variance, while $a$ and $\varepsilon$ are assigned constants. 
The model is sketched in Fig.~\ref{Causation:Fig_1}(a). 
When $\varepsilon=0$, the variable $x$ is independent of $y$ and $z$, and 
drives them in the same way: the system is thus symmetric with respect to the exchange of $y$ and $z$. 
As soon as $\varepsilon>0$, a feedback mechanism indirectly couples $z$ to $y$ (meaning that the former is influenced by the latter). 
The presence of non-reciprocal interactions drives the system out of equilibrium, and, thanks to linearity, the entropy production rate can be easily computed. Figure \ref{Causation:Fig_1}(b) shows that the entropy production rate is nonzero for every choice of $\varepsilon$. 

Although the system is still out of equilibrium at $\epsilon=0$, 
when $\varepsilon$ is increased, the entropy production rate of the system increases as well, suggesting that the dynamics is becoming more irreversible. 
The origin of this increment of the time-reversal asymmetry can be understood by looking at the causation indicators introduced in the previous Section. By computing the difference between the response function (integrated in time)
$$
\widetilde{R}_{zy}=\sum_{t=0}^{\infty}\cR_{zy}(t)
$$
and
$$
\widetilde{R}_{yz}=\sum_{t=0}^{\infty}\cR_{yz}(t)\,,
$$
one gets a quantitative estimator of the $y,z$ causation asymmetry. 
This difference is reported in Fig.\ref{Causation:Fig_1}(b); as expected, it vanishes when $\varepsilon=0$, meaning that equilibrium is broken by other mechanisms (in this case, the $x,y$ and $x,z$ asymmetries).

This simple example shows that, in the presence of non-reciprocal interactions, response functions (or any other reliable indicators of causation) provide detailed information on the origin of the time-reversal asymmetries that drive the system out of equilibrium. 
While the entropy production rate is a global quantity, which only signals to what extent the system is out of equilibrium, causation indicators are actually able to unveil the asymmetries that are responsible for the non-equilibrium state~\cite{hempel2024simple}. 
In the following Section, we will analyze a case where the scenario is enriched by the presence of temperature gradients.

\subsection{Oscillators with non-reciprocal interactions and temperature gradient} 
In this Section, we apply transfer entropy and response functions to characterize the effect of spatial symmetry-breaking in an example 
of linear out-of-equilibrium extended systems~\cite{sarra2021response}. 

Consider a system of $N$ interacting particles whose individual positions are denoted by $\{x_j\}$, with $j$ ranging between $1$ and $N$. 
Particles are coupled via nearest-neighbor elastic forces
$$
F_j=-k_L(x_j-x_{j-1})-k_R(x_j-x_{j+1})\,.
$$
Periodic boundary conditions $x_0 \equiv x_N$, $x_{N+1} \equiv x_1$ are implemented, and by setting
$$
\begin{aligned}
    k_L&=k_0-\varepsilon\\
    k_R&=k_0+\varepsilon\,,    
\end{aligned}
$$
the symmetry of the interactions is broken as soon as $\varepsilon \ne 0$.
Each particle is also subject to a restoring force $-k_0\,x_j$, and to the action of an inhomogeneous thermal bath with site-dependent temperature $T_j$. 
The stochastic dynamics of $x_j$ can thus be written, in the overdamped limit, as
\begin{equation}
    \label{eq:dyn}
    \gamma \dot{x}_j=-3k_0 x_j+ k_Lx_{j-1} + k_Rx_{j+1} +
    \sqrt{2\gamma k_BT_j}\;\xi_j
\end{equation}
where $\xi_j$ is zero-mean white Gaussian noise.
Hereafter, we adopt dimensionless units corresponding to Boltzmann constant $k_B=1$ and viscous coefficient $\gamma=1$.

The system stays in equilibrium if $\varepsilon=0$ and the bath is homogeneous, $T_j\equiv T_0$. It can be driven out of equilibrium by breaking the detailed balance,
in two ways: a) switching the interactions asymmetry on, $\varepsilon>0$, so that mechanical currents are induced across the ring; 
b) enforcing a thermal gradient 
$$
T_j=T_0+\Delta T \cbr{\Big|j-\frac{N}{2}\Big|-\frac{N}{4}}\,,
$$
where $\Delta T$ is a suitable constant.

\begin{figure*}
    \centering
    \includegraphics[width=0.98\textwidth]{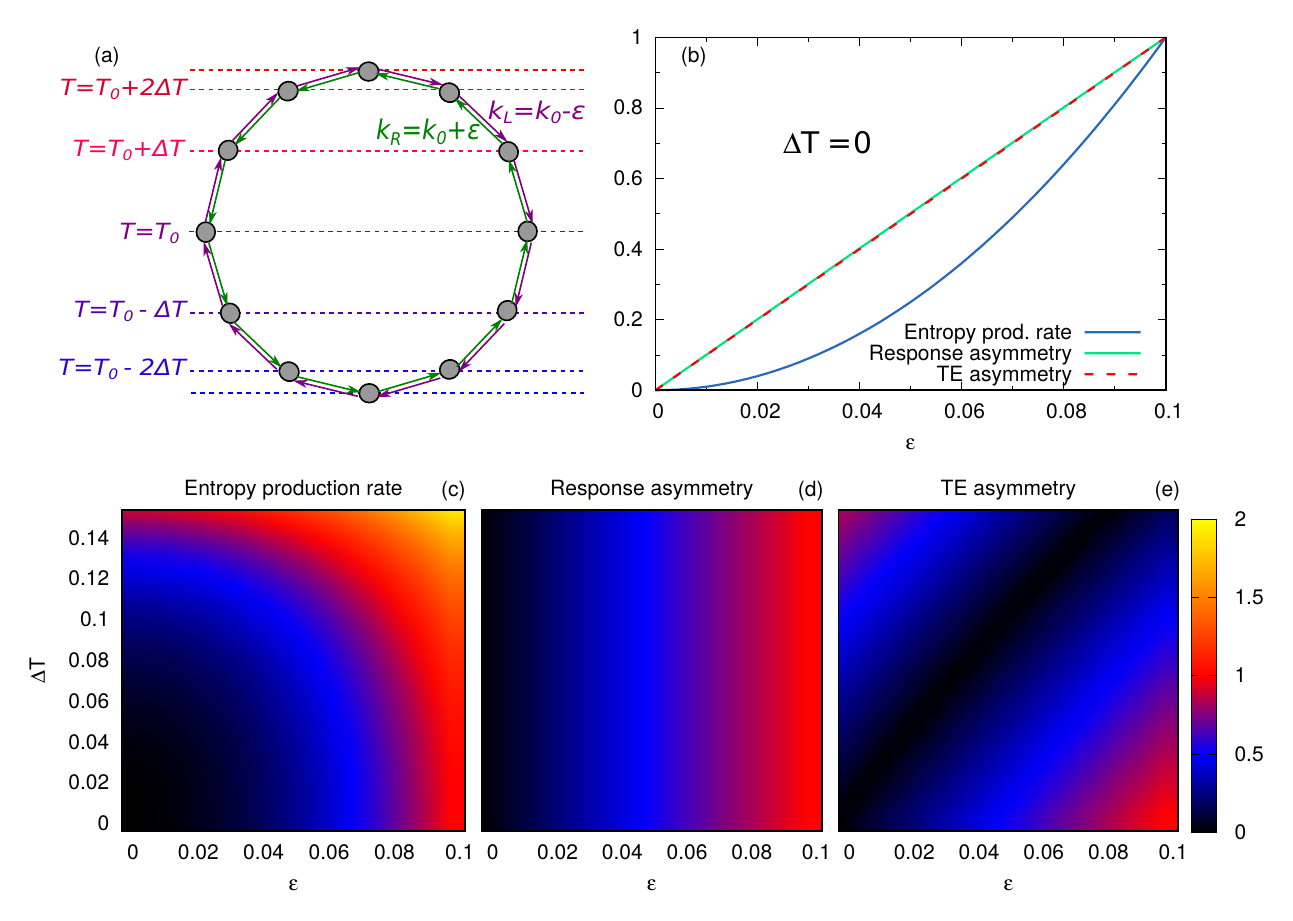}
    \caption{Asymmetry indicators in the ring model. Panel (a): scheme of the interactions between the particles, as well as the temperature gradients. Panel (b): Asymmetry indicators and entropy production rate for $\Delta T=0$, as a function of $\varepsilon$. All quantities are rescaled setting their value at $\varepsilon=0.1$ equal to unity. Panels (c)-(e): heat maps of the quantities considered in panel (b) as a function of $\varepsilon$ and $\Delta T$. Again, all quantities are rescaled with their value for $\varepsilon=0.1$ and $\Delta T=0$. Parameters: $T_0=1, k_0=1, N=20$.}
    \label{Causation:Fig_2}
\end{figure*}

In the absence of mechanical asymmetry, $\varepsilon=0$, the above choice leads to a heat flux from the hottest sites ($j=0$, $j=N$) to the coldest site ($j=N/2$). 
A schematic representation is shown in Fig.~\ref{Causation:Fig_2}(a).
The model has therefore two different sources of spatial asymmetry: 
the non-reciprocity of the interactions and the temperature gradient.

The dynamics can be written in the form of a linear stochastic process
$$
\dot{\bx}=-A \bx+\bxi, 
$$ 
where $A$ is a $N\times N$ circulant matrix and $\bxi$ is a noise with diagonal correlations, i.e. $\nu_{ij}=\ave{\xi_i \xi_j}=2T_i\delta_{ij}$. 
The properties of circulant matrices ensure that all the eigenvalues of $A$ have a positive real part.
As proven in Sec.~\ref{sec:fdr} (see also Ref.~\cite{G90}), 
for such a system the correlation and response functions are   
respectively
$$
\cC(t)=e^{-At}C\quad (\forall t>0), \qquad  \cR(t)=\Theta(t)\;e^{-tA}
$$ 
$\Theta(t)$ being the unitary step-function, and the covariance matrix $C=\cC(0)$ of the stationary state verifies $AC+CA^T=\nu$. 
Moreover, entropy production rate can be written as $\Sigma=\text{Tr}\lx(\Delta \nu^{-1} A\rx)$ where $\Delta=CA^T-AC$ measures the violation of Onsager's condition.

All the quantities previously introduced for the characterization of out-of-equilibrium systems can be explicitly computed.
In Fig.~\ref{Causation:Fig_2}, we compare the behavior of $\Sigma$ with two observables $\Delta R$ and $\Delta \mathrm{TE}$ that quantify the spatial asymmetry of the response and of the transfer entropy. 
The former is defined as
\begin{equation}
\Delta R =\lx|\sum_{j=1}^{N/4-1}\widetilde{R}_{j,N/4}-\sum_{j=N/4+1}^{N/2-1}\widetilde{R}_{j,N/4}\rx|
\end{equation}
where
$$
\widetilde{R}_{ij}=\int_0^{\infty} dt\, {\cR}_{ij} (t)\,.
$$
This ``integrated response'', inspired by the Kubo relations~\cite{kubo1966fluctuation}, takes into account the cumulative effect of $j$ on $i$. The imbalance $\Delta R$ is therefore vanishing when the effect of the particle $N/4$ on the first $N/4-1$ neighbors on the left and on the right is the same, while it is larger than zero if a spatial asymmetry is present.

Similarly, for the transfer entropy rate we introduce the unbalance
\begin{equation}
    \Delta \mathrm{TE} = \lx|\sum_{j=1}^{N/4-1}{\mathrm{TE}}_{j,N/4}-\sum_{j=N/4+1}^{N/2-1}{\mathrm{TE}}_{j,N/4}\rx|\,.
\end{equation}

Figure~\ref{Causation:Fig_2}(b) shows that, for vanishing temperature gradient, $\Delta R$ and $\Delta \mathrm{TE}$ provide the same kind of information, and their value is monotonically related to the entropy production rate. 
This is consistent with the fact that the (mechanical) asymmetry in the dynamics is the only source of equilibrium violation. 
In Fig.~\ref{Causation:Fig_2}(c)-(e) a nonzero temperature gradient is also included, and the qualitative difference between response and transfer entropy becomes evident: while the former only depends on the interaction forces between the particles, the latter is dramatically affected by the presence of heat flow.

The above example clarifies that the information coming from the analysis of the entropy production has to be complemented by other asymmetry indicators in order to provide an exhaustive description of the local currents leading the system out of equilibrium.

\section{Turbulence: a case study \label{sec:turbulence}}
In the previous sections we have mostly focused on simple, mainly
stochastic, models; in this section, we end our tour considering an
important instance of persistent non-equilibrium phenomenon, namely
the turbulent state realized in incompressible flows at high Reynolds
number \cite{falkovich2006lessons}.

The evolution of the velocity field,
$\boldsymbol{u}(\boldsymbol{x},t)$, of an incompressible
($\boldsymbol{\nabla}\cdot \boldsymbol{u}=0$) fluid is ruled by
Navier-Stokes equation \cite{landau-fm}:
\begin{equation}
  D_t \boldsymbol{u} \equiv \partial_t \boldsymbol{u} +
  \boldsymbol{u}\cdot\boldsymbol{\nabla}\boldsymbol{u} =
  -\frac{\boldsymbol{\nabla}p}{\rho} + \nu \Delta \boldsymbol{u} +
  \boldsymbol{f}\ ,
  \label{eq:NS}
\end{equation}
where $D_t$ denotes the material derivative, $p=p(\boldsymbol{x},t)$
the pressure, $\rho$ the (constant) mass density, $\nu$ the kinematic
viscosity and $\boldsymbol{f}$ a stirring force.
For $\nu=0$ and $\boldsymbol{f}=0$, \eref{eq:NS} becomes
the Euler equation that, in the presence of an ultraviolet cutoff, is
known to be an equilibrium system \cite{kraichnan1973helical}.  In
viscous fluids ($\nu\neq 0$), the force $\boldsymbol{f}$ injects
energy  at a scale $L$ at a rate (per unit mass):
\begin{equation}
    \varepsilon = \langle \boldsymbol{f}\cdot\boldsymbol{u} \rangle\ ,
\end{equation}
where $\langle [\ldots] \rangle$ denotes an average over space and
time.  The nonlinear terms of Eq.~(\ref{eq:NS}) preserve the total
kinetic energy but redistribute it among the scales. In particular, in
three dimensional (3D) turbulence, energy is transferred to smaller
and smaller scales till it is dissipated by the viscous forces.  Thus,
the system eventually reaches a (non-equilibrium) statistically
stationary state, where, on average, the injection and dissipation
rates balance. Such a process, which is a peculiar feature of 3D
turbulent fluids, is dubbed ``direct energy cascade" \cite{Frisch95}.
In the turbulent state there is a large separation between the scale
of energy injection $L$ and that of dissipation
\cite{falkovich2006lessons}, which can be estimated by dimensional
argument as $\eta=(\nu^3/\varepsilon)^{1/4}$ -- i.e. the so-called
Kolmogorov scale.  In the inertial range, $\eta \ll r \ll L$, both
energy injection and dissipation are negligible and the statistical
properties are believed to be universal and display non-trivial
scaling laws \cite{Frisch95}.  Moreover, the system is characterized
by a wide spectrum of timescales from $\tau_L=L/u_{\text{rms}}$ (where
$u_{\text{rms}} = \langle |\boldsymbol{u}|^2\rangle^{1/2}$) at the
largest scale to $\tau_{\eta}=(\nu/\varepsilon)^{1/2}$ at dissipative
scales.  Formally, the limit of infinite scale separation, or infinite
Reynolds number, $Re=UL/\nu \rightarrow \infty$, ($U$ being a typical
large-scale velocity) is called ``fully developed
turbulence". Therefore, turbulence is a multi-scale phenomenon
involving many spatial scales each with its characteristic time.
 
By performing some statistical analysis it is not hard to prove that
turbulent systems are out of equilibrium: the non-zero average energy
flux across scales, when computed in wavevector space, is roughly the
sum of third-order moments of the velocity field, and this non-zero
skewness of the velocity PDFs is incompatible with statistical
equilibrium.  In the following, we will explore two less
straightforward aspects of turbulence.  First, as first demonstrated in Ref.~\cite{Xu2014flightcrash},
we will show how asymmetric correlation functions can reveal and somehow
quantify the irreversible and thus non-equilibrium
character of turbulence by looking at the motion of a single fluid
tracer (adopting the so-called Lagrangian view of turbulence), that is
by looking at the evolution of a very small, neutrally buoyant
particle transported by the flow. Clearly, a single particle
trajectory provides only a partial information on the state of the
fluid velocity field and, as previously discussed (see, e.g.,
Sec.~\ref{subsec:gyrator} and \ref{sec:tur}) inferring and quantifying
non-equilibrium properties from partial observation is, in general,
nontrivial.  Second, we will show that asymmetric correlation
functions and response functions, thoroughly employed in the previous
sections, can be used to reveal the non-equilibrium (equilibrium)
character of the physics at scales smaller (larger) than the energy
injection scale, as shown in Ref.~\cite{cocciaglia2024nonequilibrium}
in the context of shell models for turbulence, which constitute a simplified 
laboratory for turbulent phenomenology \cite{bohr1998,biferale2003shell, ditlevsen2010}.

\subsection{Lagrangian irreversibility}
The dynamics of a fluid tracer, which in a laboratory can be realized by considering
a very small particle having the same density of the fluid, is ruled by the following equation
\begin{eqnarray}\label{eq:fludyn}
    \dot{\boldsymbol{X}}(t) = \boldsymbol{V}(t) = \boldsymbol{u}(\boldsymbol{X}(t),t)\ , 
\end{eqnarray}
where $\boldsymbol{X}(t)$ denotes the position of the Lagrangian
tracer at time $t$, and $\boldsymbol{V}(t)$ is the velocity field at
the particle position.  How to recognize the signature of the
non-equilibrium turbulent state by looking at the dynamics of fluid
tracers was first discussed in \cite{Xu2014flightcrash,
  Pumir2014redistribution} where, using both experimental and
numerical Lagrangian trajectories, it was shown that tracers
experience slow buildups of kinetic energy followed by sudden
discharges, a phenomenon dubbed ``flight-crash" event: a clear sign of
time irreversibility at the level of a single tracer particle, and
yields breaking of the detailed balance condition since forward and
backward transitions are not equiprobable (see
Sec.~\ref{sec:perneq}). This can be revealed by studying the
statistics of the tracer kinetic energy per unit mass, $E(t) =
\frac{1}{2} |\boldsymbol{V}(t)|^2$.

\begin{figure}[t!]
\centering
\includegraphics[width=0.98\textwidth]{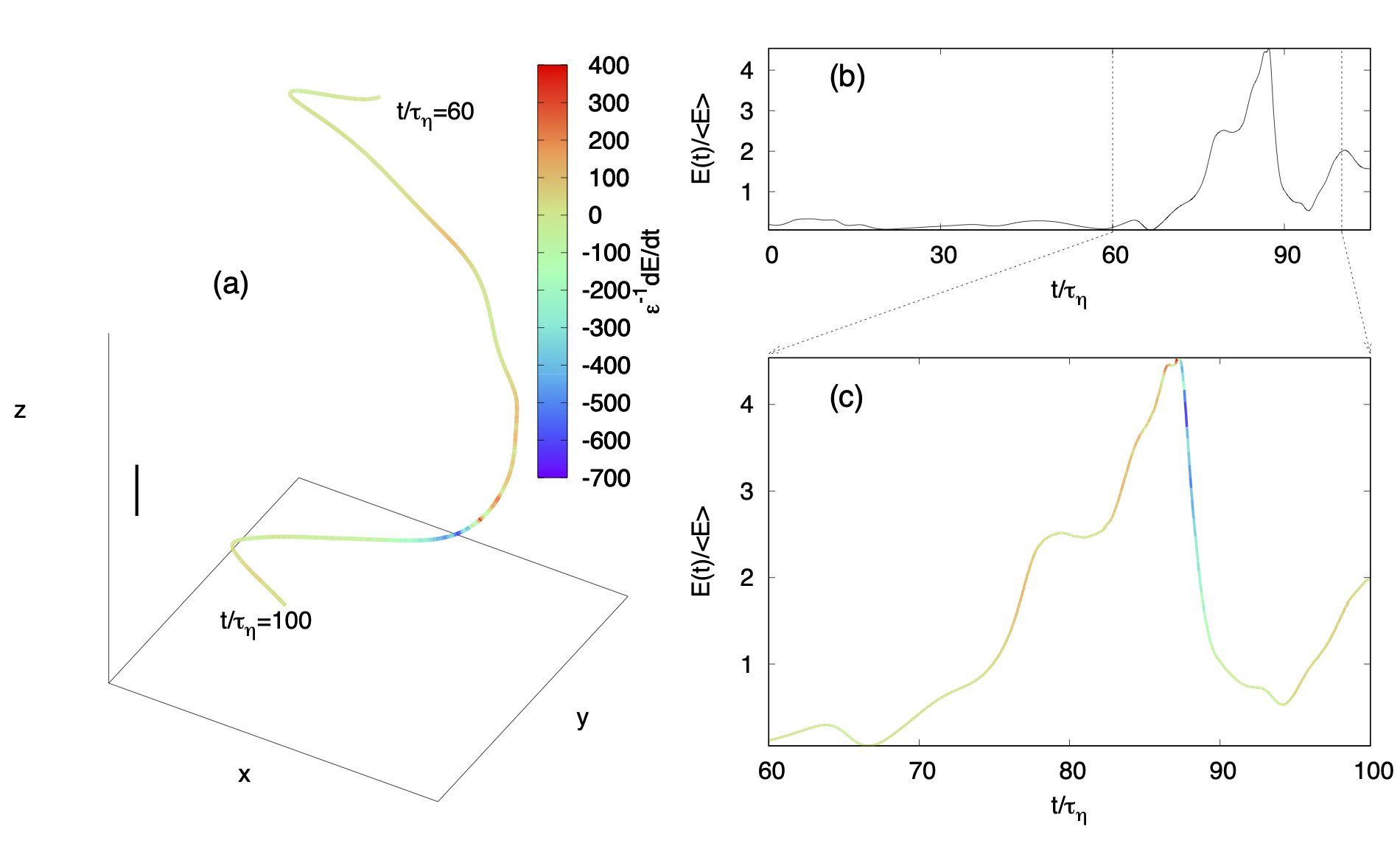}
\caption{Qualitative explanation of the asymmetry in the statistics of
  energy increments. (a) Cut of a tracer trajectory in a turbulent
  flow. Color codes the instantaneous power $p(t) = dE(t)/dt =
  \boldsymbol{A}(t) \cdot \boldsymbol{V}(t)$. (b) Time evolution of
  the tracer kinetic energy $E(t)$. (c) Enlargement of (b) with the
  same color coding of (a) to show the so-called ``flight-crash" event
  where energy grows slowly and decreases quickly. After
  \cite{Xu2014flightcrash}. The trajectory was taken from the database
  \protect\cite{databaseLagrangiano} and refers to tracers evolving in
  turbulent flow at $Re_\lambda\approx \tau_L/\tau_\eta\approx 400$
  obtained by direct numerical simulations with $2048^3$ collocation points, as described in
  \protect\cite{BEC2010}.}
\label{Turbulence:Fig_1}
\end{figure}
\begin{figure}[t]
\centering
\includegraphics[width=0.98\textwidth]{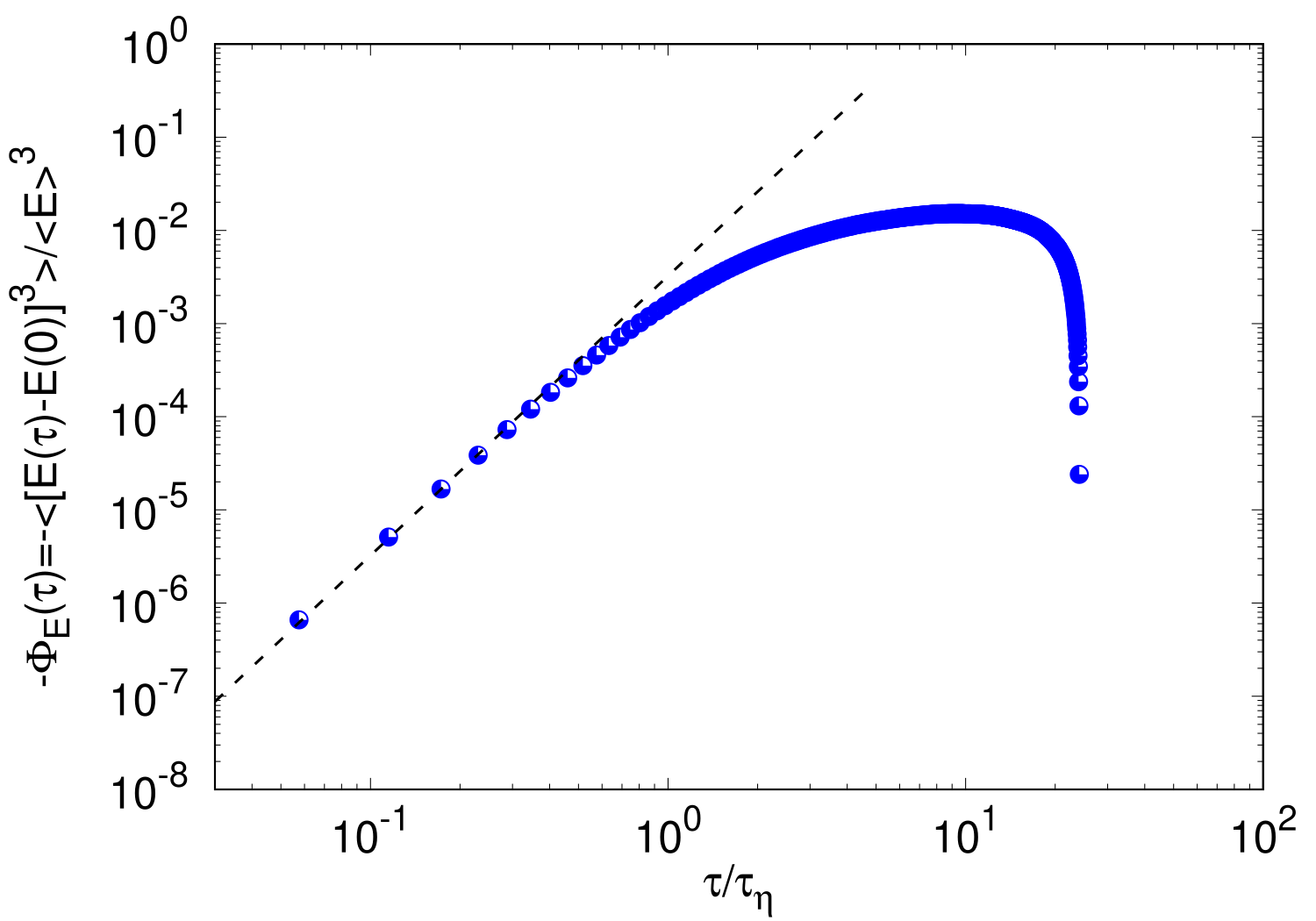}
\caption{Correlation function of the particle kinetic energy computed
  as the third moment of the energy increment, as a function of
  $\tau$, with inverted sign and time axis normalized by the
  Kolmogorov time $\tau_\eta$. The small time behavior is cubic in
  $\tau$, as shown by the dashed line. Data were taken from
  \protect\cite{databaseLagrangiano}, see also caption of
  Fig.~\ref{Turbulence:Fig_1}.}
\label{Turbulence:Fig_2}
\end{figure}

Figure~\ref{Turbulence:Fig_1} displays a qualitative demonstration of
a flight-crash event. In particular, Fig.~\ref{Turbulence:Fig_1}(a)
displays the 3D evolution of a tracer trajectory with color-coded the
rate of energy change, i.e. Lagrangian power $p(t) = dE(t)/dt =
\boldsymbol{A}(t)\cdot\boldsymbol{V}(t)$ ($\boldsymbol{A}(t) =
\dot{\boldsymbol{V}}(t)$). While Fig.~\ref{Turbulence:Fig_1}(b) shows
the time evolution of the kinetic energy with a zoom
(Fig.~\ref{Turbulence:Fig_1}(c)) in the region where energy slowly
increases and suddenly decreases. The dissimilarity between the rate of
increase and decrease of kinetic energy can be measured statistically
by the probability of observing an energy change after a
time increment $\tau$, $E(t+\tau)-E(t)$.  According to the
mechanism described above, the PDF of this quantity is expected to be skewed,
so that a good measure of irreversibility could be the non-dimensional third-order
moment 
\begin{equation}
    \Phi_E(\tau) =  \frac{\langle[E(t+\tau)-E(t)]^3 \rangle}{\langle E(t) \rangle^3}\,.
    \label{eq:phi_lagr}
\end{equation}
The above quantity, whose numerator, owing to statistical stationarity, can be expressed as the
asymmetric correlation function $\langle E(t+\tau)E^2(t)\rangle-\langle E^2(t+\tau)E(t)\rangle$
\cite{pomeau1982symetrie} (see also Sec.~\ref{sec:perneq}), is negative (Fig.~\ref{Turbulence:Fig_2})
and its modulus grows at short time separation as $\tau^3$. The short-time
behavior is indeed controlled by the third moment of the Lagrangian
power, $p(t)$, and the statistics of the latter has been shown to
display a non-trivially power law  dependence on the Reynolds number \cite{Xu2014flightcrash},
which was rationalized in terms of the multifractal model of turbulence \cite{cencini2017time}.

The statistical analysis of flight-crash events can thus be connected
with other statistical properties of turbulent flows, and it is quite
significant that they allow for detecting non-equilibrium by looking
at a very small portion (a single fluid element) of a turbulent flow.

\subsection{Coexistence of equilibrium and non-equilibrium in a shell model of turbulence}
In 3D turbulence, 
owing to the direct energy cascade as briefly summarized in the introduction to this
section, the energy flow from the scale of injection toward that of dissipation so that the range of scales in between has a clear non-equilibrium character. What happens at scales larger than that of injection is  less clear.
Numerical and experimental studies \cite{alexakis2019thermal,
  gorce2022statistical} support the view \cite{Frisch95} according to
which the statistics of these large scales are compatible with
statistical equilibrium, as for the Euler equation with ultraviolet
cutoff \cite{kraichnan1973helical}.  This is evidenced, for instance, by
energy equipartition taking place between these modes.  Yet recent
works \cite{hosking_schekochihin_2023, ding2023departure} showed that
an equilibrium description is not completely correct because
long-range interactions exist between small and large scales. Such
feature is responsible, e.g., for a dependence of the spectrum at
wavenumber smaller than the forced one from the specific forcing
employed \cite{ding2023departure}.

In principle, an answer to this riddle could be given by analyzing
suitable time correlation functions.  However, this is not an easy
task in direct numerical simulations of Eq.~(\ref{eq:NS}) for several
reasons including the fact that the phenomenon of sweeping (i.e. the
advection of small-scale eddies by larger-scale eddies) can complicate the behavior of
time correlations \cite{sweeping}.  For this reason, in
Ref.~\cite{cocciaglia2024nonequilibrium}, this problem was approached
in the framework of shell models in which, while retaining the main features of the
turbulent energy cascade, sweeping is absent by construction.

The key idea of shell models \cite{bohr1998,biferale2003shell,
  ditlevsen2010} is to consider a collection of $N$ interacting complex variables
$u_n$, $n=1,\dots,N$, 
associated with wavenumbers $k_n = k_0 2^{n-1}$ describing a sequence
of spherical shells in $k$-space with exponentially-growing radii.
The shell variables $u_n$ roughly represent velocity fluctuations at
length scales $\sim 1/k_n$.  Mimicking the Fourier representation of the
Navier-Stokes equation,  each shell variable evolves with an equation of the form
\begin{equation}
\dot{u}_n = ik_n Q[u,u] - \nu k_n^2u_n+ f_n\,,
\label{eq:shell}
\end{equation}
where the last two terms represent dissipation and forcing, while 
quadratic term $Q(u,u)$ models the nonlinear (advection) term of Eq.~\ref{eq:NS}.
In principle, $Q(u,u)$ should couple each mode to all the others in a way to preserve
the conservation laws of the Euler equation. In shell models, owing to the idea of locality
\cite{rosesulem} (i.e. that the most relevant interaction involves close-by wavenumbers),
only  nearest
and next-to-nearest neighbors interactions are retained (we
shall comment this assumption at the end of the Section). For Sabra shell
model \cite{Sabra}, the quadratic term reads
\begin{eqnarray}
Q[u,u]=2u_{n+2}u_{n+1}^*\!-\!\tfrac{1}{2}u_{n+1}u_{n-1}^*\!+\!\tfrac{1}{4}u_{n-1}u_{n-2}\,,
\label{eq:sabra}
\end{eqnarray} 
where $*$ denotes complex conjugation, and boundary conditions
$u_{k}=0$ for $k<1$ and $k>N$. The coefficients are chosen such that,
as for the Navier-Stokes equation, in the inviscid unforced case
($\nu=f_n=0$), two global conserved quantities exist, the total energy
$E = \sum_{n=1}^{N} e_n$ and total helicity $H = \sum_{n=1}^{N}
(-1)^{n} k_n e_n$, where:
\begin{equation}
    e_n=|u_n|^2/2
    \label{eq:e_n}
\end{equation} 
is the energy content of shell $n$.  Shell models have a multiscale
character, each shell variable having its own typical timescale. This
is needed to reproduce the large hierarchy of time- and length-scales
present in real turbulent flows.

As many studies have made clear,  shell models share many features with turbulent flows, such as
a well defined direct energy cascade with an energy flux $\varepsilon$, in
the shells in between forcing and dissipation, which on average equal the energy injection and dissipation rates. The flux out of shell $M$, $\Pi_M^{(E)}$, due to the nonlinear term can be computed as 
\begin{equation}
    \Pi_M^{(E)} =
    -\sum_{n=1}^M\frac{de_n}{dt} = k_M
    \text{Im}\left[2 u^*_M u^*_{M+1} u^{}_{M+2} + u^*_{M-1} u^*_{M}
      u^{}_{M+1} / 2\right]\,.
    \label{eq:flux}
\end{equation}
\begin{figure}[t]
\centering
\includegraphics[width=0.98\textwidth]{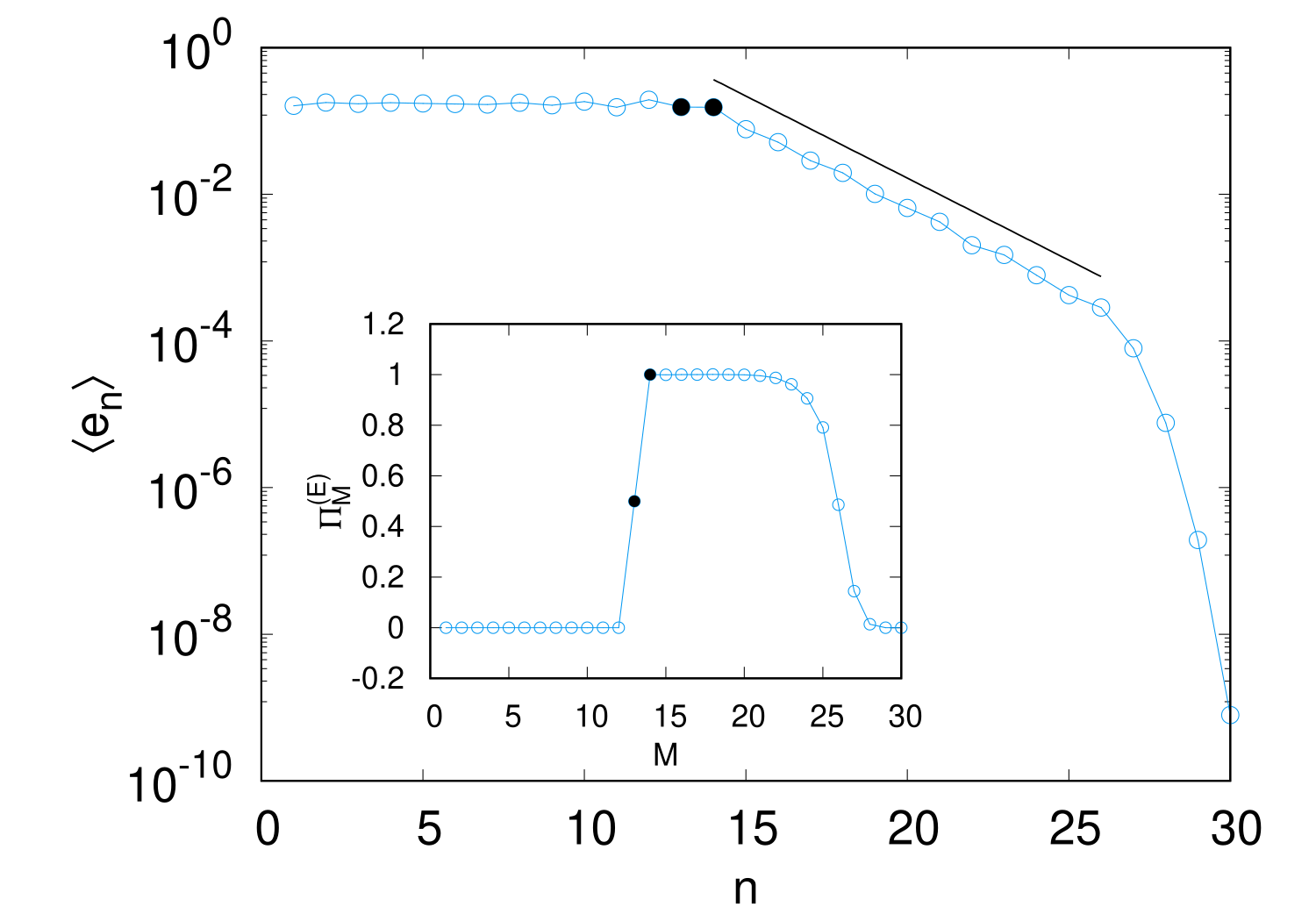}
\caption{Numerical simulation of Sabra shell model with $N=30$ shells
  and forcing on shells $n_f=\{13,14\}$ (filled circles). Main: energy
  spectrum, showing equipartition for $n<n_f$ and anomalous scaling
  $\langle e_n \rangle \sim k_n^{-\zeta(2)} \sim 2^{-n\zeta(2)}$ for
  $n>n_f$. $\zeta(2)=0.74(4)$ is the second-order anomalous exponent,
  and the black solid line indicates this scaling law. Inset: average
  energy flux out of shell $M$, which is zero for $n<n_f$ and equal to
  $\varepsilon=1$ for $n>n_f$.}
\label{Turbulence:Fig_3}
\end{figure}

\begin{figure}[t]
\centering
\includegraphics[width=0.98\textwidth]{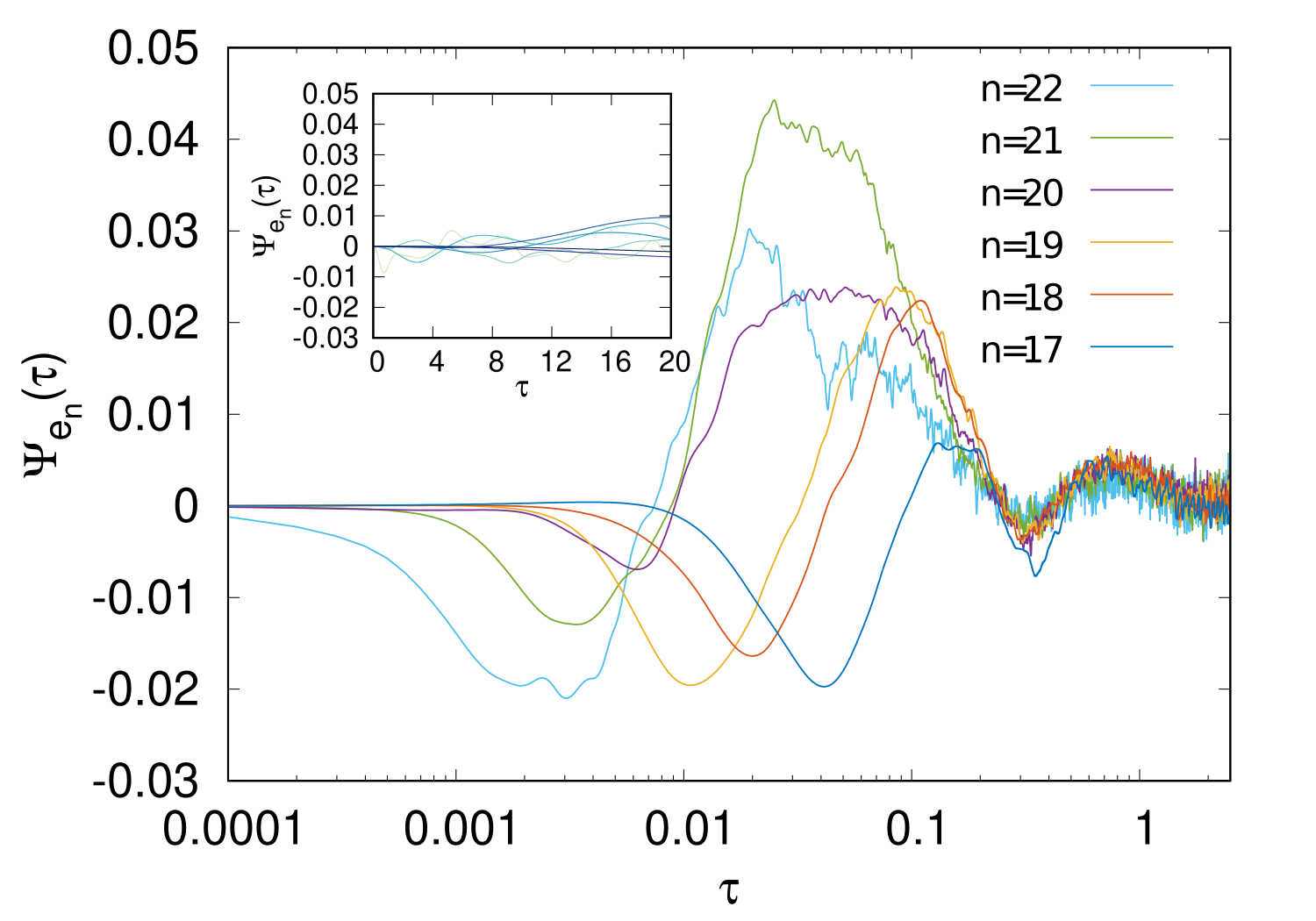}
\caption{Asymmetric time correlation functions \eqref{psi_en} for
  shells with index larger (main panel) and smaller (inset) than
  forced shells, which represent smaller and larger scales,
  respectively. The functions in the inset range from $n=5$ (darker)
  to $n=11$ (lighter).}
\label{Turbulence:Fig_4}
\end{figure}

Figure~\ref{Turbulence:Fig_3} shows the average energy per shell
$\langle e_n \rangle$, which in shell models is also the energy
spectrum. As clear from the figure, there is a clear difference
between shells larger or smaller (i.e. scales below or above) than the
two forced shells (indicated by filled symbols): larger ones are in
the range of direct energy cascade, and show the expected scaling
behavior of the energy spectrum (main panel) and a constant positive
energy flux (inset). Conversely, shells smaller than the forced ones
display energy equipartition and zero energy flux.
Asymmetric time correlation functions \cite{pomeau1982symetrie} have been then
studied to further inquire about the statistical character at scales above and below
the forcing scale. In particular, in Ref.~\cite{cocciaglia2024nonequilibrium}
it was studied
\begin{equation}
    \Psi_{e_n}(\tau)=\frac{\langle e_n^2(t)e_n(t+\tau) \rangle - 
    \langle e_n(t)e_n^2(t+\tau) \rangle}{\langle e_n^3(t)\rangle}\ ,
    \label{psi_en}
\end{equation}
in order to quantify time irreversible fluctuations shell by shell. Its behavior for shells
larger or smaller than forcing is shown in Fig.~\ref{Turbulence:Fig_4}.
The correlation functions in the cascade range (main panel), i.e. at
scales smaller than the forced ones, display a clear non-zero signal a
nontrivial dependence on $\tau$, which is linked to energy gains and
losses experienced by these shells
\cite{cocciaglia2024nonequilibrium}. Conversely, at scales larger than
the forced ones the correlations are compatible with zero as expected
for a reversible dynamics, i.e. at equilibrium.

Further insights into the different physics at scales larger/smaller
than the forcing scale can be obtained by studying how a perturbation
on the energy of a given shell influences the energy content of nearby
shells. This can be realized by inspecting the following ``energy
response functions":
\begin{equation}
    R_{m,n}(t) = \frac{\overline{\delta e_n}(t)}{\delta e_m}\ ,
    \label{eq:resp_en}
\end{equation}
in which the initial impulsive perturbation on shell $m$ is of the
order of typical energy fluctuations measured on that shell.
\begin{figure}[t]
\centering
\includegraphics[width=0.98\textwidth]{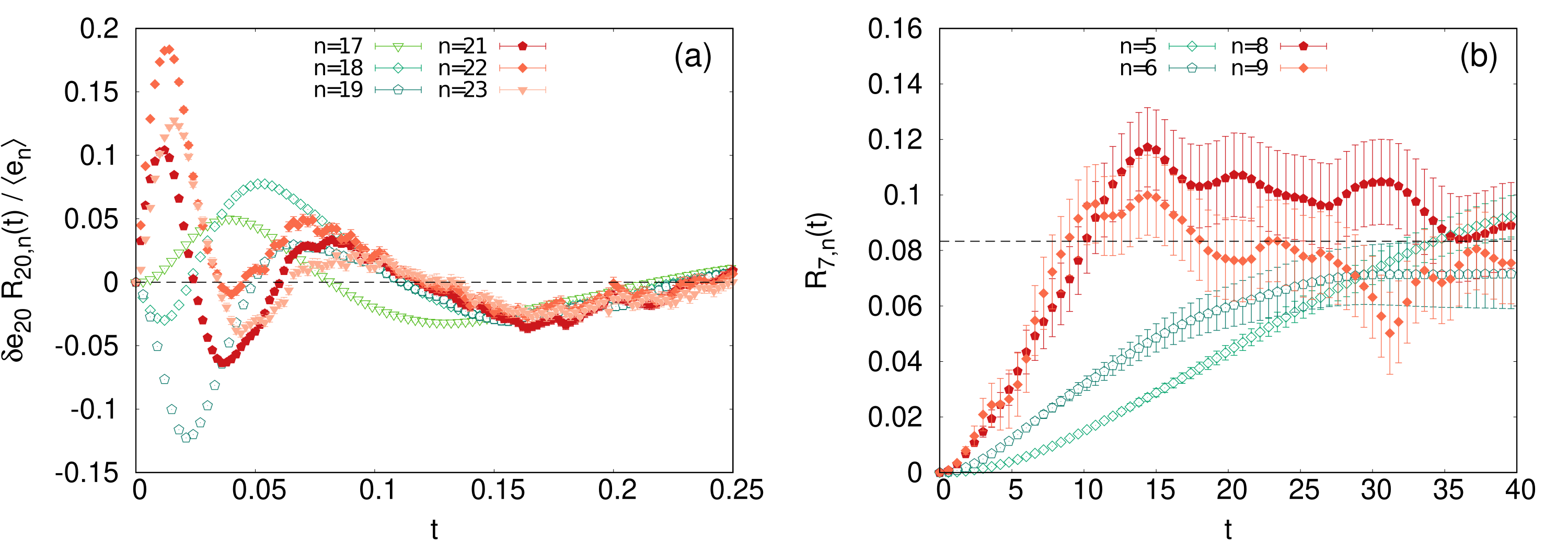}
\caption{Energy response functions \eqref{eq:resp_en} measured at
  shells larger (filled symbols, shades of red) or smaller (empty
  symbols, shades of green) than the perturbed shell.
(a) Rescaled response functions for shells in the cascading range:
  the perturbation acts at shell $m=20$ and the response is shown
  for three shells smaller (greenish curves) and larger (reddish curves) within the inertial range; (b) Response functions for
  shells in the equipartition range: the perturbation acts at shell $m=7$ and the response is shown
  for two shells smaller (greenish curves) and larger (reddish curves) within the equipartition range.}
\label{Turbulence:Fig_5}
\end{figure}
Fig.~\ref{Turbulence:Fig_5} displays a comparison between response
functions when perturbation and responses are in the cascade range
(panel (a)) or in the equipartition range (panel (b)).  The former
functions are rescaled as, $\delta e_m R_{m,n}(t) / \langle e_n
\rangle = \overline{\delta e_n}(t) / \langle e_n \rangle$, thereby
measuring the average energy deviation relative to the corresponding
steady-state value.  Dissimilarities between panels (a) and (b) are
recognizable: while the functions in the cascade range show initially
both positive and negative values, then followed by a decay to zero,
those in the equipartition range are all positive and reach a common
positive asymptotic value.  In a few words, the former behavior is
connected to the presence of an underlying energy flux directed toward
small scales, while the latter is easily explainable with equilibrium
statistical mechanics, assuming the small scales to form a
conservative (sub)system. For the cascade range,
it is also interesting to note the differences between the responses
at shell  larger and smaller than the perturbed shell, while the former initially gain in energy
the latter lose energy, an asymmetry which is absent in the equipartition range.
Such an asymmetry is reminiscent of those that have been discussed in Sec.~\ref{sec:causation}.
More details can be found in Ref.~\cite{cocciaglia2024nonequilibrium}. 

These results reveal that the shell model with
intermediate-scale forcing shows an interesting coexistence of
equilibrium and non-equilibrium properties, at scales respectively
larger and smaller than forcing.  We conclude with a word of caution
in extending these findings to real turbulent flows.  As discussed
earlier one of the assumptions used to build the shell model is that
in the quadratic term of\eref{eq:sabra} only local-in-scale
interactions are retained.  As discussed in
\cite{hosking_schekochihin_2023, ding2023departure} the main obstacle
in having statistical equilibrium at scales larger than the forcing
one is the presence of non-local interactions.  It would thus be very
desirable to devise a numerical study to inspect properly defined
correlation and/or response functions in direct numerical simulations
of the Navier-Stokes equation to test the equilibrium/non-equilibrium
character of these scales.

\section{Final Remarks and Conclusions
\label{sec:conclusion}}

In this review, we discussed several tools and methods for the characterization of different facets of the statistical features in 
non-equilibrium systems and some readers might feel that the matter was excessively scattered.
Such a feeling is somehow correct, however, it is not entirely our responsibility.
Rephrasing Lev Tolstoy famous incipit, we can say that
{\it  all the equilibrium systems share quite similar features, each non-equilibrium system has its own peculiarity}.
\\
In a nutshell, all the properties of an equilibrium system are embodied in the dependence of its partition function on temperature, pressure, applied fields, and other state variables.  
Conversely and unfortunately, in non-equilibrium systems, there is nothing similar to the simple and elegant recipes one can use in the equilibrium case. 
It is natural to wonder about the reasons behind such a significant difference.
\\
Let us  remind  the grounds of the (apparent) simplicity of the equilibrium statistical mechanics, we  have rather strong arguments:\footnote{Actually there is not general consensus on the conceptual aspects of the statistical mechanics, but this is not relevant for our discussion \cite{Chibbaro2014}}
once  the proper invariant distribution, i.e. the microcanonical one, has been chosen one has a way to  build  of a consistent mathematical theory.
On the contrary, in non-equilibrium system it is not easy at all, apart in few special cases, to determine the invariant distribution.
\\
By considering the features of a chaotic dissipative system, e.g., the Navier-Stokes equation, we can understand the great difficulties in building a theory for the statistical theory of NE systems.
Unlike the Hamiltonian cases, the asymptotic dynamics of a dissipative chaotic system will take place on a strange attractor whose invariant measure cannot be smooth, i.e. it is not continuous with respect to the Lebesgue measure,  and usually, it is described by a multifractal measure \cite{Cencini2009}.
This unpleasant technical problem can be prevented by introducing a small noise, such that the problem of the singularity of the invariant measure is removed.
Such a procedure is not only a mathematical trick, because it is quite natural to assume that any system is inherently noisy, e.g. due to the influence of the external environment~\cite{eckmann1985ergodic}. 
\\
Once the mathematical difficulties about the non-singular character of the invariant measure are removed, one has to find the invariant distribution of a corresponding Fokker-Planck equation of the system.
 This is surely a well-defined problem, but rather difficult even in low-dimensional systems. 
For instance, up to now, nobody has been able to find the invariant probability distribution of the Lorenz model with noisy terms. 
Therefore, we have that even for noisy chaotic dissipative systems there is not a well-defined (operative) protocol to determine the invariant probability. 
This is the first clear difference with Hamiltonian systems.
\\
In addition, the stationary distribution provides answers to some questions only, {\em i.e.} mean values; 
while to understand the dynamical features, such as correlation functions and responses, a more detailed ability is needed to master the system under investigation.
\\
Beyond the obvious technical difficulties of dealing with nonequilibrium, one has to treat a plethora of different problems that are absent in equilibrium cases.
In particular, in out-of-equilibrium high-dimensional systems, the interplay among different degrees of freedom or in general among their parts 
generates a wealth of features that cannot be described in a simple 
framework.
\\
The need of studying and measuring dynamical quantities implies a knowledge of the system along all of its characteristic time scales: such time scales can grow in number and reduce in separation, when the dimensionality of the system increases, making the analysis of irreversibility difficult or even impossible. Building upon this general problem, in this Review we have shown with several examples and approaches, the elusive nature of entropy production, which is usually considered a central quantity in characterising irreversibility.  Entropy production is a global quantity which requires knowledge of all important currents in a system, and can be severely underestimated if some part of the system is not accessible, for instance for lack of space or time resolution.  On the contrary, we have shown how a ‘skillful’ use of correlation and response functions in certain cases can be more informative and characterise non-equillibrium in the lack of a precise knowledge of the system. There are of course possible pitfalls, particularly in the case of linear systems where the coarse-graining may completely remove any signature of irreversibility, but there are also several interesting recent results which suggest how inference of the so-called “distance from non-equilibrium” in a system can be improved.

\section{Appendices} \label{sec:appendix}
\subsection{Entropy production rate for Markov processes}
The definition of entropy production rate $\Sigma$ for a continuous process is 
\begin{equation} \label{eq:entproddef2}
\Sigma = \lim_{\cT\to\infty} \ave{ \frac{1}{\cT} \ln{\frac{\text{Prob}(\Tdir)}{\text{Prob}(\Tinv)}} } = \lim_{\cT\to\infty} \Sigma^{(\cT)}
\end{equation}
where $\Tdir$ and $\Tinv$ denote the direct and inverse path respectively, and the average $\ave{\cdot}$ is taken with respect to the probability $\text{Prob}(\Tdir)$ of the direct path.
In the case of Markov processes it is possible to provide an expression for $\Sigma$ in terms of
the transition probability $\cW_t(\bx|\by)$ of going from state $\by$ to $\bx$ in a time $t$, normalized as
$$
\int d\bx\, \cW_{ t}(\bx|\by) = 1\,, 
$$
and of the stationary measure
$$
\pi(\bx) = \lim_{ t\to\infty}\cW_{ t}(\bx|\by)\,.
$$
In order to see this, we first divide the interval $[0,\cT)$ into $n$ sub-intervals of length $\Delta t=\cT/n$, and we approximate the probability of the direct $\Tdir$ and the inverse $\Tinv$ continuous-time trajectories with the probabilities of the discrete-time paths
\begin{align*}
    \text{Prob}(\Tdir) &\simeq P(\bx(0)=\bx_0,\bx(\Delta t)=\bx_1,...,\bx(\cT)=\bx_n) \\
    &= \pi(\bx_0)\prod_{i=1}^n \cW_{\Delta t}(\bx_i|\bx_{i-1})
\end{align*}
and
\begin{align*}    
    \text{Prob}(\Tinv) &\simeq  P\lx(\bx(0)=\bx_n,\bx(\Delta t)=\bx_{n-1},...,\bx(\cT)=\bx_0\rx) \\
    &= \pi(\bx_n)\prod_{i=1}^n \cW_{\Delta t}(\bx_{i-1}|\bx_i)\,.
\end{align*}
By replacing the above expressions into~\eref{eq:entproddef2}, we can compute a discrete-time approximation of $\Sigma^{(\cT)}$, which will in general depend on $\Delta t$. The exact value will be recovered at the end of the calculation by taking the limit $\Delta t\to 0$, keeping $\cT$ fixed. We get
\begin{align} \label{eq:entprodmarkov2}
\Sigma^{(\cT)}_{\Delta t} &= \frac{1}{n \Delta t}\int d\bx_0...d\bx_n  \pi(\bx_0)\prod_{i=1}^n \cW_{\Delta t}(\bx_i|\bx_{i-1}) \lx\{ \ln\frac{\pi(\bx_0)}{\pi(\bx_n)} + \sum_{i=1}^n \ln{\frac{\cW_{\Delta t}(\bx_i|\bx_{i-1})}{\cW_{\Delta t}(\bx_{i-1}|\bx_i)}} \rx\} = \nonumber \\
&= \frac{1}{\Delta t}\int d\bx \pi(\bx) \int d\by \cW_{\Delta t}(\by|\bx)\ln{\frac{\cW_{\Delta t}(\by|\bx)}{\cW_{\Delta t}(\bx|\by)}} = \\
&= \frac{1}{\Delta t}\int d\bx d\by\, P_{\Delta t}(\bx,\by) \ln\frac{P_{\Delta t}(\bx,\by)}{P_{\Delta t}(\by,\bx)} = \Sigma_{\Delta t} \nonumber
\end{align}
where $P_{\Delta t}(\bx,\by)=\pi(\bx)\cW_{\Delta t}(\by|\bx)$ is the joint probability of having $\bx$ at time $0$ and $\by$ at time $t$. In the computation above, we exploited the identity
$$
\int d\bx d\by\, \pi(\bx) \cW_{\Delta t}(\by|\bx) \lx( \ln{\pi(\by)}-\ln{\pi(\bx)} \rx) = 0
$$
and the chain rule $\pi(\bx)=\int d\bx\,\pi(\by) \cW_{\Delta t}(\bx|\by)$.
Note that, as expected, in the case of Markov processes $\Sigma^{(\cT)}_{\Delta t}$ does not actually depend on $\cT$. This means that the $\cT\to\infty$ limit appearing in the definition~\eqref{eq:entproddef2} is trivial.
The value of $\Sigma$ for the continuous-time case is then recovered in the limit $\Delta t\to 0$.

\subsection{Entropy Production Rate for Gaussian Linear Processes} \label{subsec:entgau}
Let us consider the $D$-dimensional linear stochastic dynamics
driven by a Gaussian noise, i.e.
\begin{align*}
\dot{\bx} + \matx{A}\bx &= \bxi \qquad \bxi \sim \cG_\matx{D}(\bxi)=\frac{e^{-\frac{1}{2}\bxi^T \matx{D}^{-1} \bxi}}{\sqrt{|2\pi \matx{D}|}}
\end{align*}
for which a direct computation allow us to write the evolution $\bx\to\by$ in a time $t$ as
\begin{align*}
\by = e^{-t \matx{A}} \bx + \bfeta\,,
\end{align*}
where $\bfeta$ is a Gaussian noise $\bfeta \sim\cG_{\matx{M}_{t}}(\bfeta)$ with covariance matrix 
$$
\matx{M}_{t}=\int_0^t ds\; e^{-s\matx{A}} \matx{D} e^{-s\matx{A}^T}\,.
$$
In this way, we have an explicit expression for the transition rate
\begin{align}
\cW_{t}(\by|\bx)= \int d\bfeta \, \cG_{\matx{M}_{t}}\lx(\bfeta\rx) \delta(\by - e^{-t \matx{A}}\bx - \bfeta) = \cG_{\matx{M}_{t}}\lx(\by-e^{-t\matx{A}}\bx\rx)
\end{align}
and then, since $\lim_{t\to\infty} \matx{M}_{t} = \matx{C}$, we have an explicit expression even for the stationary measure
\begin{align*}
\pi(\by)= \lim_{t\to\infty} \cW_{t}(\by|\bx) = \cG_\matx{C}(\by) = \frac{e^{-\frac{1}{2}\by^T \matx{C}^{-1}\by}}{\sqrt{\lx|2\pi \matx{C}\rx|}}\,.
\end{align*}
We can compute $\Sigma_{\Delta t}$ from \eref{eq:entprodmarkov2} by averaging with the joint probability
\begin{align}
\cP_{\Delta t}(\bx,\by) &= \pi(\bx)\cW_{\Delta t}(\by|\bx)=\cG_\matx{C}(\bx)\cG_{\matx{M}_{\Delta t}}\lx(\by-e^{-\Delta t\matx{A}}\bx\rx)
\end{align}
the logarithm of the ratio $\cP_{\Delta t}(\bx,\by)/\cP_{\Delta t}(\by,\bx)$ as prescribed in \eref{eq:entprodmarkov2}, i.e.
\begin{align*}
\Sigma_{\Delta t} &= 
\frac{1}{2\Delta t} \ave{\lx(\bx-e^{-\Delta t\matx{A}}\by\rx)^T \matx{M}_{\Delta t}^{-1}\lx(\bx-e^{-\Delta t \matx{A}}\by\rx) + \by^T\matx{C}^{-1}\by} + \\
&- \frac{1}{2 \Delta t}\ave{\lx(\by-e^{-\Delta t\matx{A}}\bx\rx)^T \matx{M}_{\Delta t}^{-1}\lx(\by-e^{-\Delta t\matx{A}}\bx\rx) + \bx^T\matx{C}^{-1}\bx} = \\
&= \frac{1}{\Delta t} \ave{\bx^T \lx( e^{-\Delta t\matx{A}^T} \matx{M}_{\Delta t}^{-1} -\matx{M}_{\Delta t}^{-1}e^{-\Delta t\matx{A}} \rx) \by} = \\
&= \frac{1}{\Delta t} \ave{\bx^T\lx( e^{-\Delta t\matx{A}^T} \matx{M}_{\Delta t}^{-1} -\matx{M}_{\Delta t}^{-1}e^{-\Delta t\matx{A}} \rx)e^{-\Delta t\matx{A}}\bx} = \\
&= \frac{1}{\Delta t}\text{Tr}\lx\{\lx( e^{-\Delta t\matx{A}^T} \matx{M}_{\Delta t}^{-1} -\matx{M}_{\Delta t}^{-1}e^{-\Delta t\matx{A}} \rx)e^{-\Delta t\matx{A}}\matx{C}\rx\} \,.
\end{align*}
In doing the above steps, we took advantage of the fact that the marginal distributions $\int d\by\,\cP_{\Delta t}(\bx,\by)$ and $\int d\bx\,\cP_{\Delta t}(\bx,\by)$ are identical, and they coincide with the stationary measure: this implies that several averages involving only one variable among $\bx$ and $\by$ cancel each other out, since they appear twice, with reversed signs.     
The last two steps are the consequence of
$$
\int d\by\, \cW_{\Delta t}(\by|\bx) \by = e^{-\Delta t\matx{A}}\bx
$$
and
$$
\ave{\bx^T \matx{B} \bx} = \text{Tr}(\matx{B}\matx{C})=\text{Tr}(\matx{C}\matx{B})\quad \forall \matx{B}\,.
$$
At the leading order of $\Delta t$, since $\matx{M}_{\Delta t} \simeq t \matx{D}$ and $e^{-\Delta t\matx{A}}\simeq 1-\Delta t\matx{A}$, we have
\begin{align*}
    \Sigma_{\Delta t} = \frac{1}{\Delta t}\lx\{ \text{Tr}\lx(\matx{D}^{-1}\matx{A}\matx{C}\rx) - \text{Tr}\lx(\matx{A}^T \matx{D}^{-1}\matx{C}\rx) \rx\} + \text{Tr}\lx\{(\matx{A}^T\matx{D}^{-1}-\matx{D}^{-1}\matx{A})\matx{A}\matx{C}\rx\} + \mathcal{O}(\Delta t)
\end{align*}
It can be seen that the term of order $1/\Delta t$ vanishes because of the cyclic property of the trace and the symmetry of $\matx{D}$ and $\matx{C}$:
\begin{align*}
\text{Tr}(\matx{D}^{-1}\matx{A}\matx{C})=\text{Tr}(\matx{C}\matx{D}^{-1}\matx{A})=\text{Tr}(\matx{A}^T \matx{D}^{-1}\matx{C})\,.
\end{align*}
Recalling the equality $\matx{A}\matx{C}+\matx{C}\matx{A}^T=\matx{D}$ valid for Orstein-Uhlenbeck processes~\cite{G90}, we finally get, taking the limit $\Delta t \to 0$,
\begin{align*}
\Sigma &= \text{Tr}\lx\{ \lx(\matx{C}\matx{A}^T-\matx{A}\matx{C}\rx) \matx{D}^{-1}\matx{A}\rx) = \text{Tr}\lx\{(\matx{A}^T\matx{D}^{-1}-\matx{D}^{-1}\matx{A})\matx{A}\matx{C}\rx\}\,.
\end{align*}
\subsection{Simulation algorithm for linear stochastic processes} \label{subsec:sims}
Linear stochastic processes, such as $\dot{\bx}+\matx{A}\bx=\xi+\zeta$ in which Poissonian random jumps $\zeta$ are added to a typical Gaussian noise $\xi$, can be exactly simulated, with an error due exclusively to the numerical precision of the calculator and the goodness of the random number generator.
The idea is based on the fact that, in the absence of jumps, the propagator $\cW_{t}(\bx'|\bx)$ from $\bx$ to $\bx'$ in a time $t$  is well known, it is Gaussian with mean $e^{-t\matx{A}}\bx$ and covariance matrix $\matx{M}_{t} = \int_0^t ds e^{-s\matx{A}} \matx{D} e^{-s\matx{A}^T}$.
This means that, regardless of the value of $t$ and in absence of jumps, we can obtain the value of $\bx'$ simply by adding to the value $e^{-t\matx{A}}\bx$ a random vector $\bz$ extracted from a normal distribution $\cG_{\matx{M}_{t}}(\bz)$. The presence of jumps it is only a small complication to this rule.
In fact, once the sampling frequency $f=1/\delta$ with which we want temporally discretize the trajectories of the system has been chosen, we just have to know in advance the time of the next jump and appropriately break the evolution into two pieces, before and after the jump, in the time interval at which the jump occurs.
Now we describe the algorithm in the general case in which the jumps affect only some components and the Poissonian rates and the distribution of the intensity with which these processes occur may be different between one component and another. All other cases, for example that of an equal rate for all components or a distribution of dependent jumps, should be simpler.
Let $\delta$ the sampling time, $\matx{A}$ the drift matrix, $\matx{D}$ the covariance matrix of Gaussian noise, $\cJ$ the set of component afflicted by jumps and $\lx\{\lambda_i,P_i(u)\rx\}_{i\in\cJ}$ the set of rates and intensity distribution of such jumps. First of all, we note that, for any value of $t$, we are able to numerically compute $\cR(t)=e^{-t\matx{A}}$ and $\matx{M}_{t}=\int_0^t ds\, e^{-s\matx{A}}\matx{D} e^{-s \matx{A}^T}$ through appropriate matrix diagonalizations and analytical simplification. Then we assume that we have the appropriate random number generators for all distribution we consider: the multi-normal one whatever the covariance matrix $\matx{M}_{t}$ is, the Poissonian one $\lambda e^{-\lambda t}$ and the set of $\lx\{P_i(u)\rx\}_{i\in\cJ}$ for jumps intensity.
With such assumptions the flow of the algorithm can be organized into following routines.
\begin{enumerate}
    \item[INIT] Set the initial state $\bx$, extract with probability $\lambda_i e^{-\lambda_i t_i}$ the times of the next jumps for all the components in $\cJ$, store such times in a set 
    $\cT = \lx\{ t_i \rx \}_{i \in \cJ}$, call UPDATE, then set $t=\delta$ and call SELECT ;
    \item[UPDATE] Find the index $k$ of the smallest time $\tau=t_k$ in the set $\cT$, delete $t_k$ from such set and shift all other remains times $t_i \to t_i-\tau$ ;
    \item[SELECT] if $\tau > \delta$ call GAUSS, print $\bx$ and set $t=\delta$, otherwise call JUMP ;
    \item[GAUSS] Iterate with recursion rule $\cR(t)\bx + \bz_{t} \to \bx$ with $\bz_{t} \sim \cG_{\matx{M}_{t}}(\bz_{t})$, set $\tau = \tau-t$ and call SELECT ;
    \item[JUMP] iterate with $\cR(\tau)\bx + \bz_\tau \to \bx$, then set $x_k =x_k + u$ where $u$ is extract from the distribution $P_k(u)$, set $t = t - \tau$ then extract a new time $t_k \sim \lambda_k e^{-\lambda_k t}$ for the index $k$, add this time to set $\cT$, call UPDATE then call SELECT. 
\end{enumerate}

\section*{Acknowledgments}
We thanks our collaborators on the issues here discussed: A. Baldassarri, L. Caprini, L. Cerino, A. Gnoli, U. Marini Bettolo Marconi, L. Peliti, A. Plati, C. Sarra and A. Sarracino.
In addition, we are grateful to L. Caprini and U. Marini Bettolo Marconi for their careful reading of the manuscript and useful remarks. 
M.B. and M.V. were supported by ERC Advanced Grant RG.BIO (Contract No. 785932). M.V. was also supported by MIUR FARE 2020 (project “INFO.BIO” No. R18JNYYMEY). A.P., A.V., F.C. and D.L. acknowledge the financial support from the MIUR PRIN 2017 (project “CO-NEST” No. 201798CZLJ). 

\bibliographystyle{ieeetr.bst}
\bibliography{bibliography}
\end{document}